\documentclass[a4paper,fleqn,usenatbib,useAMS]{mnras}

\usepackage{graphicx}	
\usepackage{amsmath}	
\usepackage{amssymb}	
\usepackage{multicol}        
\usepackage{bm}		
\usepackage{pdflscape}	

\newcommand{\kms}{km\,s$^{-1}$} 
\newcommand{\cmss}{cm\,s$^{-2}$}
\newcommand{\fluxcgs}{erg\,s$^{-1}$\,cm$^{-2}$}
\newcommand{\hdstar}{HD\,122563}
\newcommand{\stagger}{{\sc Stagger}}
\newcommand{\scate}{{\sc SCATE}}
\newcommand{\atmo}{{\sc Atmo}}
\newcommand{\vald}{{\sc VALD}}
\newcommand{\hipparcos}{{\sc Hipparcos}}
\newcommand{\teff}{{T_\text{eff}}}
\newcommand{\diff}{{\text{d}}}
\newcommand{\loggf}{{\log{g\!f}}}
\newcommand{\abund}[1]{{\log\epsilon_\text{\,#1}}}

\usepackage[T1]{fontenc}
\usepackage{ae,aecompl}

\usepackage{newtxtext,newtxmath}

\title[{\hdstar}: A 3D abundance analysis of CNO]{The benchmark halo giant {\hdstar}: CNO abundances revisited with three-dimensional hydrodynamic model stellar atmospheres}

\author[R. Collet et al.]{R. Collet,$^{1,2}$\thanks{E-mail: remo@phys.au.dk (RC)}
{\AA}. Nordlund,$^{3}$ M. Asplund,$^{2}$ W. Hayek$^{4}$ {and} R. Trampedach$^{1,5}$ 
\\
$^{1}$ Stellar Astrophysics Centre, Department of Physics and Astronomy, Ny Munkegade 120, Aarhus University, DK-8000 Aarhus C, Denmark \\
$^{2}$ Research School of Astronomy \& Astrophysics, Australian National University, Cotter Road, Canberra ACT 2611, Australia \\
$^{3}$ Centre for Star and Planet Formation, Niels Bohr Institute and Natural History Museum of Denmark, University of Copenhagen, {\O}ster Voldgade 5-7, \\ DK-1350 Copenhagen, Denmark \\
$^{4}$ National Institute of Water and Atmospheric Research, 301 Evans Bay Parade, Hataitai, Wellington 6021, New Zealand \\
$^{5}$ Space Science Institute, 4750 Walnut Street, Suite 205, Boulder, CO 80301, USA}

\date{Last updated 2017 December 21; in original form 2017 August 18}

\pubyear{2017}

\begin{document}
\label{firstpage}
\pagerange{\pageref{firstpage}--\pageref{lastpage}}
\maketitle

\begin{abstract}
We present an abundance analysis of the low-metallicity benchmark red giant star {\hdstar} based on realistic, state-of-the-art, high-resolution, three-dimensional (3D) model stellar atmospheres including non-grey radiative transfer through opacity binning with four, twelve, and 48 bins.
The 48-bin 3D simulation reaches temperatures lower by $\sim 300-500$\,K than the corresponding 1D model in the upper atmosphere. Small variations in the opacity binning, adopted line opacities, or chemical mixture can cool the photospheric layers by a further ${\sim}100-300$\,K and alter the effective temperature by ${\sim}100$\,K.

A 3D local thermodynamic equilibrium (LTE) spectroscopic analysis of \ion{Fe}{i} and \ion{Fe}{ii} lines gives discrepant results in terms of derived Fe abundance, which we ascribe to non-LTE effects and systematic errors on the stellar parameters.
We also determine C, N, and O abundances by simultaneously fitting CH, OH, NH, and CN molecular bands and lines in the ultraviolet, visible, and infrared.
We find a small positive 3D$-$1D abundance correction for carbon ($+0.03$\,dex) and negative ones for nitrogen ($-0.07$\,dex) and oxygen ($-0.34$\,dex).
From the analysis of the [\ion{O}{i}] line at $6300.3$\,{\AA}, we derive a significantly higher oxygen abundance than from molecular lines
($+0.46$\,dex in 3D and $+0.15$\,dex in 1D).
We rule out important OH photodissociation effects as possible explanation for the discrepancy and note that lowering the surface gravity would reduce the oxygen abundance difference between molecular and atomic indicators.
\end{abstract}



\section{Introduction}
\label{sect:intro}
{\hdstar} is the brightest known metal-poor halo giant star ($V = 6.19$, \citealt{Ducati:2002}; [Fe/H] $\approx -2.7$,\footnote{[A/B] $\equiv \log \left( \frac{n_\text{A}}{n_\text{B}} \right) - \log \left( \frac{n_\text{A}}{n_\text{B}} \right)_\odot$, where $n_\text{A}/n_\text{B}$ and $(n_\text{A}/n_\text{B})_\odot$ are the number densities of element A relative to element B in the star and the Sun, respectively.} \citealt{Barbuy:2003}). 
Since its discovery by \citet{Wallerstein:1963}, {\hdstar} has been the subject of numerous spectroscopic analyses \citep[e.g.][]{Westin:2000,Barbuy:2003,Cowan:2005,Aoki:2007,Afsar:2016,Dobrovolskas:2015,Prakapavicius:2017}; a wealth of photometric data is also available for this star in a variety of filter systems \citep[e.g.][]{Hauck:1998,Ducati:2002,Cutri:2003}. 
Owing to its vicinity to the Sun, {\hdstar} has accurate parallax measurements \citep{van-Leeuwen:2007} and interferometric stellar diameter determinations \citep{Creevey:2012}. 
As a result, it is generally included in lists of benchmark stars for the calibration and validation of stellar parameter and abundance determinations in stellar surveys \citep{Casagrande:2014,Heiter:2015}.

For the above listed reasons, {\hdstar} has de facto become a template for low-metallicity giants in stellar structure and evolution studies \citep{VandenBerg:2014}. 
An accurate determination of its composition is of great significance in the context of Galactic chemical evolution where elemental abundances in {\hdstar} are effectively regarded as a yardstick for spectroscopic analyses of metal-poor red giant stars.

The vast majority of abundance analyses of {\hdstar} are based on classical, one-dimensional (1D), stationary, hydrostatic model stellar atmospheres. 
These kind of models are especially limited in their description of convective energy transport in the deeper layers of late-type stellar atmospheres as well as of the effects of convective overshooting on the temperature-density stratification and associated velocity field in the upper photosphere. 
Convection is typically treated by means of approximate recipes such as mixing-length theory (MLT, \citealt{Bohm-Vitense:1958}) or full-spectrum of turbulence models (FST, \citealt{Canuto:1991}), all dependent on a variety of tuneable free parameters. 
Atmospheric velocity fields and temperature and density inhomogeneities are also not contemplated in the upper atmospheric layers of such models because of the assumption of 1D geometry and hydrostatic equilibrium.
When using 1D model stellar atmospheres for computing synthetic spectra and for abundance analysis purposes it is therefore necessary to introduce additional tuneable broadening parameters such as micro- and macro-turbulence in order to compensate for the lack of velocity fields in the theoretical description and to be able to reproduce the observed strengths of spectral lines.

In recent years, a more comprehensive approach to modelling late-type stellar atmospheres that relies on realistic ab initio time-dependent, three-dimensional (3D), hydrodynamic simulations of stellar surface convection has steadily been gaining ground \citep{Asplund:2005}.
The primary advantage of such simulations is that convection and atmospheric flows emerge naturally from the numerical solution of the conservation equations without the need to introduce dedicated adjustable parameters.
Grids of 3D model atmospheres of late-type stars are currently being developed with direct applications to 3D spectral line formation and abundance analyses \citep{Magic:2013,Tremblay:2013,Trampedach:2013,Beeck:2013a}.
Such 3D model atmospheres can be used as the foundation for more realistic stellar abundance analyses, overcoming many of the limitations of traditional studies based on 1D models.

One of the main results from 3D hydrodynamic stellar surface convection simulations is the prediction of systematically cooler mean photospheric temperature stratifications (up to several hundred kelvins) for low-metallicity late-type stars compared with 1D models with the same atmospheric parameters \citep{Asplund:1999,Collet:2007}.
The low temperatures in the upper layers of metal-poor 3D simulations have the effect to boost the overall number density of molecules under chemical equilibrium conditions and produce generally stronger molecular spectral lines with respect to calculations based on classical models.
As a consequence, spectroscopic analyses of metal-poor stars based on 3D models typically result in significantly lower abundances for the trace elements forming those molecules \citep{Asplund:2001,Collet:2007,Ivanauskas:2010}.
The large 3D$-$1D abundance corrections call for verification and validation tests of 3D model stellar atmospheres and their spectra, which is best carried out with an analysis of benchmark stars with reliable stellar parameter determinations.

Few spectroscopic analyses specifically aimed at the template metal-poor giant {\hdstar} and relying on 3D model stellar atmospheres have been carried out up to now.
\citet{Collet:2009} performed a differential 3D$-$1D abundance analysis of Fe and infrared (IR) OH lines in {\hdstar} using a tailored 3D simulation of the star.
\citet{Ramirez:2010} adopted an updated version of that 3D simulation to compute asymmetries and core wavelength shifts of \ion{Fe}{i} lines to probe surface granulation signatures and found excellent agreement with high-resolution spectroscopic observations.
\citet{Bergemann:2012} extracted the mean temperature and density stratification from the 3D simulation of {\hdstar} and applied it to a 1D non-local thermodynamic equilibrium (non-LTE) Fe abundance analysis.
More recently, \citet{Dobrovolskas:2015} derived oxygen abundances in four very metal-poor stars including {\hdstar} by means of an analysis of OH IR lines based on 3D model stellar atmospheres.
\citet{Prakapavicius:2017} used an updated version of the 3D model by \citet{Dobrovolskas:2015} to derive the oxygen abundance in {\hdstar} by means of a spectroscopic analysis of OH lines in the ultraviolet (UV).

Here, we present a revisited in-depth spectroscopic analysis of {\hdstar}'s composition in the light of realistic 3D stellar surface convection simulations and 3D local thermodynamic (LTE) spectral line formation. We also perform a differential comparisons with the results of a classical derivation of elemental abundances based on 1D model atmospheres.
In particular, we focus on the determination of carbon, nitrogen and oxygen abundances based on the simultanoeus fitting of molecular bands and individual molecular lines in {\hdstar}'s spectrum.

\section{3D hydrodynamic model stellar atmospheres}
\label{sect:hydro}

\subsection{Code}
\label{sect:code}
We carry out a series of numerical 3D hydrodynamic simulations of convection at the surface of the metal-poor red giant {\hdstar}.
We use a custom version of the 3D, time-dependent, radiation-magnetohydrodynamic (MHD) {\stagger} code \citep{Nordlund:1994,Nordlund:1995,Nordlund:2009,Collet:2011a}. 
In the code, the equations for the conservation of mass, momentum, and energy for compressible flows together with the induction equation are discretised using a high-order finite-difference scheme (sixth-order numerical derivatives, fifth-order interpolations) and solved on a rectangular Eulerian mesh, extending over a representative volume of stellar envelope around the optical surface and including the photospheric layers as well as the upper part of the stellar convection zone.
The solution of the equations is advanced in time using a third-order Runge-Kutta integration scheme.
The adopted mesh is \emph{staggered}: scalar, thermodynamic variables --density, internal energy per unit volume, and temperature-- are cell-centred, while the momentum (mass density flux) and magnetic field components are face-centred. 
By design, the {\stagger} derivative operators return the value of derivatives half a cell up or down relative to the input variables: this is where the values of the derivatives of the various quantities are often needed in the MHD equations expressed in conservative form, and using a staggered mesh ensures the accuracy of the calculations is highest at these locations.  
In all other cases, the high-order {\stagger} interpolation operator can be used to return the value of a variables or its numerical derivative half a cell back at the cost of losing only little accuracy.
While the {\stagger} code is capable of solving the full MHD equations, we have ignored for the time being all MHD effects and set the magnetic field components to zero.
Boundary conditions are periodic in the horizontal directions and open in the vertical.
At the bottom boundary, we enforce constant gas pressure (in both time and space) and assume the inflowing gas to have constant entropy per unit mass. Outflowing gas at the top and bottom is free to carry entropy fluctuations out of the simulation domain.

\subsection{Stellar parameters}
\label{sect:stellar-param}

\begin{table}
\centering
\caption{Adopted background chemical composition of {\hdstar} (17 most abundant elements) for computing ionisation and molecular equilibria and continuous opacities in spectral synthesis calculations. 
Elemental abundances are expressed on the conventional logarithmic scale where $\abund{H} = 12.00$, by definition. 
Abundances are taken from \citep{Barbuy:2003} when available, complemented with values from a scaled solar chemical composition with [Fe/H] $= -2.7$ for other elements. 
Note that in the case of molecular lines and bands, the abundances of C, N, and O are varied consistently with the values adopted in line formation calculations. }
\label{tab:bg-abund}
\begin{tabular}{lccccc}
	\hline
	Element  & $\log\epsilon$ & Element  & $\log\epsilon$ & Element  & $\log\epsilon$ \\
	\hline
	H  & $12.00$  &  Na & $3.68$  &  K  & $2.79$ \\
	He & $10.93$  &  Mg & $5.14$  &  Ca & $3.82$ \\
	C  & $5.12$   &  Al & $3.14$  &  Cr & $2.40$ \\
	N  & $6.32$   &  Si & $5.22$  &  Fe & $4.79$ \\ 
	O  & $6.63$   &  S  & $4.33$  &  Ni & $3.50$ \\
	Ne & $5.08$   &  Ar & $3.40$  &  &  \\
	\hline
\end{tabular}
\end{table}

\subsubsection{Effective temperature}
\label{sect:teff}
We adopt an effective temperature of ${\teff} = 4\,600 \pm 47$\,K for {\hdstar}, as determined by \citet{Casagrande:2014} using the infrared flux method (IRFM).
This value is in excellent agreement with the effective temperature of ${\teff} = 4\,598 \pm 41$\,K derived by \citet{Creevey:2012} and based on a combination of accurate interferometric stellar diameter information and photometry, and is consistent with the value ${\teff} = 4\,587 \pm 60$\,K from the compilation of fundamental stellar parameters by \citet{Heiter:2015}, also based on stellar diameter and bolometric flux measurements.
It is also in very good agreement with the values in the range ${\teff} \approx 4\,600$--$4\,665$\,K adopted by previous spectroscopic analyses \citep[e.g.][]{Barbuy:2003,Aoki:2007,Bergemann:2012}.

\subsubsection{Surface gravity}
We derive {\hdstar}'s surface gravity $g$ using the fundamental relation \citep[e.g.][]{Nissen:1997}:
\begin{equation}
\log \frac{g}{g_{\sun}} = \log\frac{M}{M_{\sun}} +4\log\frac{{\teff}}{{\teff}_{\sun}} -  \log\frac{\mathcal{F}_\text{bol}}{\mathcal{F}_{\text{bol,}\sun}}+2\log{\hat{\pi}}
\label{eq:logg}
\end{equation}
where $M$ is the stellar mass, ${\teff}$ the effective temperature, $\mathcal{F}_\text{bol}$ the star's bolometric flux, and $\hat{\pi}$ the star's parallax in radians; the subscript $\sun$ refers to the solar values of the variables.
We adopt $M = 0.855 \pm 0.025$\,$M_{\sun}$ for the mass of {\hdstar}, as estimated by \citet{Creevey:2012}. 
We derive the star's bolometric flux using the fundamental relation $\mathcal{F}_\text{bol} = \sigma{\teff}^4{\cdot}(\theta/2)^2$ where $\sigma$ is the Stefan-Boltzmann constant and $\theta$ is the angular diameter of the star in radians as seen from Earth.
We combine the value $\theta = 0.940 \pm 0.011$\,mas for the angular diameter as derived by \citet{Creevey:2012} using interferometric observations and a limb-darkening relation from 3D convection simulations with the value of the effective temperature ${\teff} = 4\,600$\,K determined by \citet{Casagrande:2014} to obtain a bolometric flux $\mathcal{F}_\text{bol} = 1.318{\cdot}10^{-7}$\,{\fluxcgs} for {\hdstar}.
We adopt a value of $\hat{\pi} = 4.22 \pm 0.35$\,mas for {\hdstar}'s parallax as listed in the revised {\hipparcos} data catalogue \citep{van-Leeuwen:2007}.
Finally, for the solar bolometric flux and the Sun's effective temperature we adopt a total solar irradiance at $1$\,au of $\mathcal{F}_{\text{bol},\sun} = 1.361{\cdot}10^{6}$\,{\fluxcgs} \citep{Kopp:2011} and ${\teff}_{\sun} = 5\,771.8$\,K,\footnote{This value for the solar effective temperature is derived consistently with the adopted solar bolometric flux and with the solar radius determination of $R_{\sun} = 695.66 \pm 0.14$\,Mm by \citet{Haberreiter:2008}} respectively.
From Eq.~\ref{eq:logg} we obtain a value of $\log{g} = 1.61 \pm 0.07$\,({\cmss}) for {\hdstar}'s surface gravity, consistent with interferometric \citep{Creevey:2012} and spectroscopic \citep[e.g.][]{Mashonkina:2011,Bergemann:2012} derivations and with the recommended value by \citet{Heiter:2015}.
At present, no asteroseismic determinations of the surface gravity of {\hdstar} are available with which we can compare our derived value.

\subsubsection{Reference metallicity from the literature}
\label{sect:metallicity}
The star has an overall low metallicity of [Fe/H] $\approx -2.7$ \citep{Jofre:2014}, a carbon under-abundance of [C/Fe] $\approx -0.6$, and an oxygen over-abundance of [O/Fe] $\approx +0.6$ \citep{Barbuy:2003}.
It is also enhanced in nitrogen by a factor of [N/Fe] $\approx +0.6$ \citep{Aoki:2007}.
For our spectroscopic analysis of {\hdstar}, we adopt the chemical composition derived by \citet{Barbuy:2003} as reference when computing ionisation and molecular equilibria and background continuous opacities in the line formation calculations.
The adopted abundances of the 17 most important elements are listed in Table~\ref{tab:bg-abund}.

\begin{table*}
	\centering
	\caption{List of parameters of the different {\hdstar} simulations considered in the present work. The binning criterion used to quantify opacity strength at wavelength $\lambda$ is either the Rosseland optical depth $\tau_\text{Ross}$ or the ratio of monochromatic and Rosseland extinction coefficients at which monochromatic optical depth $\tau_\lambda = 1$. 
	Two different solar mixtures are considered, from \citet{Asplund:2009} (AGSS09) and from \citet{Grevesse:1998} (GS98).
	The underlying line opacity data used in the case of the scaled AGSS09 and GS98 mixtures are based on opacity sampling (OS) and opacity distribution functions (ODFs), respectively. The time-averaged effective temperatures of the simulations and associated standard deviations are given in the last column. (As explained in the main text, the emergent radiative flux --hence the effective temperature-- of a relaxed 3D stellar surface convection simulation of the kind considered in this work is not constant over time but fluctuates around a mean value.) }
	\label{tab:params}
	\begin{tabular}{lccccccr}
		\hline
		Resolution & Solar mix & [Fe/H] & [$\alpha$/Fe] & \# of bins & binning criterion & time [h] & ${\teff}$ [K] \\
		\hline
		$504^2{\times}252$  & AGSS09 & $-2.5$ & $+0.4$ & 48 & $\tau_\text{Ross}$ & $600$ & $4\,662 \pm 11$ \\
		$504^2{\times}252$  & AGSS09 & $-2.5$ & $+0.4$ & 12 & $\tau_\text{Ross}$ & $560$ & $4\,663 \pm 12$ \\
		$504^2{\times}252$  & AGSS09 & $-2.5$ & $+0.4$ & 12 (alt) & $\tau_\text{Ross}$ & $600$ & $4\,663 \pm 10$ \\
		$504^2{\times}252$  & AGSS09 & $-2.5$ & $+0.4$ & 4 & $\tau_\text{Ross}$ & $310$ & $4\,689 \pm 11$ \\
		$504^2{\times}252$  &  GS98  & $-2.5$ & $+0.0$ & 4 & $\tau_\text{Ross}$ & $280$ & $4\,752 \pm 15$ \\
		$504^2{\times}252$  &  GS98  & $-2.5$ & $+0.0$ & 4 & $\chi_{\!\lambda}/\chi_{\text{Ross}}$ & $270$ & $4\,754 \pm 15$ \\
		$504^2{\times}252$  & AGSS09 & $-3.0$ & $+0.4$ & 4 & $\chi_{\!\lambda}/\chi_{\text{Ross}}$ & $410$ & $4\,785 \pm 12$ \\
		$504^2{\times}252$  &  GS98  & $-3.0$ & $+0.0$ & 4 & $\chi_{\!\lambda}/\chi_{\text{Ross}}$& $330$ & 4\,809$ \pm $11 \\
		$504^2{\times}252$  &  GS98  & $-3.0$ & $+0.0$ & 4 & $\tau_\text{Ross}$ & $330$ & 4\,805$ \pm $11 \\
		$1008^2{\times}504$ & AGSS09 & $-2.5$ & $+0.4$ & 48 & $\tau_\text{Ross}$ & $90$ & $4\,667 \pm 8$ \\
		$1008^2{\times}504$ & AGSS09 & $-2.5$ & $+0.4$ & 12 & $\tau_\text{Ross}$ & $320$ & $4\,658 \pm 11$ \\
		$1008^2{\times}504$ & AGSS09 & $-2.5$ & $+0.4$ & 12 (alt) & $\tau_\text{Ross}$ & $200$ & $4\,658 \pm 11$ \\
		\hline
	\end{tabular}
\end{table*}

\subsection{Simulation parameters}
For the main hydrodynamic simulations, as default, we have adopted a scaled solar composition \citep{Asplund:2009} with [Fe/H] $= -2.5$ and an $\alpha$-enhancement of [$\alpha$/Fe] $= +0.4$, very close to the \citet{Barbuy:2003} reference mixture.
We have also run a number of test simulations at [Fe/H] $= -3.0$ to study the sensitivity of the temperature stratification to changes in metallicity (more details are given in \S~\ref{sect:microphys}).
We have assumed a gravitational acceleration with a constant value of $\log{g} = 1.6$\,({\cmss}) along the vertical direction throughout the simulation box.
The selected simulation domain is $3\,750$\,Mm~$\times$~$3\,750$\,Mm wide and $1\,230$\,Mm deep, i.e. large enough to allow for about ten granules to coexist at any one time and deep enough to span six pressure scale heights above the stellar optical surface and four pressure scale heights below in the case of a red giant with the stellar parameters of {\hdstar}. 
In terms of Rosseland optical depth, this corresponds to the range $-6 \la \log\tau_\text{Ross} \la 7$.
Across the depth of the simulation, which corresponds to $7$\,\% of {\hdstar}'s radius\footnote{The estimated radius for {\hdstar} is $R = \sqrt{{G M}/{g}} \approx \,1.7{\cdot}10^4$\,Mm, while the thickness of the simulation domain and of the photosphere are approximately $1.2{\cdot}10^3$\,Mm and $4.5{\cdot}10^2$\,Mm, respectively.}, the actual gravitational acceleration is expected to vary by approximately $14$\,\%. 
However, we consider this variation negligible for the purpose of spectral line formation calculations with the present simulations: over the thickness of the visible photosphere, where the emergent stellar spectrum forms, the gravitational acceleration varies only by about $5$\,\%.
Finally, for the present simulations, we also neglect sphericity effects.

We have adequately adjusted the bottom boundary values for the internal energy and density of the inflowing gas to target an effective temperature of approximately $4\,600$\,K.
We would like to note here that the effective temperature of the simulations is not a fixed input parameter, but, on the contrary, a function of the entropy of the inflowing gas at the bottom of the simulations as well as of the stellar chemical composition and surface gravity. 
Furthermore, the total emergent radiative flux at the stellar surface --i.e. the effective temperature--  is not constant in time but fluctuates slightly around a mean value as a result of the atmospheric hydrodynamic flows, p-mode-like oscillations, and limited extent of the simulation domain that only covers a small fraction of the entire stellar surface.
Attaining a particular value for a simulation's mean effective temperature to within a desired accuracy requires a significant degree of fine tuning of the bottom boundary entropy value followed by long simulation runs to achieve full relaxation\footnote{We consider a simulation \emph{relaxed} when it no longer shows any variation over time of in key physical quantities such as the mean effective temperature and mean internal energy per unit mass and mean density at the bottom boundary, and when $p$-mode-like oscillations have been naturally damped to their minimum level \citep[see also][]{Magic:2013}.}, which is computationally a very time-consuming procedure.
All our simulations have relaxed at slightly hotter effective temperatures than the targeted $4\,600$\,K (see Table~\ref{tab:params}); 
however, given that for the main simulations used in our abundance analysis the differences in ${\teff}$ from the targeted value are comparable to the observational errors, we regard such deviations as minor.
Also, these differences become less important when our abundance results are compared differentially with the ones from analyses based on 1D model atmospheres computed for the same stellar parameters.

\subsubsection{Initial simulation state}
We generate the initial simulation states by scaling snapshots from an earlier surface convection simulation of {\hdstar} by \citet{Collet:2009} to the updated stellar parameters and targeted effective temperature.
Scaling relations for the mean stratification of the various physical variables with depth as well as for the spatial dimensions are derived with the aid of 1D model envelopes constructed for the same stellar parameters as the simulations. To first order, the level of 3D inhomogeneities with respect to the mean stratification in the scaled snapshot is assumed to be the same as in the original simulation.
After the scaling, bottom boundary conditions are adjusted consistently with the scaled 3D structure. 
We then run the simulations for several convective turn-over time-scales until thermal and dynamic relaxation is achieved. 
If a simulation's effective temperatures differs significantly from the targeted one, the scaling process is repeated anew until the desired accuracy is reached.

\subsubsection{Numerical resolution}
We have run simulations at increasingly higher numerical resolutions, from ${480}^2{\times}240$ for the initial simulations after scaling to ${504}^2{\times}{252}$ and ${1008}^2{\times}{504}$ for the production runs.
While more computationally expensive, the higher numerical resolution simulations allow us to capture finer turbulent structures and produce a better representation of the distribution of flow velocities in the stellar atmosphere \citep[e.g.][]{Asplund:2000c}.
The main motivation behind running such high-resolution simulations is to verify whether we are able to resolve the relevant  scales and velocity structures that affect the strength of the predicted spectral diagnostics.

\subsection{Radiative transfer}
\label{sect:radtran}
An accurate representation of the energy exchange between radiation and gas is crucial to properly model the thermal stratification in the stellar atmosphere and in the transition region between the photosphere and the upper convective zone.
We account for the energy exchange between radiation and matter via a radiative heating term in the equations. 
More precisely, the gas internal energy $e$ per unit volume varies with time according to the conservation equation
\begin{equation}
\frac{\partial e}{\partial t} = -\nabla\!\cdot\!e\,\mathbfit{u} - P\,\nabla\!\cdot\!\mathbfit{u} + Q_\text{rad} + Q_\text{visc},
\label{eq:dedt}
\end{equation}
where $\mathbfit{u}$ is the bulk gas velocity, $P$ is the gas pressure, and $Q_\text{visc}$ and $Q_\text{rad}$ are the viscous dissipation and radiative heating rates, per unit volume respectively.
The radiative heating rate at a given physical location is computed by integrating the difference between monochromatic intensity $I_{\!\lambda}$ and source function $S_{\!\lambda}$ weighed by the monochromatic extinction (absorption plus scattering) coefficient $\chi_{\!\lambda}$, in units of inverse length, over the whole solid angle and over all wavelengths:
\begin{equation}
Q_\text{rad} = \int_\Omega \! \int_0^\infty \!  \chi_{\!\lambda} \, (I_{\!\lambda}-S_{\!\lambda}) \, \diff \lambda \, \diff \Omega,
\label{eq:qrad}
\end{equation}
Under the assumption of isotropic source function and extinction coefficient, Eq.~\ref{eq:qrad} can be rewritten as
\begin{equation}
Q_\text{rad} = 4 \pi \! \int_0^\infty \!  \chi_{\!\lambda} \, (J_{\!\lambda}-S_{\!\lambda}) \, \diff \lambda,
\label{eq:qrad2}
\end{equation}
where $J_{\!\lambda} = \frac{1}{4\pi}\int_\Omega\,I_{\!\lambda}\,\diff \Omega$ is the mean monochromatic intensity averaged over the solid angle.
The necessary wavelength-dependent mean intensities are computed by solving the radiative transfer equation at each time step and along a representative set of inclined rays cast through the 3D simulation domain
\begin{equation}
\frac{\diff I_{\!\lambda}}{\diff \tau_{\!\lambda}} = I_{\!\lambda} - S_{\!\lambda},
\label{eq:radtran}
\end{equation} 
where $I_{\!\lambda} = I_{\!\lambda}(\Omega)$ is the monochromatic intensity at wavelength $\lambda$ along a given direction $\Omega$, $\tau_{\!\lambda}$ the optical depth along that direction, and $S_{\!\lambda}$ is the monochromatic source function.

\subsubsection{Input microphysics}
\label{sect:microphys}
For the simulations, we strive to implement microphysics that is as realistic as possible.
In particular, we adopt an updated version of the equation of state by \citet{Mihalas:1988,Hummer:1988,Dappen:1988}, accounting for the effects of ionisation, excitation, and dissociation of the 15 most abundant elements and of H$_2$ and H$_2^+$. 
Continuous opacities are based on an updated and revised version of the compilation by \citet{Gustafsson:1975} (Trampedach, priv. comm.; see also \citealt{Trampedach:2013,Hayek:2010} for a comprehensive list of the adopted continuous opacity sources), while sampled line opacities for wavelengths from $900$ to $200\,000$\,{\AA} are taken from B. Plez (priv. comm.) and \citet{Gustafsson:2008}. 
Equation of state and opacities for the main hydrodynamic simulations are computed assuming a solar composition \citep{Asplund:2009} with all metal abundances scaled proportionally to an iron abundance of [Fe/H] $= -2.5$ and with the abundances of $\alpha$-elements enhanced by [$\alpha$/Fe] $= +0.4$.
We have also run a number of simulations with slightly different chemical compositions, namely with [Fe/H]$= -3.0$, with and without alpha-enhancement, or assuming a scaled solar mixture by \citet{Grevesse:1998}.
As we do not have readily available sampled line opacity data for the latter composition, we rely instead on opacity distribution functions (ODFs) by \citet{Kurucz:1992,Kurucz:1993}.
The full list of compositions adopted for the simulations is given in Table~\ref{tab:params}.

\begin{figure*}
   \centering
   \resizebox{\hsize}{!}{
   	\includegraphics{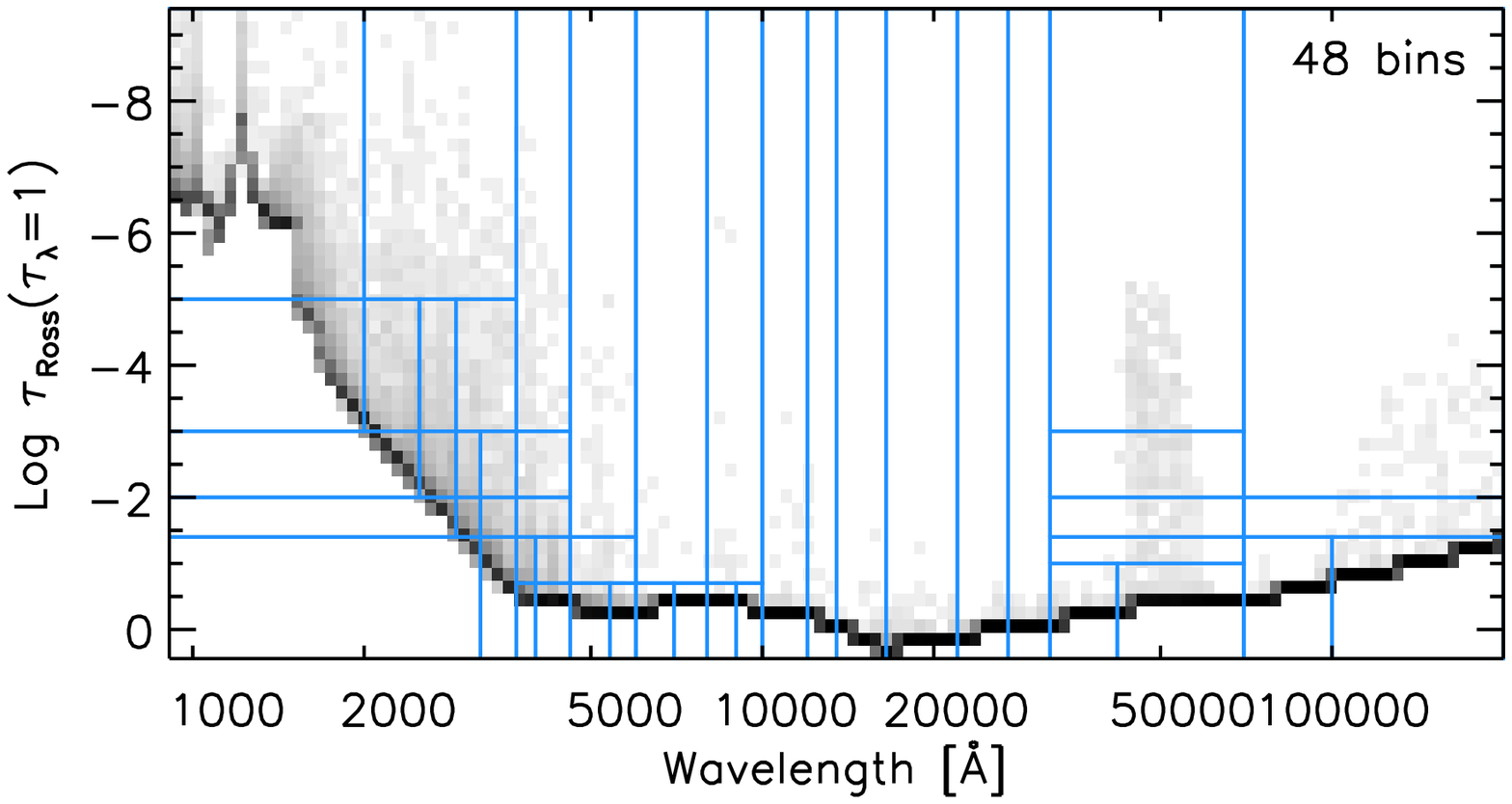}
   	\includegraphics{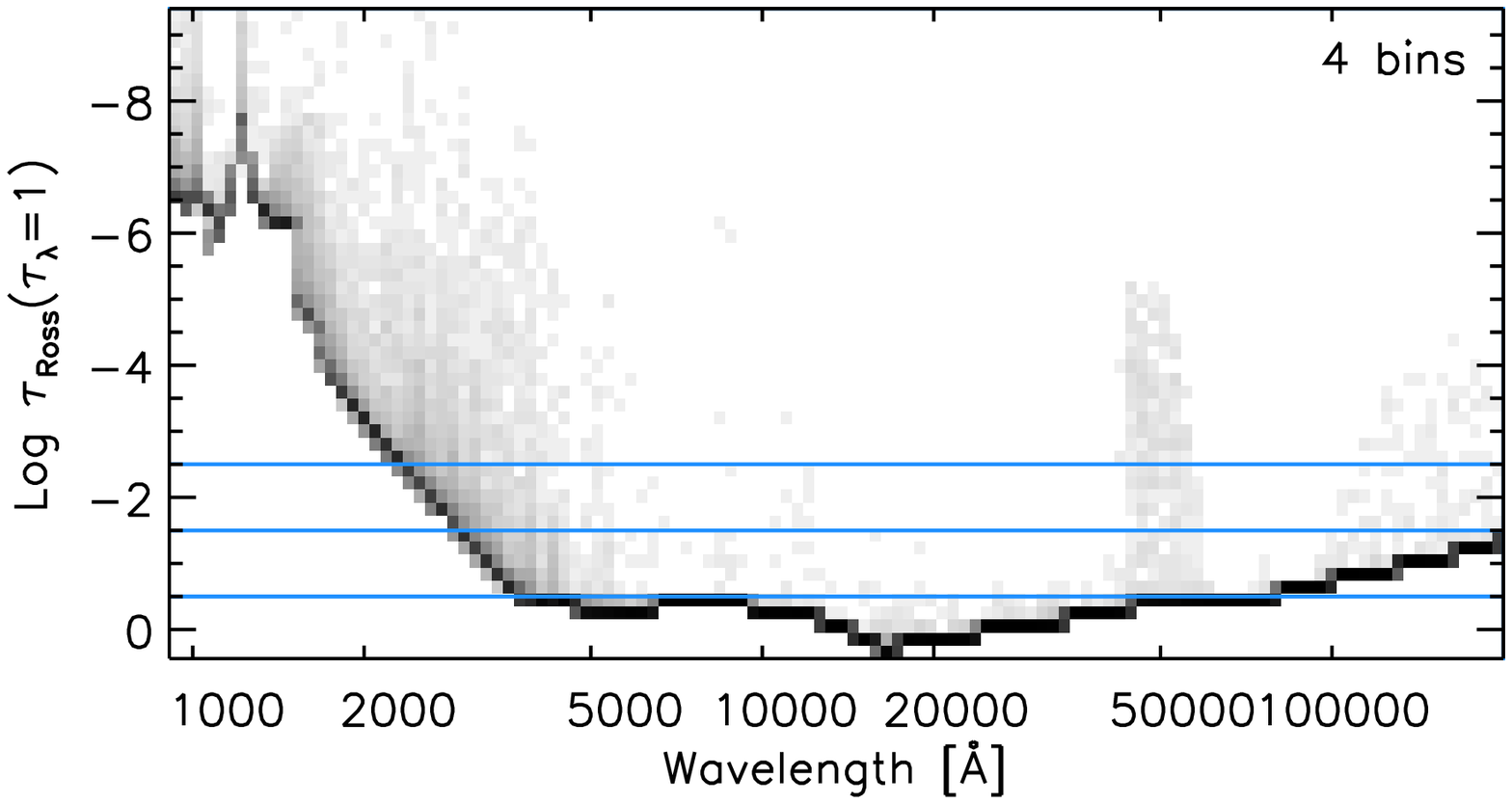} }
   \resizebox{\hsize}{!}{
   	\includegraphics{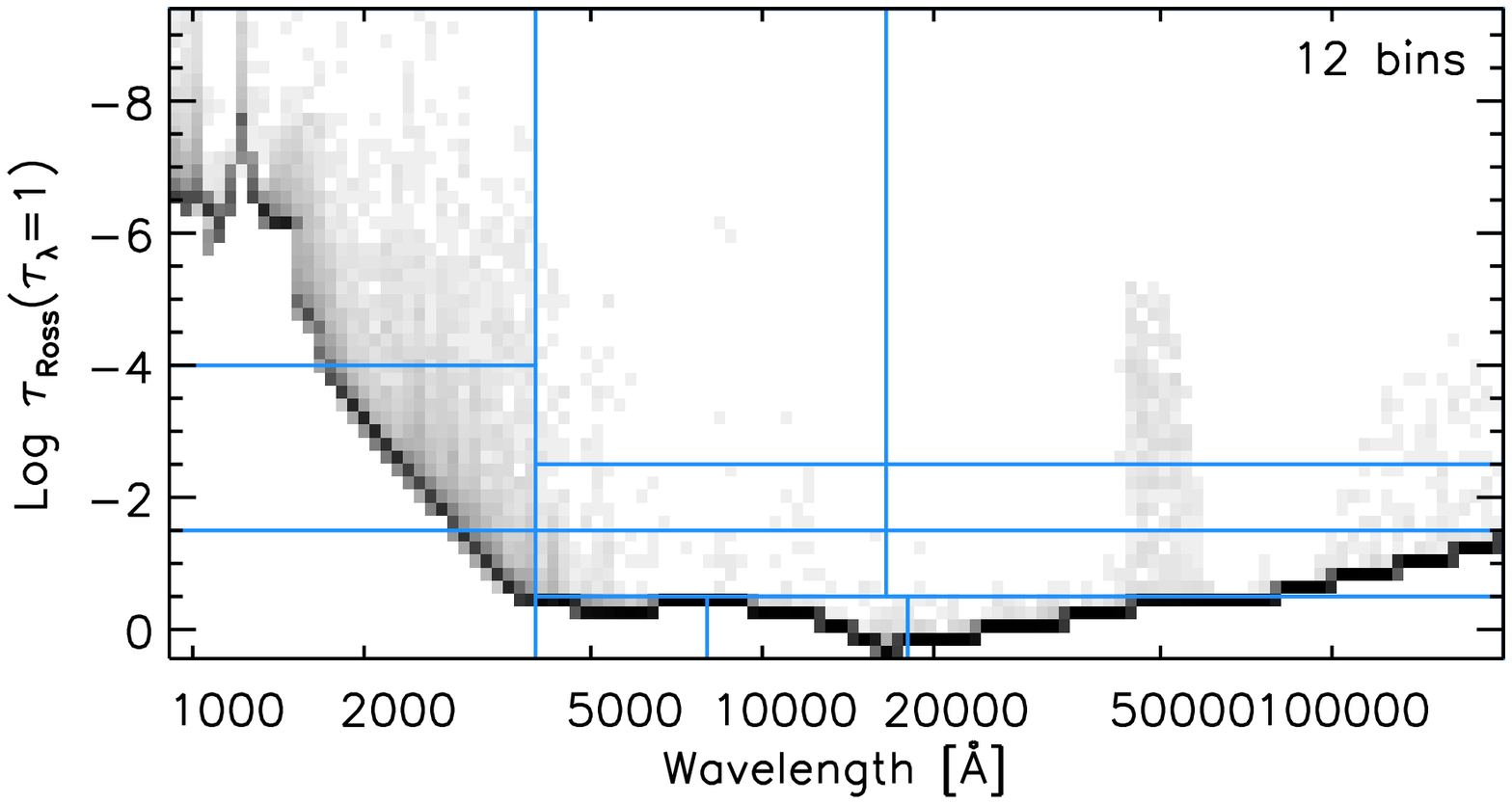}
   	\includegraphics{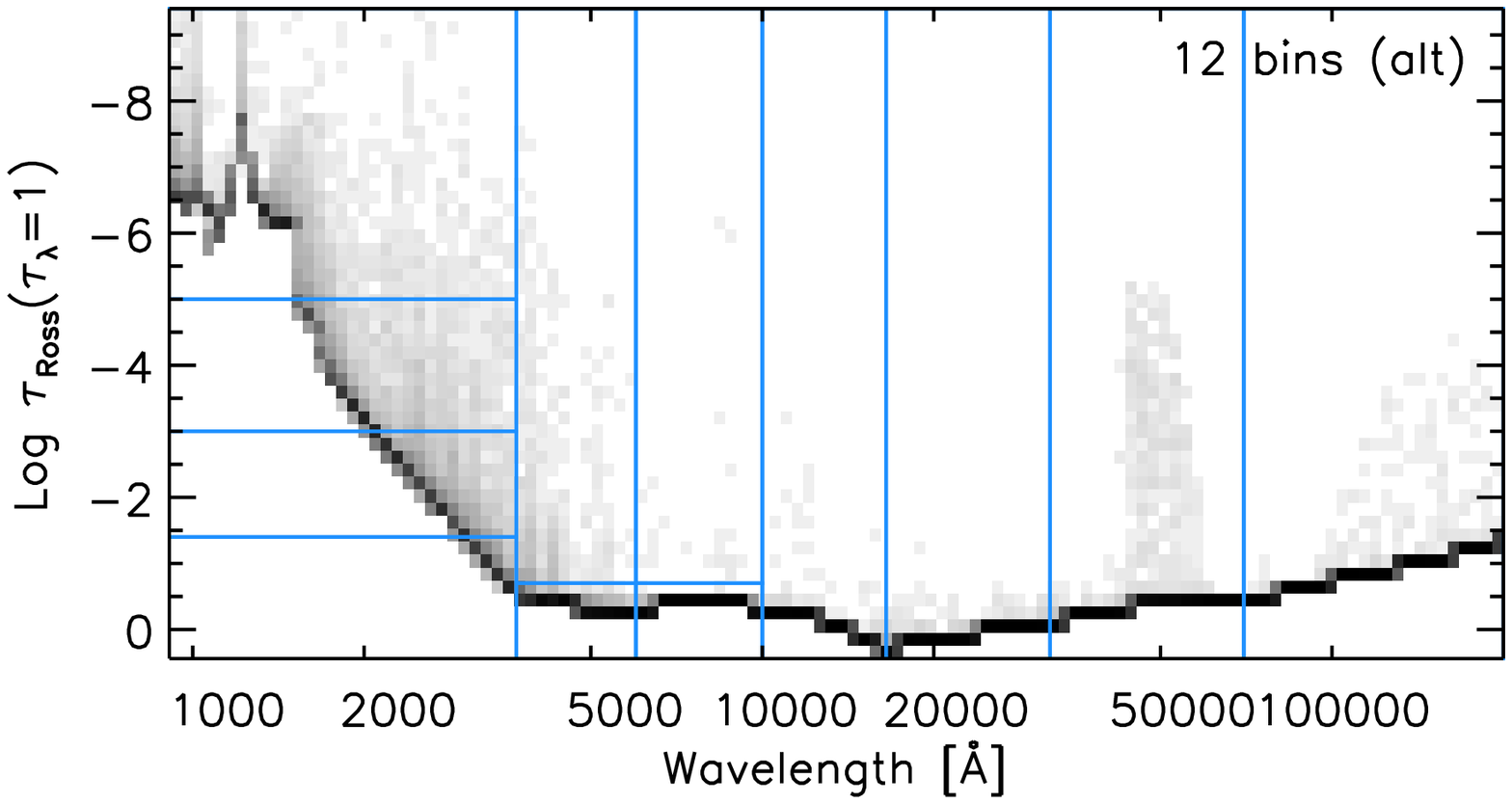} }	
   \caption{\emph{Grey shaded areas}: histogram of formation depth as a function of wavelength for {\hdstar}'s 3D model's mean temperature-density stratification. The distribution is constructed by computing the Rosseland optical depth at which monochromatic optical depth equals one for all wavelength points in the original opacity sampling data (computed for the \citealt{Asplund:2009} solar abundance mix scaled down to [Fe/H] $= -2.5$ and with [$\alpha$/Fe] $= +0.4$). Darker shades indicate higher frequency of occurrence. \emph{Blue lines}: adopted opacity bin boundaries for the 48-bin (\emph{top left panel}), four-bin (\emph{top right panel}), and two twelve-bin (\emph{bottom panels}) realisations.}
   \label{fig:binning}
\end{figure*}

\subsubsection{Opacity binning}
\label{sect:binning}
Solving Eq.~\ref{eq:radtran} at each time step in the 3D simulations for a large number of wavelengths that cover the relevant spectral range to comprehensively model radiative energy transport in stellar atmospheres is at present still too computationally demanding.
We therefore resort to using the \emph{opacity binning} (or \emph{multigroup}) method \citep{Nordlund:1982,Skartlien:2000} to represent the wavelength-dependence of the opacities in the radiative transfer calculations in an approximate but significantly less costly way.
The basic idea is to reduce the size of the radiative transfer problem from the more than $100\,000$ wavelengths of the original opacity sampling (OS) data to a tractable number of representative \emph{groups} of wavelengths (or \emph{bins}).
The sampled wavelengths are sorted into bins according to the strength of the corresponding monochromatic continuous plus line opacity and to the spectral interval to which they belong.
In each bin, opacities are appropriately averaged and source functions are integrated to obtain mean and cumulative functions of density and temperature for the two quantities, respectively.
The radiative heating rate integral in Eq.~\ref{eq:qrad2} is then reduced to a sum over all representative opacity bins:
\begin{equation}
Q_\text{rad} \approx 4 \pi \! \sum_i \!  \chi_{i} \, (J_{i}-S_{i})
\label{eq:qrad3}
\end{equation}
where $\chi_{i}$ is the mean extinction coefficient of all wavelengths that are members of bin $i$, $S_{i}$ is the total source function, and $J_{i}$ is the total mean (angle-averaged) intensity of that bin. 
The latter is evaluated by solving the radiative transfer equation for the intensity $I_{i}$ in bin $i$
\begin{equation}
\frac{{\diff}{I_{i}}}{{\diff}\tau_{i}} = I_{i} - S_{i},
\label{eq:radtran2}
\end{equation}
where $\tau_{i}$ is the optical depth in bin $i$, then by averaging the solution over the full solid angle, as in the monochromatic case.

For the specific opacity binning implementation, we adopt here the \emph{no-scattering-in-streaming-regime}  approximation proposed by \citet{Collet:2011}: we assume the source function to be Planckian ($S_{\!\lambda} = B_{\!\lambda}$) everywhere in the simulation domain and ignore the contribution of scattering when computing the mean extinction coefficients in the optically thin stellar surface layers.
In surface convection simulations of metal-poor red giants, such approximation reproduces very closely but at a much lower computational cost the results of radiative transfer calculations based on opacity binning that self-consistently accounts for the contribution of scattering to the mean extinction coefficients and to the integrated source functions. 
We refer to the paper by \citet{Collet:2011} for a more detailed description of the adopted version of the opacity binning method and of the calculation of the mean bin opacities.

When solving the radiative transfer equation in the simulations, we neglect macroscopic stellar velocity fields and the resulting Doppler shifts on the wavelengths of the extinction profiles.
It is by means of this approximation that we may treat source functions and extinction coefficients as isotropic and compute radiative heating rates using Eqs.~\ref{eq:qrad2} and \ref{eq:qrad3}.
However, we do take into account Doppler broadening of spectral lines by photospheric velocity fields in the original sampling data which were computed assuming a constant micro-turbulence parameter of $1.0$\,{\kms} for the line extinction profiles.
Typical bulk velocities and velocity dispersions in stellar atmospheric layers of red giants like {\hdstar} vary with depth and are of the order of a few {\kms}; 
on the other hand, given the coarse wavelength resolution of the opacity binning implementation, we deem this micro-turbulence value sufficient for the purposes of including the main effects of Doppler broadening on opacities by convective velocities in the present simulations.

As a rule, we define bin membership (opacity strength) of a given wavelength as the height where monochromatic optical depth $\tau_\lambda$ equals one on the simulations' mean temperature-density stratifications.
The mean stratifications have been computed from thermally and dynamically relaxed simulation sequences, by averaging the 3D temperature and density structures over time and over surfaces of constant column mass density.
The opacity binning affects the radiative heating and, in turn, the temperature-density structure in the simulations. 
Therefore, during the relaxation phase of the simulations, we iteratively recompute the average stratifications and rebin the wavelengths until we reach convergence. 

As reference height scale for each simulation, we adopt the Rosseland optical depth computed for the mean temperature-density stratification.
For some of our test simulation runs, however, we also consider an alternative criterion in which bin membership is determined by the ratio of monochromatic to Rosseland opacity at the height where $\tau_\lambda = 1$.
This approach is akin to the one taken by \citet{Collet:2009}, which allows us to directly compare with previous results and quantitatively understand the differences between old and new simulations in terms of predicted temperature stratifications.

We have considered various binning configurations with different numbers of bins and partitions in wavelength and opacity strength.
The main hydrodynamic simulations have been computed solving the radiative transfer for 48 opacity bins: we regard this as the reference case.
We have also generated simulation sequences using two different realisations of the opacity binning partitions with twelve bins to estimate the sensitivity of the resulting simulation's temperature stratification on a particular selection of opacity bins.
Finally, we have run a series of test simulations using classical binning with four bins in which opacities are selected based on their strength only and not according to wavelength.
Except for the four-bin case, we have defined all opacity binning partitions manually.

Figure~\ref{fig:binning} illustrates four examples of opacity binning selections based on the opacity sampling (OS) data for the scaled solar composition at [Fe/H] $= -2.5$ with 48, four, and twelve bins (two cases).
The figure shows the distribution of formation heights as a function of wavelength for the {\hdstar} simulation's mean stratification and the bin configurations we have explored.
The majority of wavelength points are distributed near the continuum-forming layers and it is therefore a sensible choice to try to resolve those regions with an adequate number of opacity bins.
At the same time, it is desirable to allocate some opacity bins for distinguishing among spectral lines of different strength: this is especially important for a realistic modelling of the radiative heating of the upper layers of the stellar atmosphere.
With 48 opacity bins, (Fig.~\ref{fig:binning}, top left panel,) we can achieve a reasonably fine resolution of the main wavelength regions and spectral features in the formation height diagram.
Obviously, the capabilities of the four-bin case (Fig.~\ref{fig:binning}, top right panel) are more limited in this respect; besides the fact that the adopted four-bin scheme does not rely on wavelength as selection criterion, all wavelengths with associated formation height $\log\tau_\text{Ross}{\leq}-2.5$ are gathered into one single bin, rendering the modelling of radiative heating in those layers in the simulations more uncertain.
For the 48-bin case, we also opt for a coarser resolution of the optically thin layers with $\log\tau_\text{Ross} \la -5$; there, however, modelling becomes anyway less realistic due to limitations in the basic physical assumptions of the present simulations, such as the neglect of magnetic fields and absence of a possible chromospheric temperature rise.

The bottom panel of Fig.~\ref{fig:binning} shows for comparison the two twelve-bin opacity binning representations considered in our work. 
For the first representation, (bottom left panel, or case ``12'' in Table~\ref{tab:params},) we have identified three main wavelength regions ($\lambda < 4\,000$\,{\AA}, $4\,000$\,{\AA} $\leq \lambda < 18\,000$\,{\AA}, and $\lambda \geq 18\,000$\,{\AA}) and sorted opacities according to typically four levels of opacity strength within each region; the only exceptions are the short-wavelength region, which we map with three opacity bins only, and the region between $4\,000$ and $18\,000$\,{\AA} around the continuum-forming layers, which we split in two bins.
For the second one, (bottom right panel in Fig.~\ref{fig:binning}, or case ``12 alt'' in Table~\ref{tab:params},) we have considered instead seven wavelength regions for the opacity binning; the four regions with $\lambda \geq 10\,000$\,{\AA} are mapped with one single unsplit opacity bin each. The two regions between $\lambda = 3\,700$\,{\AA} and $\lambda = 10\,000$\,{\AA} are both split into two opacity bins, covering continuum and weak line opacities and moderate to strong line opacities, respectively. Finally, the short-wavelength region is resolved with four different levels of opacity strength.
These two representations exemplify two rather different approaches to opacity binning: the first aims at mapping opacities according to an as regular as possible combination of wavelength interval membership and levels of opacity strength, while the second focuses on grouping the opacities based almost solely on spectral region membership.
As we show in Sect.~\ref{sect:temp3d}, when implemented into the simulations, the two opacity binning strategies result in slightly different mean temperature stratifications.

\subsection{Numerical radiative transfer solver}
The assumption of isotropic source functions and extinction coefficients allows us to reformulate Eq.~\ref{eq:radtran2} in terms of the average of in-coming and out-going intensities along a given direction $\Omega$:
\begin{equation}
\frac{{\diff}^2{P_{i}}}{{\diff}\tau_{i}^2} = P_{i} - S_{i},
\label{eq:radtran3}
\end{equation}
where $P_{i} = \frac{1}{2}(I_{i}(\Omega)+I_{i}(-\Omega))$ .
We discretise and solve Eq.~\ref{eq:radtran3} using a modified version \citep{Nordlund:1982} of the second-order finite-difference long-characteristics \citet{Feautrier:1964} method which is particularly advantageous in terms of speed and stability.
To maximise computational efficiency, we only solve radiative transfer for the region of the simulation with $\tau_\text{Ross}{\leq}500$. 
For optically thicker layers, we simply assume no radiative heating. 
As boundary conditions for the intensities, we assume no incoming radiation at the top of the simulation's domain, a Planckian radiation field at the bottom, and periodic boundaries in the horizontal directions.

The radiative transfer equation is solved along a set of rays that pass through each grid-point at the simulation's surface.
We consider nine different directions for the rays, a combination of two $\mu$-angles\footnote{$\mu = \cos{\theta}$, where $\theta$ is the ray inclination from the vertical direction.} and four azimuthal $\phi$-angles plus the vertical. 
The integral of the radiative transfer solution over solid angle is evaluated using the Radau quadrature scheme \citep{Radau:1880}.

\subsection{Parallelisation}
The {\stagger} code is parallelised using Message Passing Interface (MPI) and domain decomposition which allows efficient calculations on distributed supercomputing systems.
The original simulation domain is decomposed into smaller volumes of equal physical and numerical size assuming a Cartesian topology; subdomains are then assigned to individual MPI processes which are then distributed to a large number of computing units.
The solution to the hydrodynamic equations is evaluated and advanced in time locally by each process in its subdomain; the necessary information about boundary values is exchanged between neighbouring subdomains at each time step using standard MPI communication.

\begin{figure}
   \centering
   \includegraphics[width=\columnwidth]{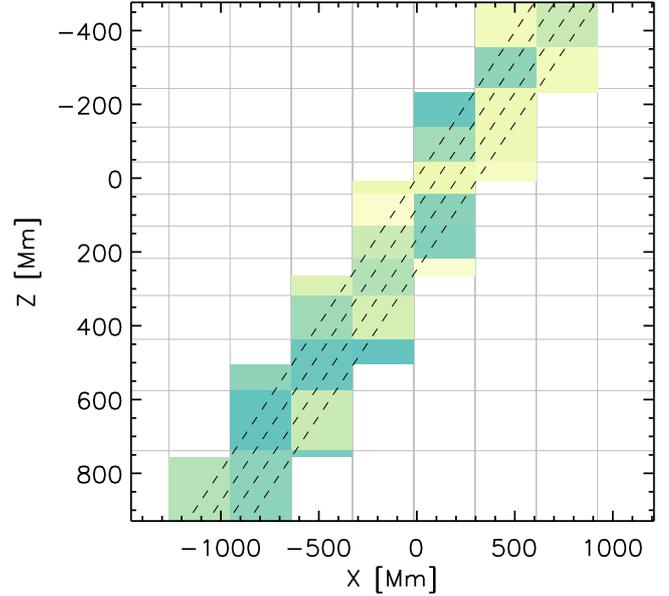}
   \caption{Schematic representation of ray reconstruction in the {\stagger} code. \emph{Grey contours:} individual MPI subdomains considered by the hydrodynamics solver. \emph{Dashed lines:} bundle of rays assigned to a given MPI process. \emph{Coloured boxes:} portions of the hydrodynamics subdomains that are communicated to the given MPI process to reconstruct the full path of the rays in the bundle; different colours indicate different MPI subdomains.}
   \label{fig:rays}
\end{figure}

\subsubsection{Radiative transfer and ray reconstruction}
In general, the rays considered for the solution of the radiative transfer problem cross multiple subdomains between the bottom and top boundary of the simulation; consequently, each MPI subdomain locally includes only partial ray segments.
Due to the non-local nature of radiative transfer, information about the radiation's intensity needs to be propagated along a ray across all these subdomains in order to compute the full radiative transfer solution.
Efficient algorithms exist for solving the radiative transfer problem with domain decomposition using the integral method and MPI \citep{Heinemann:2006}.
We explore here a different approach in which we distinguish between ``hydrodynamics'' and ``ray'' domains and use both at each time step.
In the ray domain, for any given direction, we assign to each MPI process a bundle of parallel rays extending the whole way from bottom to top boundary.
Source functions and extinction coefficients are then communicated from all MPI hydrodynamics subdomains intersected by a given bundle of rays to the corresponding MPI process in the ray domain (see also Fig.~\ref{fig:rays}).
This way, all the relevant information to solve the radiative transfer along those rays, from bottom to top boundary, becomes self-contained and controlled by a single  MPI process in the ray domain.
Once the radiative transfer equation has been solved in the ray domain, partial heating rates are computed and distributed back from each MPI process in the ray domain to the MPI subdomains in the hydrodynamics domain so that the solution to the conservation equations can be advanced in time.

\begin{figure*}
  \centering
   \resizebox{\hsize}{!}{
   \includegraphics{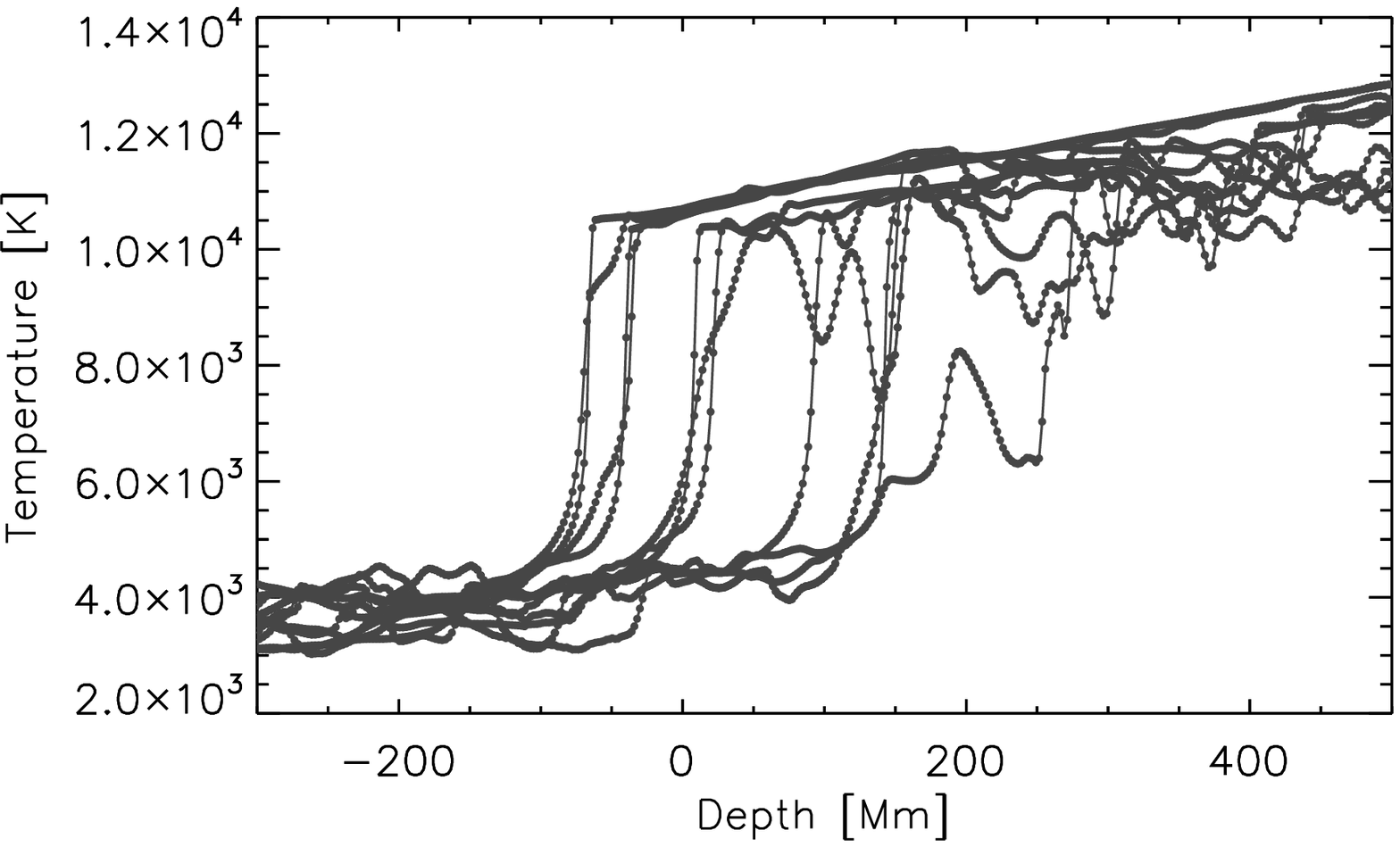} 
   \includegraphics{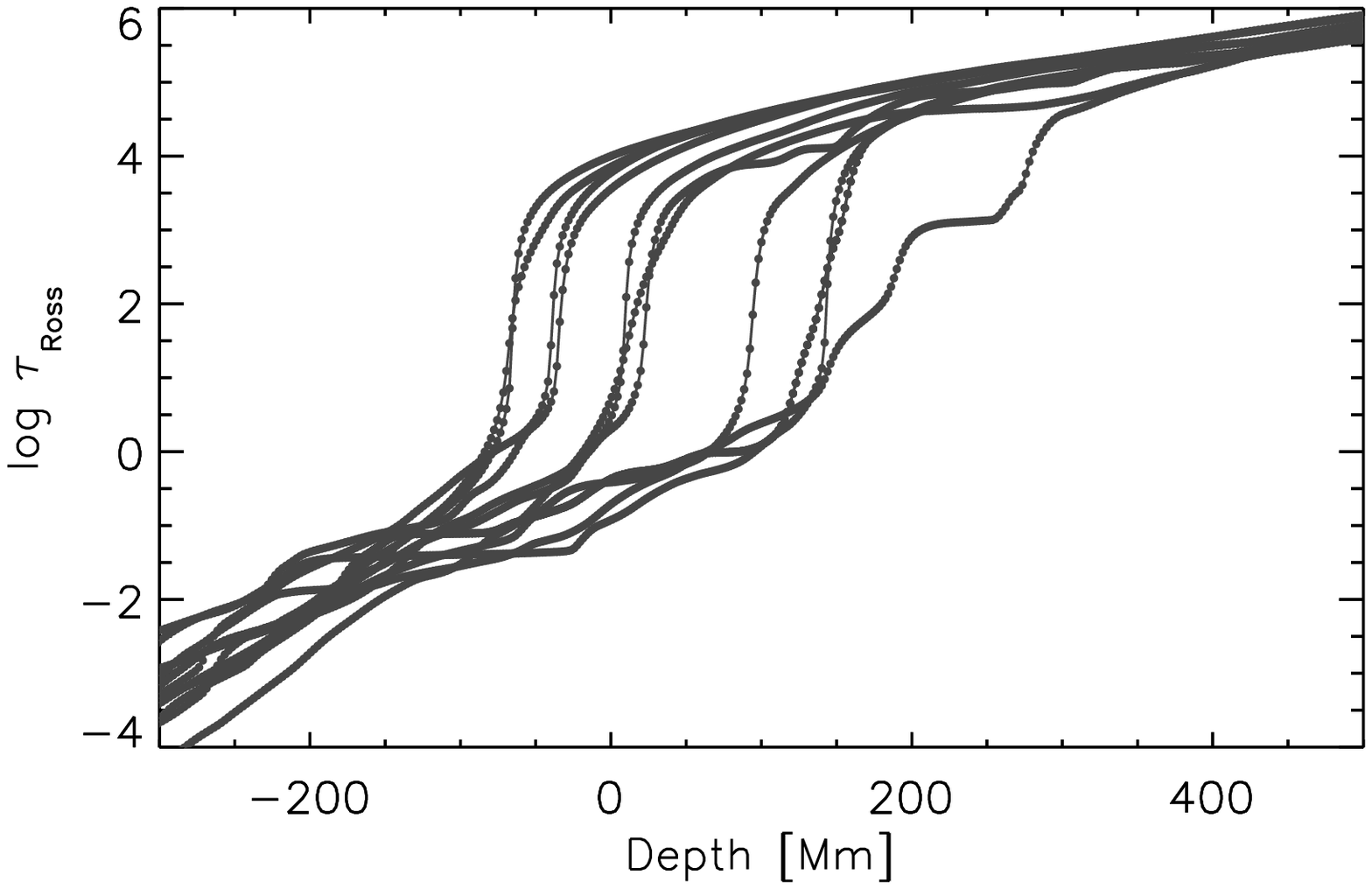} }
   \caption{Temperature and optical depth gradients in the outer layers of {\hdstar}. \emph{Left:} Temperature as a function of geometrical depth for a few selected columns of a typical surface convection simulation snapshot of the metal-poor red giant star. \emph{Right:} Rosseland optical depth as a function of geometrical depth for the same columns.}
   \label{fig:grad}
\end{figure*}

\begin{figure}
   \centering
   \includegraphics[width=\columnwidth]{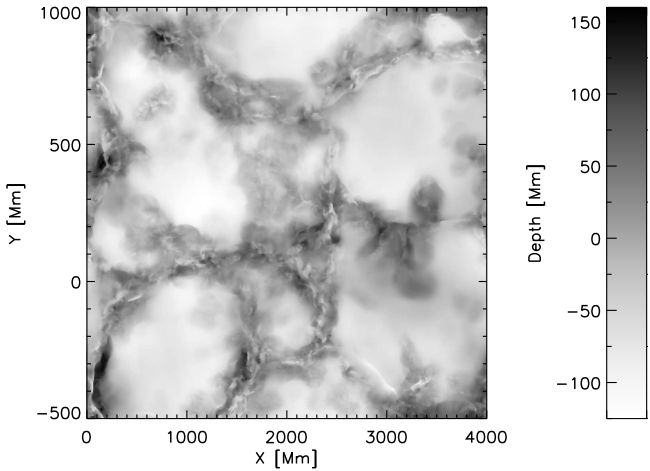}
   \caption{Map showing the geometrical depths at which $\tau_\text{Ross} = 1$ across the surface of a typical snapshot from our 48-bin very high-resolution simulation of {\hdstar}. Darker shades indicate larger geometrical depths.}
   \label{fig:optsurf}
\end{figure}

\begin{figure*}
  \centering
   \includegraphics[width=0.92\textwidth]{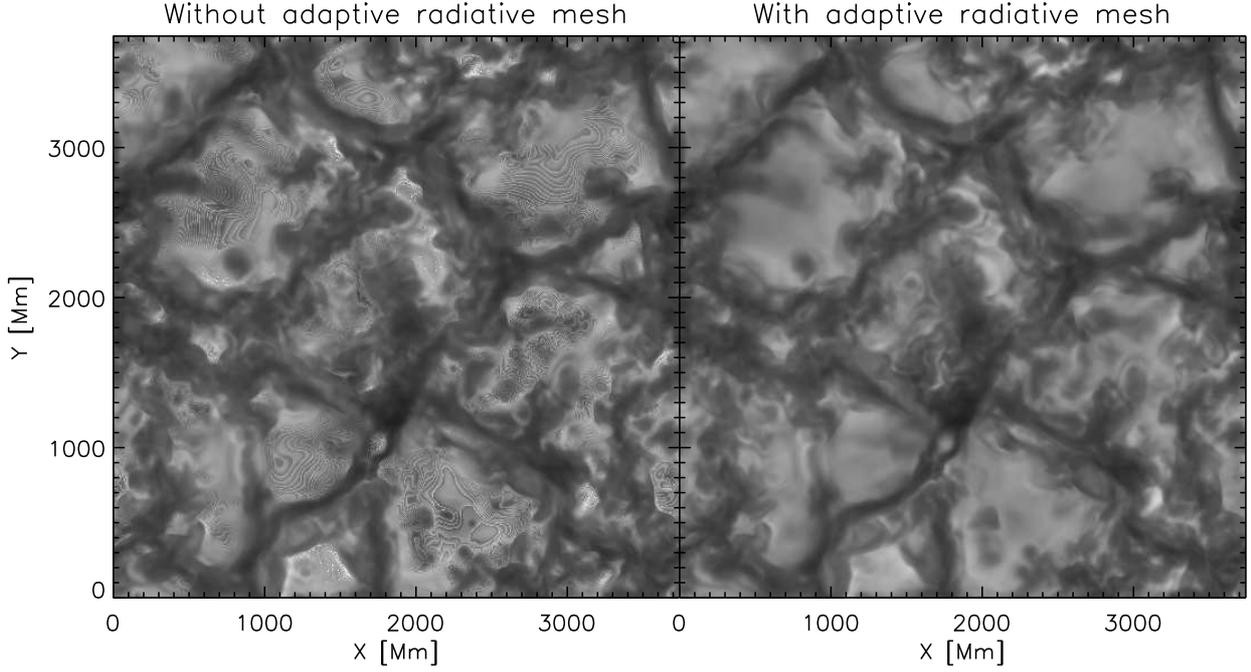}
   \caption{Comparison between the spatially resolved surface intensities resulting from two simulations nearly identical to each other except for the adopted radiative transfer depth scale: one uses the same single fixed depth scale as the hydrodynamics solver (\emph{left panel}), while the other uses an adaptive mesh (\emph{right panel}). 
Ringing (e.g. at around $x \!=\! 1\,200$\,Mm, $y \!=\! 900$\,Mm, or $x \!=\! 2\,900$\,Mm, $y \!=\! 1\,600$\,Mm) develops rapidly in the simulation without adaptive radiative mesh owing to insufficient numerical resolution of steep opacity gradients in the photosphere.}
   \label{fig:ringing}
\end{figure*}

\begin{figure*}
  \centering
   \resizebox{\hsize}{!}{
   \includegraphics{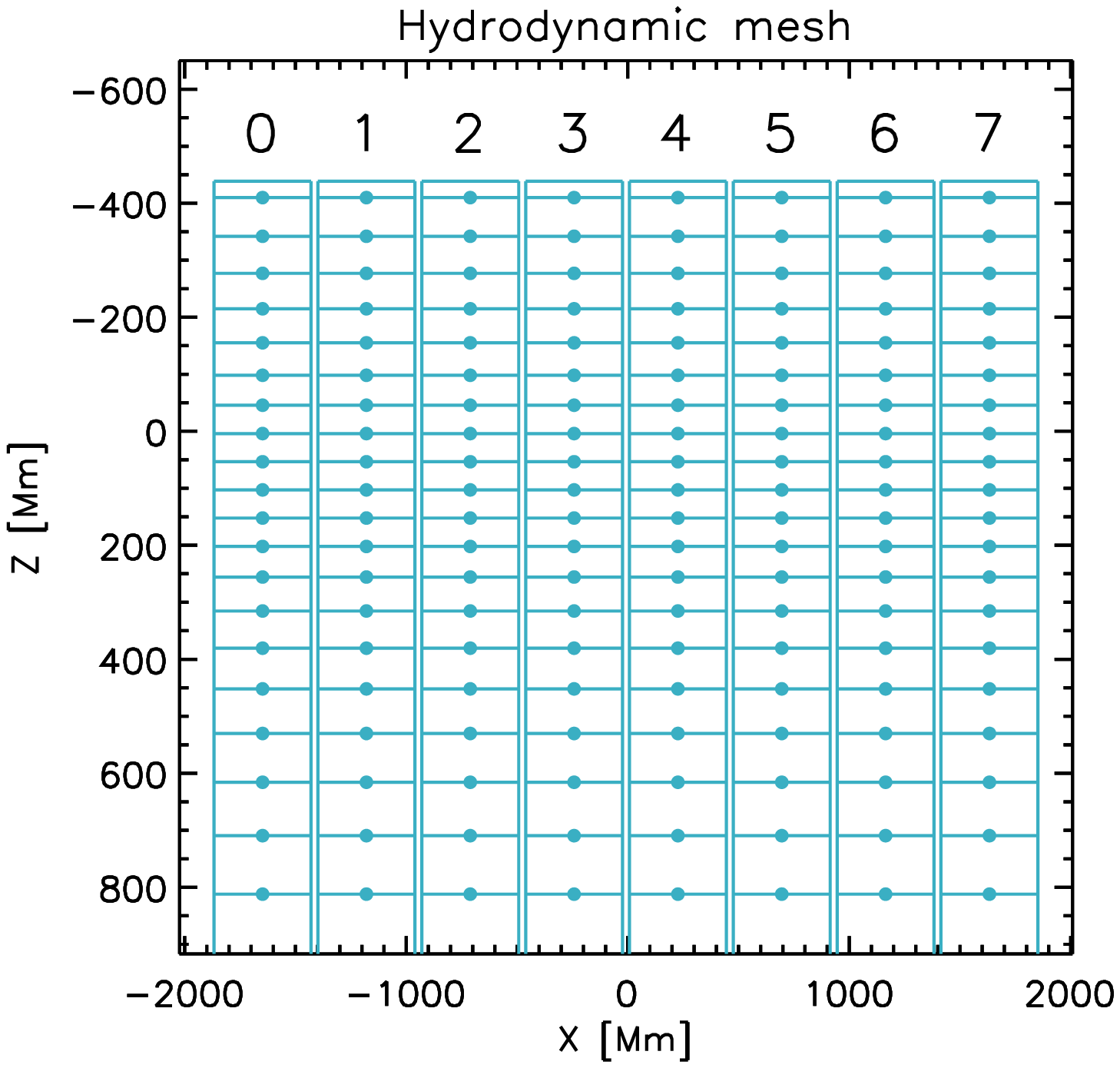} 
   \includegraphics{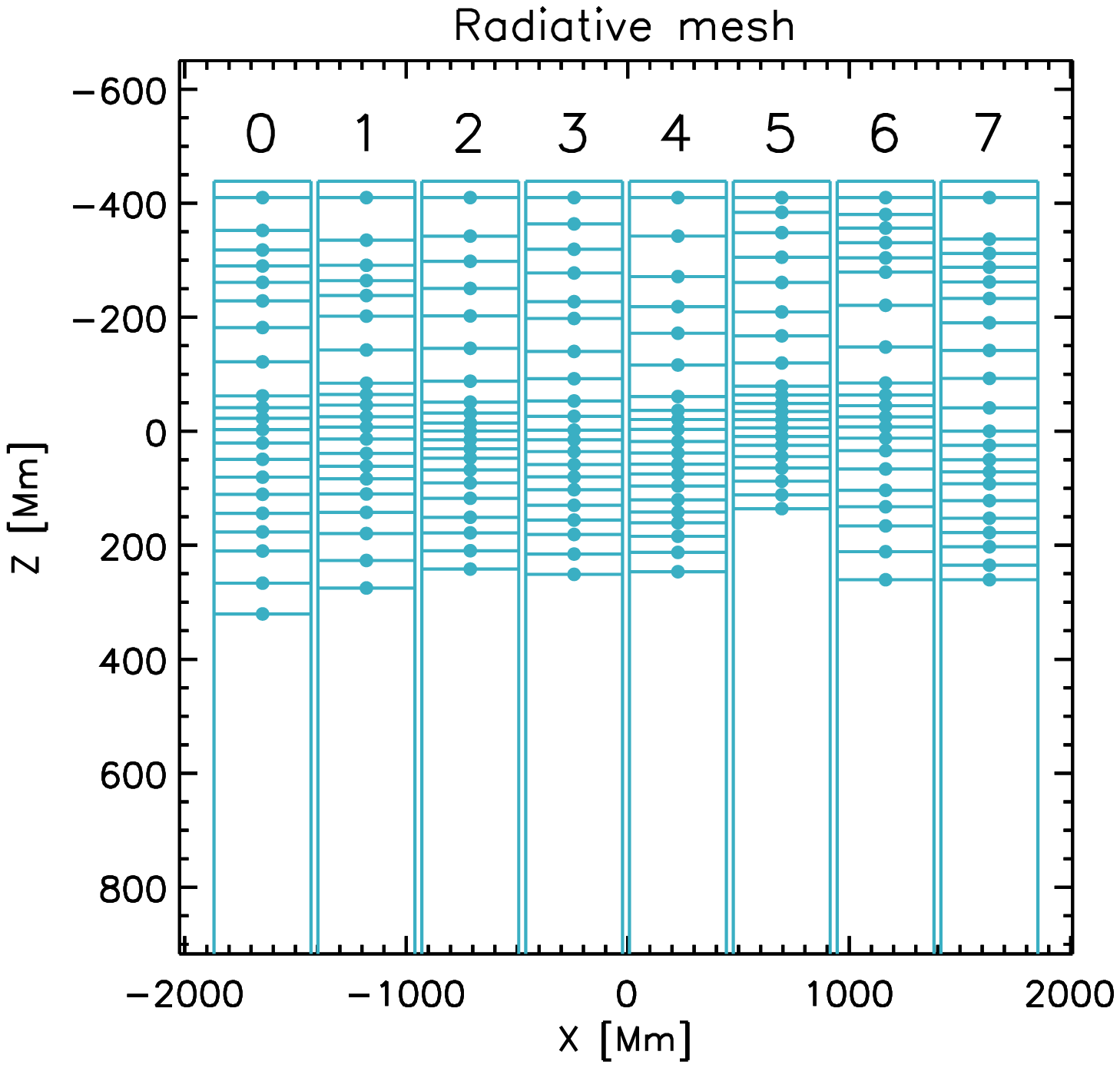} }
   \caption{Schematic comparison between the hydrodynamic (\emph{left}) and adaptive radiative mesh (\emph{right}). Each vertical rectangular block in the diagram represents a set of columns in the simulation box. The labels at the top indicate the various MPI processes to which the blocks are assigned. The large dots represent the depth points in each block. In the hydrodynamic mesh (\emph{left}), the location of the depth points is fixed and is the same for all blocks; in the adaptive radiative mesh (\emph{right}, the depth points are rearranged in each individual block to optimally resolve steep opacity gradients.}
   \label{fig:radscale}
\end{figure*}

\subsection{Adaptive radiative mesh}
Steep temperature and opacity gradients occur near the optical surface of our metal-poor red giant convection simulations.
In the simulations, in most columns corresponding to stellar granules, the gas temperature drops from $10\,000$\,K to $5\,000$\,K within ${\sim}10$\,Mm near the surface, i.e. less than $1$\,\% of the simulation's geometrical depth (Fig.~\ref{fig:grad}).
Furthermore, according to the simulations, the optical surface of a metal-poor red giant stellar atmosphere appears very corrugated: the geometrical depth where $\tau_\text{Ross} = 1$ varies across the stellar surface spanning a range of ${\sim}280$\,Mm (Fig.~\ref{fig:optsurf}).
With insufficient vertical resolution, using the same mesh for the radiative transfer as for the hydrodynamics solver would result in the emergence of a spurious ringing pattern in the radiative intensities and radiative heating rates at various depths throughout the atmosphere as well as in the spatially resolved surface intensity as can be seen in Fig.~\ref{fig:ringing}.
The problem is not merely an aesthetic one: the pattern can eventually be imprinted in the internal energy structure and, consequently, amplified via positive feedback, leading to numerical artefacts in the photospheric temperature and density stratification.

Accurately solving the radiative transfer equation with one single fixed depth scale that captures the above steep gradients and prevents the emergence of the ringing pattern would require a mesh with extremely high numerical resolution.
Here, we have opted instead to implement a separate adaptive mesh for computing the solution to the radiative transfer.
At each time step, the depth points from the original grid are repositioned and concentrated around regions where vertical opacity gradients are steepest. Also, the depth points are distributed only across the region where the numerical solution to the radiative transfer equation is effectively computed ($\tau_\text{Ross}{\leq}500$).
The relevant physical variables are interpolated to the new mesh on which the radiative transfer is then solved and the heating rates are computed. 
The latter are then distributed back to the original mesh in an energy conserving way before advancing in time the solution to the hydrodynamic equations.
Previous 3D simulations of surface convection in red giants by \citet[e.g.][]{Collet:2007} and \citet{Trampedach:2013} adopted at each time step a single radiative depth scale for the full domain when solving the radiative transfer equation. This solution is generally sufficient in order to resolve temperature gradients in the atmospheres of red giant stars with surface gravities $\log g \ga 2$\,({\cmss}).
However, in light of the high level of corrugation of the optical surface in low-surface-gravity metal-poor red giant stellar atmospheres, it is not beneficial to use a single radiative depth scale for the entire simulation domain. Instead, we subdivide the domain in vertical blocks and compute a separate radiative depth scale for each one of them. For all practical purposes, with typical MPI runs with a sufficient number of parallel processes we can simply define the blocks by grouping together MPI subdomains that are vertically aligned in the MPI Cartesian topology.
Within each vertical block, the new radiative depth scale is defined so that grid points become equidistant in $\int_{z_0}^z\!\sqrt{f(z)}\,{\diff}{z}$ where 
\begin{equation}
f(z) = \max_{z=\text{const}} \left| \frac{1}{\chi_\text{Ross}}\frac{\partial^2 \tau_\text{Ross}}{\partial z^2} \right| = \max_{z=\text{const}} \left| \frac{\partial \ln\!\chi_\text{Ross}}{\partial z} \right|
\label{eq:radscale}
\end{equation}
denotes the maximum absolute value of the Rosseland extinction coefficient's vertical derivative on a given horizontal layer.\footnote{For our implementation, we actually apply some smoothing to $f(z)$ prior to evaluating the integral $\int\!\sqrt{f(z)}\,{\diff}{z}$ to increase numerical stability of the method, although this is not strictly necessary in most situations.}
This definition is based on a modified version of the method by \citet[][p. 44]{De-Boor:1978} for the construction of the optimal mesh to resolve a given function, which, in our case, is the Rosseland optical depth.
Ideally, the adaptive radiative mesh should be computed anew for every inclined direction considered during the solution of the radiative transfer problem. In practice, based on the results of test calculations, for the present simulations we find it sufficient to define the radiative mesh only for the vertical rays, then simply tilt it and stretch it for the inclined directions.
Figure~\ref{fig:radscale} illustrates schematically our adaptive radiative mesh concept in relation to the fixed hydrodynamic mesh.

\section{Spectral line formation and abundance analysis}
\label{sect:lineform}
We use the 3D stellar surface convection simulations as time-dependent hydrodynamic model atmospheres to compute flux profiles for spectral lines of various atomic and molecular species and derive elemental abundances from the analysis of {\hdstar}'s spectrum.
To this purpose, we employ here the line formation code {\scate} by \citet{Hayek:2011}. 
The code assumes local thermodynamic equilibrium (LTE) but self-consistently accounts for coherent isotropic continuum scattering in the extinction coefficient and source function for the solution of the radiative transfer equation.
We adopt the 48-bin stellar surface convection simulation of {\hdstar} at ${504}^2{\times}{252}$ numerical resolution as primary sequence for the line formation calculations.

\subsection{Atomic lines}
\label{sect:atomic}
We synthesise over 400 individual atomic lines of 16 elements (O, Na, Mg, Al, Si, Ca, Sc, Ti, V, Cr, Mn, Fe, Co, Ni, Zn, Sr) in neutral and singly ionised stages.
We adopt the line list by \citet{Aoki:2007} but with updated line parameters from the Vienna Atomic Line Database ({\vald}) \citep{Kupka:1999,Piskunov:1995} and NIST \citep{Kramida:2014} databases.
For \ion{Fe}{i} and \ion{Fe}{ii} lines, we use ${\loggf}$-values and excitation potentials extracted from the critical compilation by \citet{Fuhr:2006}, whenever available.
Partition functions for the various atoms and ions come from the compilation by \cite{Barklem:2016}.

For the line formation calculations, we select a subset of 50 snapshots taken every twelve hours of stellar time from the original simulation sequence
We compute the emergent intensities for each spectral line profile at various angles across the stellar surface by solving the radiative transfer equation along representative sets of rays cast through the 3D domain with 13 different inclinations (three $\mu$-angles and four $\phi$-angles plus the vertical). 
We then perform a disk integration via Gaussian quadrature, a spatial average of the line profiles across the stellar surface and, finally, a time average over all selected snapshots. 
Line profiles are sampled using 101 wavelength points with a resolution $\Delta\text{v} = 0.5$\,{\kms} in velocity space  ($1.0$\,{\kms} for wide lines).
To reduce the computational load, we consider only rays that cross the surface of the simulation domain every fourth grid point in each horizontal direction.
Test calculations on a subset of lines using 201 wavelength points and higher resolution in velocity space to resolve line profiles or in which rays are cast through every grid point at the simulation surface indicate that the errors due to the above two approximations are negligible and, in both cases, translate into uncertainties of $0.004$\,dex or less on derived elemental abundances.
Ionisation, excitation, and molecular equilibria as well as continuous opacities for the line formation calculations are computed assuming a fixed background chemical composition, as given in Table~\ref{tab:bg-abund}  (see also Sect.~\ref{sect:metallicity}) and input micro-physics consistent with the 3D simulations.
When computing the emergent intensities, we account for Doppler shifts and broadening of the absorption profiles induced by thermal motions and projected macroscopic velocity fields in the simulations along the various line of sights.
For the collisional broadening by hydrogen and helium atoms, we implement the quantum mechanical description by \citet{Barklem:2000} whenever applicable. When not available, we adopt the classical description of van der Waals broadening by \citet{Unsold:1955} with an enhancement factor of 2.0.

We use the same numerical code and input physics to carry out spectral line formation calculations with 1D plane-parallel model atmospheres generated for the same stellar parameters as the 3D simulations. 
The plane-parallel models are constructed with the custom 1D model atmosphere code {\atmo} by W. Hayek \citep[see also][]{Magic:2013} that uses the same equation of state and opacity binning data as the simulations but includes an implementation of the mixing-length theory to account for convective energy transport.
In the 1D calculations with {\scate}, in order to compensate for the lack of self-consistent velocity fields and yet account for convective-like broadening, we compute spectral line profiles assuming a range of micro-turbulence values between $\xi = 2.0$\,{\kms} and $2.5$\,{\kms}, covering the typical values adopted in previous spectroscopic analyses of {\hdstar} \citep[e.g][]{Barbuy:2003,Aoki:2007}.
We would like to stress once again here that no micro-turbulence enters the 3D line formation calculations: in 3D, only the velocity fields predicted by the simulations are used to model non-thermal line broadening and shifts associated with convective macroscopic flows \citep[e.g.][]{Asplund:2000}.

We compute flux profiles in 3D and 1D for each atomic spectral line for a range of elemental abundances and interpolate in line strength to find the value that fits the equivalent widths measured by \citet{Aoki:2007}.
We refer to the differences between the so derived 3D and 1D abundances as the 3D$-$1D abundance corrections.

\subsection{Molecular lines and bands}
\label{sect:molecular}
We use {\scate} to also synthesise flux profiles for a number of bands from CNO-based diatomic molecules: the OH A-X and CH C-X bands between $3100$ and $3200$\,{\AA}, the CH A-X $G$ and OH A-X bands between $4290$ and $4315$\,{\AA}, the NH A-X band between $3350$ and $3390$\,{\AA}, and the CN B-X band between $3860$ and $3890$\,{\AA}.
For the CH bands, we implement the line list by \citet{Masseron:2014}, while for the OH and NH systems we adopt the excitation potentials and ${\loggf}$ values recently recomputed by T. Masseron (priv. comm.). 
The OH oscillator strengths predicted by T. Masseron are overall in good agreement with those by \citet{Gillis:2001}, but the NH ${\loggf}$ values are $0.3-0.7$\,dex smaller than the widely used ones from \citet{Kurucz:2011}\footnote{\url{http://kurucz.harvard.edu/linelists/linesmol/nh.asc}}:
the existence of a systematic offset in ${\loggf}$ values for the NH UV lines is supported by evidence from various spectroscopic analyses that \citeauthor{Kurucz:2011}'s ${\loggf}$ values for NH UV lines may be overestimated by ${\sim}0.4$\,dex \citep[e.g.][]{Spite:2005,Aoki:2006}.
Finally, for the CN band, we implement a line list by B. Plez (priv. comm.) which is an updated version of the one used by \citet{Hill:2002} and \citet{Hedrosa:2013}.
Molecular partition functions and equilibrium constants are taken from \citet{Barklem:2016}.
As we compute the molecular bands over relatively extended wavelength regions, we need to account for blends by atomic lines too. 
For the spectral synthesis calculations, we therefore include atomic lines in the relevant wavelength ranges as extracted from {\vald}.

Owing to the non-linearities involved in the chemical equilibrium and the interdependencies among the various molecules in the reaction network, it is generally necessary to account for the simultaneous variation of the CNO abundances when fitting observations.
This is the case, in particular, for CH and OH, whose equilibrium number densities in the atmospheres of late-type stars are strongly controlled by the formation of the tightly bound CO molecule.
In order to properly fit the observed CH and OH bands simultaneously, we therefore have to consider a whole range of possible combinations of C and O abundances for the spectral synthesis calculations. 
Similarly, fitting the observed CN and NH bands simultaneously requires that we take into account various combinations of C and N abundances when computing the synthetic spectra. 
More details about the choice of CNO abundances range and the fitting procedure are given in Sect.\,\ref{sect:fitting}.

We adopt a resolution $\Delta\text{v} = 1.0$\,{\kms} in velocity space for computing the molecular bands.
For the 3D syntheses, we select a subset of ten snapshots taken at regular intervals from the original simulation sequence. 
We have carried out test calculations with 100 snapshots for a few selected combinations of CNO abundances and found no noticeable differences in the resulting synthetic bands and in the derived abundances: the uncertainty on derived abundances due to the lower number of snapshots is about $0.01$\,dex.
To further reduce the computational time, we also scale down the numerical resolution of the original snapshots to $240^3$.
For the 1D calculations, we also assume a micro-turbulence of $2.0$\,{\kms}.
As in the case of the atomic line calculations, all synthetic molecular bands are computed accounting for coherent continuum scattering in both 1D and 3D.
We would like to stress the importance of including scattering in the spectral synthesis calculations, particularly in the UV region: at these short wavelengths, neglecting the effects of scattering would result in weaker spectral lines and require higher oxygen abundances to fit the observed OH UV band.
More specifically, in 3D, this would translate into an error $\Delta\abund{O} \approx 0.3$\,dex.

Besides the above molecular bands, we also synthesise 16 individual OH $X^2\Pi$ vibration-rotational lines in the infrared (IR) region around $1.5$--$1.7$\,{$\mu$m} to provide an additional constraint to the O abundance determination {\hdstar}.
We adopt the list of line parameters and measured equivalent widths from \citet{Barbuy:2003}.
As in the case of the molecular band calculations, we consider a wide range of O and C abundance combinations when synthesising the OH IR line profiles.
For each given value of the C abundance, we then vary the O abundance individually for each OH IR line, until the measured equivalent width is matched. 

\subsection{Molecular band fitting and CNO abundance determination}
\label{sect:fitting}
We determine the CNO abundances in {\hdstar} by directly fitting the synthetic molecular bands to an observed spectrum of the star as well as matching computed and measured equivalent widths of individual OH IR lines.
We use the spectrum of {\hdstar} from the Ultraviolet Echelle Spectrograph Paranal Observatory Project (UVES POP, \citealt{Bagnulo:2003})\footnote{European Southern Observatory (ESO) Director Discretionary Time (DDT) Program ID 266.D-5655, \url{http://www.eso.org/sci/observing/tools/uvespop.html}} acquired at the Kueyen unit of the Very Large Telescope (VLT).
The merged and co-added spectrum has a resolution $R = \lambda/\Delta\lambda \approx \,80\,000$. 
The signal-to-noise ratio ($S/N$) is about $400$ in the G band around $\lambda = 4300$\,{\AA}, but it is appreciably lower in the near UV, being about $100$ at $3350$\,{\AA} and $30$ at $3150$\,{\AA}.
Before proceeding with our analysis of molecular bands, we have normalised the spectrum by fitting a smooth cubic spline to the continuum regions.

We first proceed with synthesising and fitting the OH and CH bands between $3100$ and $3200$\,{\AA} and between $4290$ and $4315$\,{\AA}. 
We compute synthetic molecular bands for different combinations of a specific number of C and O abundances in the ranges $\abund{C} = 5.20$--$5.45$\,dex and $\abund{O} = 6.10$--$6.70$\,dex for the 3D calculations and $\abund{C} = 5.20$--$5.45$\,dex and $\abund{O} = 6.40$--$7.10$\,dex in 1D. 
Spacing between abundance values varies between $0.02$ and $0.05$\,dex, with finer resolution being used around values closer to the best fitting abundances.
Obviously, the refinement of the grid of (C,O) abundances around the best fitting values requires a few iterations of our analysis.
Fortunately, the OH and CH equilibrium number densities and, consequently, the strength of the associated molecular bands are effectively insensitive to the N abundance, which reduces the size of the parameter space to be explored in our analysis. 
Without loss of generality, we therefore assume a fixed nitrogen abundance of $\abund{N} = 6.40$\,dex in both 3D and 1D calculations.
For the atomic lines, we adopt the elemental abundances derived by means of the equivalent width analysis in Sect.~\ref{sect:atomic}, with few minor adjustments to compensate for uncertainties in $\loggf$-values and to fine-tune the strength of emergent atomic features to improve the fit to the observations. 
We would like to stress here that our purpose is not to model these atomic lines to derive accurate abundances for elements other than C, N, and O but only to reproduce the effects of blends on molecular bands.
 
Since we aim at directly comparing the synthetic molecular bands with the observed spectra, in the calculations we need to account for additional broadening mechanisms that alter the shape of spectral lines while leaving their strength unaffected, namely, instrumental, rotational, and macro-turbulent broadening. 
The first two mechanisms have to be included in both 3D and 1D calculations of synthetic spectral bands. 
Instrumental broadening is related to the optical properties of the telescope and spectrograph and is usually determined via calibration of the instrument system itself. 
Rotational broadening has to be treated as a free parameter as the stellar rotation rate cannot be derived from ab initio modelling of stellar atmospheres and stellar convection. 
It can be calibrated to a certain extent, however, by fitting spectral lines to observational spectra with sufficiently high resolution.
The third mechanism is necessary only for 1D calculations  and serves to compensate, together with micro-turbulence, for the lack of a self-consistent description of macroscopic flows in the stellar atmosphere and associated Doppler broadening of spectral lines.
Current time-dependent 3D models appear to successfully reproduce the bulk of the distribution of flow velocities in the stellar atmosphere \citep[e.g.][]{Asplund:2000,Ramirez:2009,Ramirez:2010}, hence there is no need for additional macro-turbulent broadening in 3D spectral synthesis calculations.

Instrumental and macro-turbulent broadening are usually approximated by convolving the synthetic spectra with a Gaussian profile. 
Strictly, rotational broadening is not Gaussian, but can be implemented as such for small stellar rotation rates for the purpose of our fitting, without appreciable differences (\citealt{Ramirez:2010} find a projected rotational velocity $V{\sin{i}} \approx \,3.2$\,{\kms}).
In order to account for the above three broadening mechanisms, we therefore simply convolve the 1D and 3D synthetic spectra with Gaussian profiles.
We determine the widths of the Gaussians by fitting 1D and 3D synthetic line profiles to a selection of observed \ion{Fe}{i} lines. 
We find a full width at half maximum (FWHM) of $8.5$\,{\kms} and of $11$\,{\kms} for the 3D and 1D cases, respectively.

Once the 3D and 1D synthetic spectra have been computed and broadened, we determine the quality of the fit to the observed molecular bands for each combination of C and O abundances by calculating the corresponding $\chi^2$.
We mask out the more heavily blended portions of the molecular bands where modelling is more uncertain.
The best fitting $\abund{C}$ and $\abund{O}$ values are found by minimising $\chi^2$. Prior to the $\chi^2$ minimisation, in order to improve the accuracy of the procedure, we fit a high-order polynomial function of two variables to the actual $\chi^2$ surface and evaluate it on a finer grid of C and O abundances.

We follow a similar procedure for fitting the NH and CN bands. 
We compute the synthetic bands for different combinations of N and C abundances in the ranges $\abund{C} = 5.10$--$5.50$\,dex and $\abund{N} = 5.50$--$5.80$\,dex  for the 3D calculations and  $\abund{C} = 5.15$--$5.70$\,dex and $\abund{N} = 5.60$--$5.90$\,dex in 1D, with a resolution of $0.05$\,dex.
Sensitivity of the NH and CN bands to O abundance is low, nonetheless we adopt fixed values for $\abund{O}$ of $6.40$ and $6.72$\,dex, in 3D and 1D, respectively, that are very close to the O abundances we derive. The choice of these particular values is based on iterating the whole abundance analysis a number of times.
As for the other molecular bands, we include the effects of blends by atomic lines in the calculations and account for additional broadening mechanisms by convolving  the synthetic spectra with Gaussians.
The best fitting C and N abundances are then determined via a minimisation procedure similar to the one used for the analysis of the OH and CH features.

\section{Results}
\label{sect:results}

\begin{figure*}
  \centering
   \resizebox{\hsize}{!}{
   \includegraphics{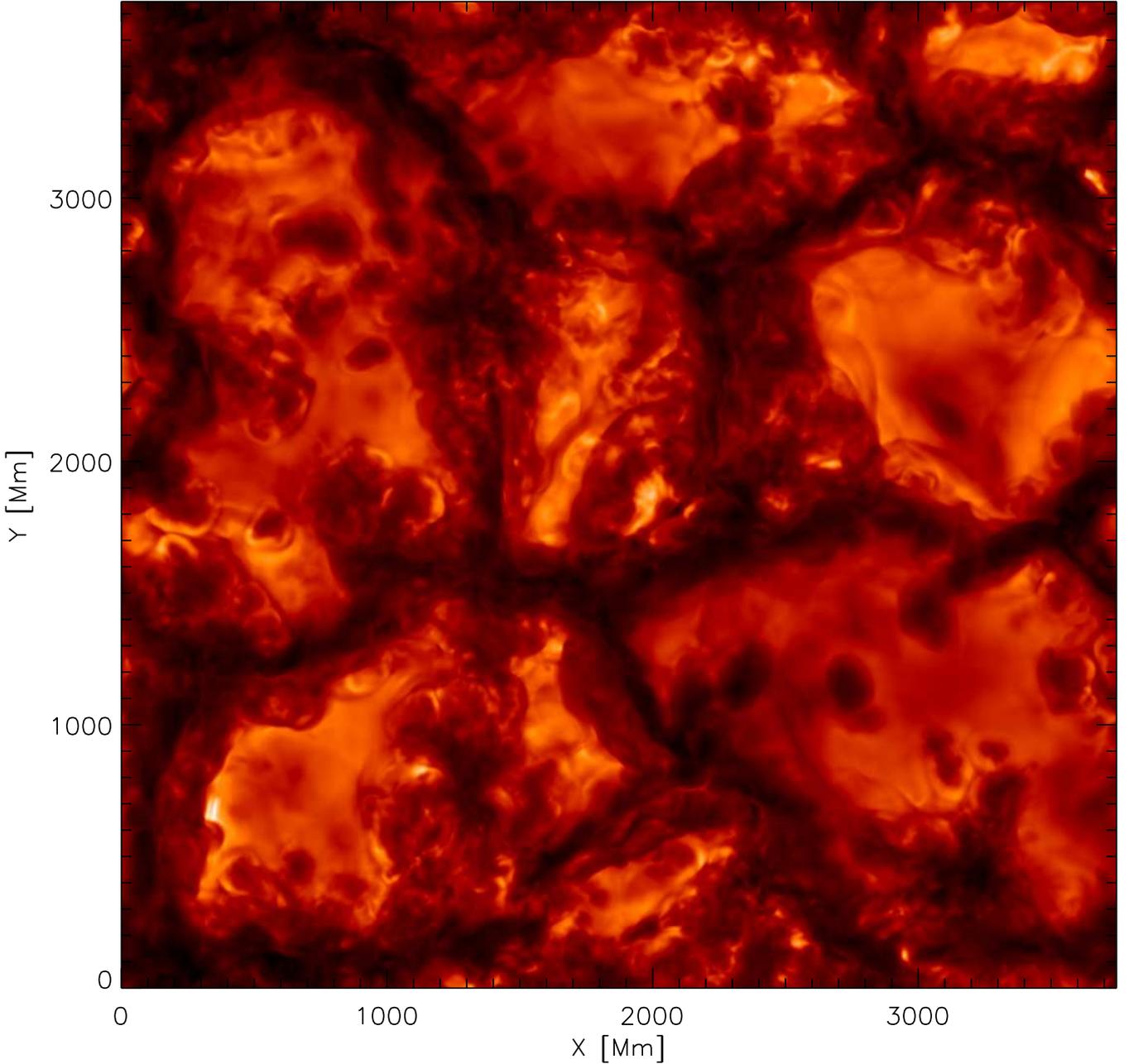}}
   \caption{Spatially resolved bolometric intensity pattern from the 48-bin $1008^2{\times}504$ surface convection simulation of {\hdstar}. 
    }
   \label{fig:surfint}
\end{figure*}

\begin{figure*}
  \centering
   \resizebox{\hsize}{!}{
   \includegraphics{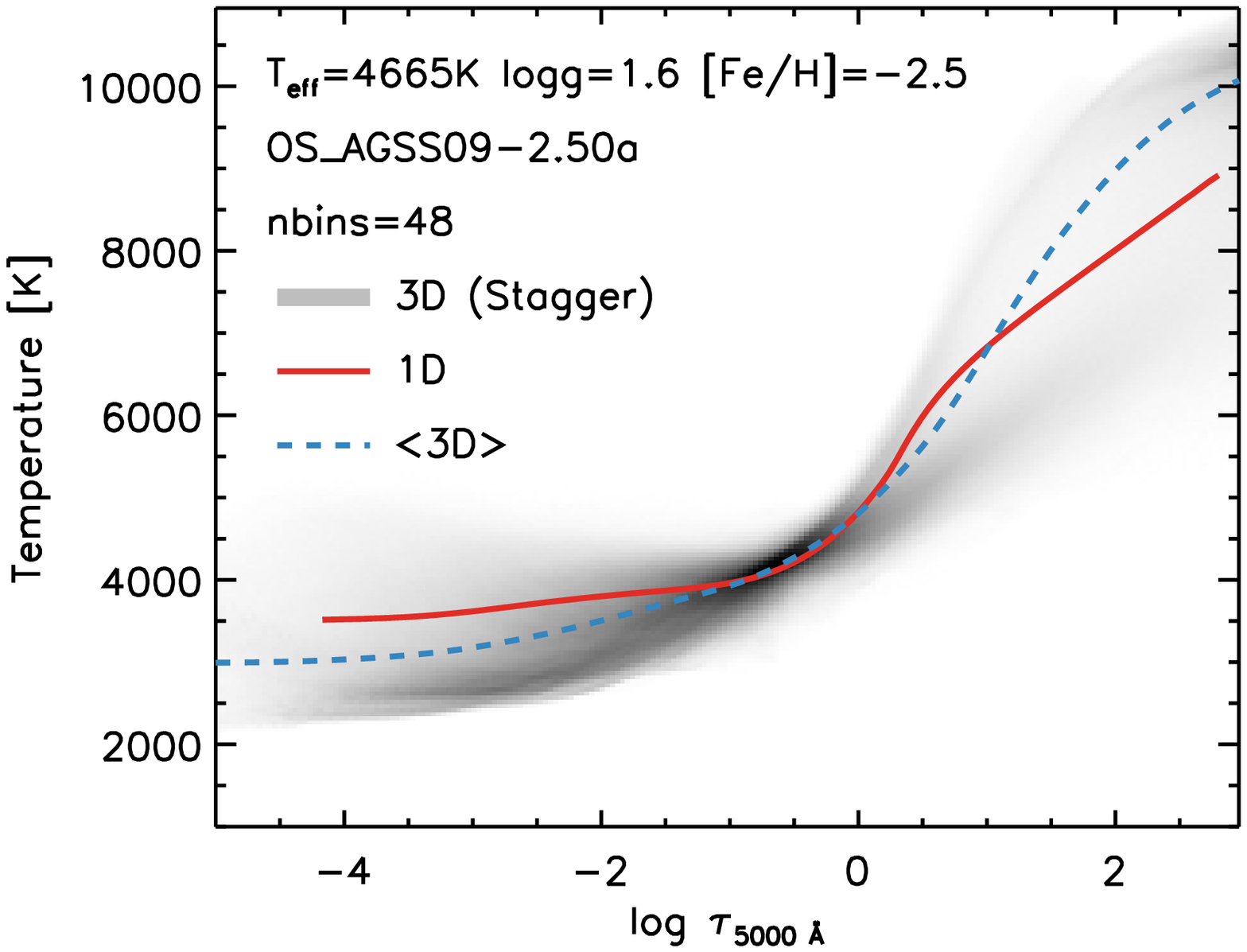}
   \includegraphics{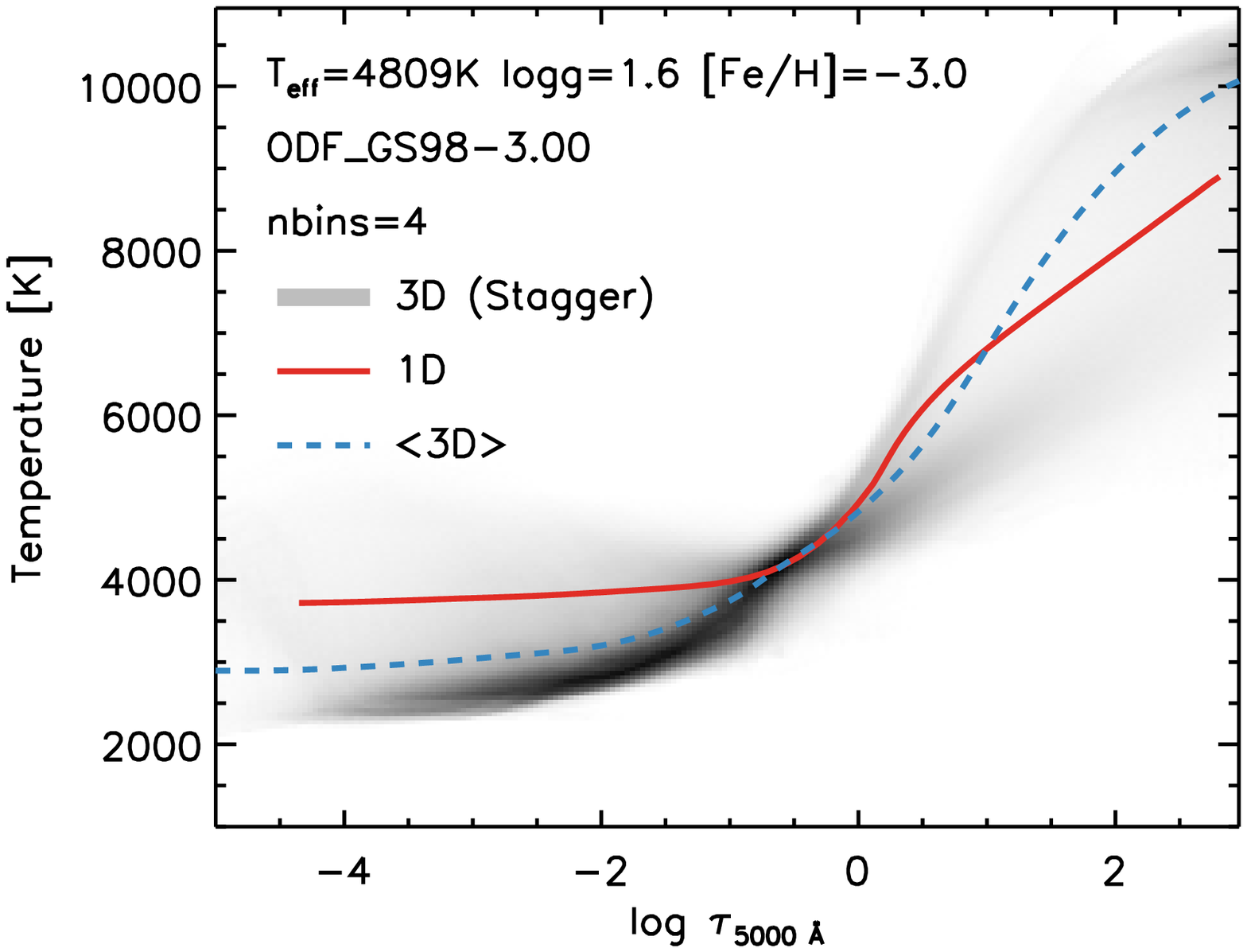} }
   \caption{\emph{Grey shaded areas}: histograms of the temperature distribution versus standard optical depth at $5000$\,{\AA} in the atmospheric layers of two $504^2{\times}252$ 3D surface convection simulations of {\hdstar}. Darker shades of grey indicate higher frequency of occurrence. \emph{Left}: temperature distribution as a function of standard optical depth for the reference 48-bin simulation at [Fe/H] $= -2.5$ and with [$\alpha$/Fe] $= +0.4$. \emph{Right}: temperature distribution for the 4-bin simulation based on an equation of state and ODF data computed for a scaled solar composition by \citet{Grevesse:1998} with [Fe/H] $= -3.0$. \emph{Dashed blue lines}: mean 3D temperature stratifications averaged on surfaces of constant optical depth. \emph{Continuous red lines}: corresponding stratifications from 1D model atmospheres computed for the same stellar parameters and with the same input physics as the simulations.}
   \label{fig:lgtautt}
\end{figure*}

\begin{figure*}
  \centering
   \resizebox{\hsize}{!}{
   \includegraphics{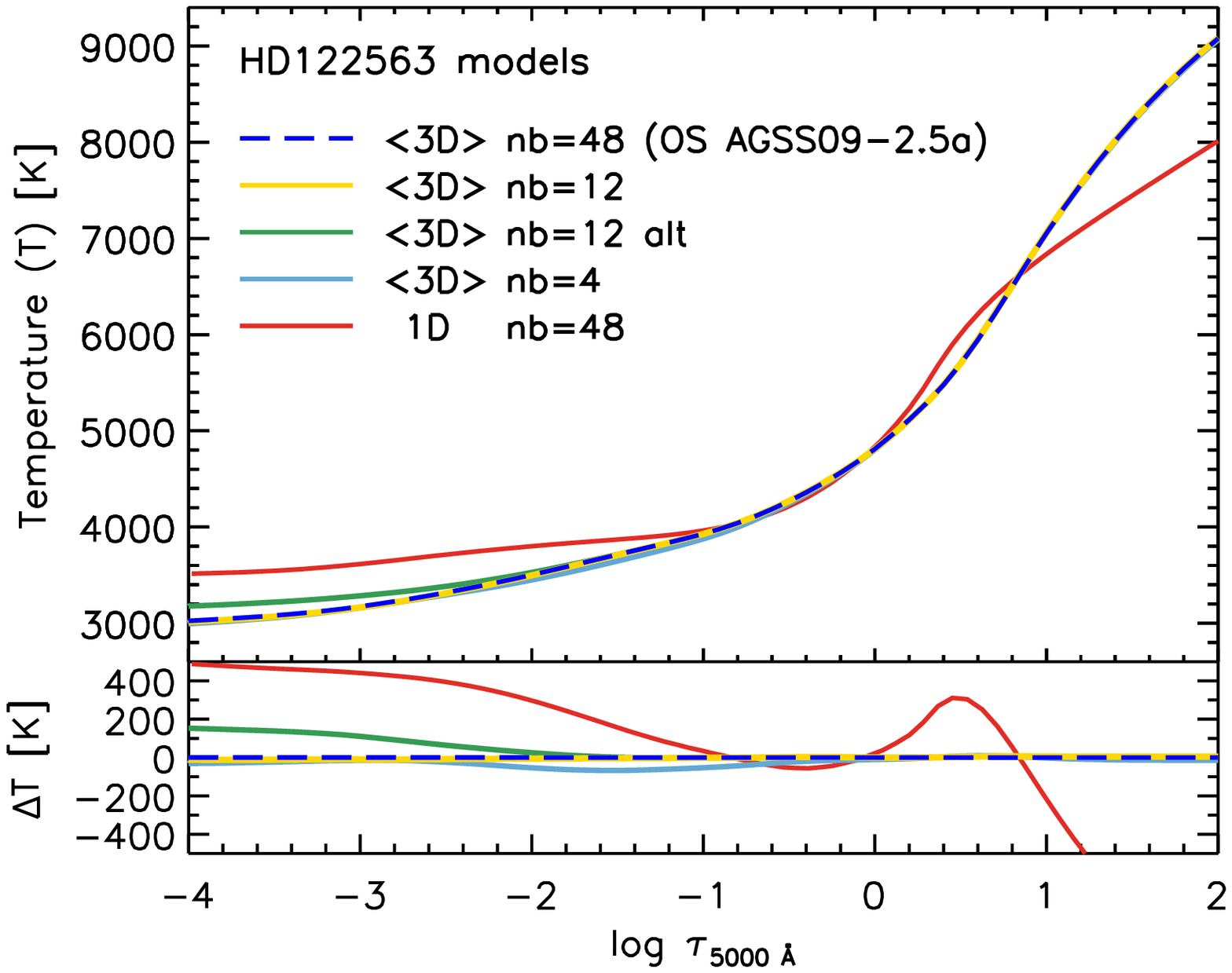}
   \includegraphics{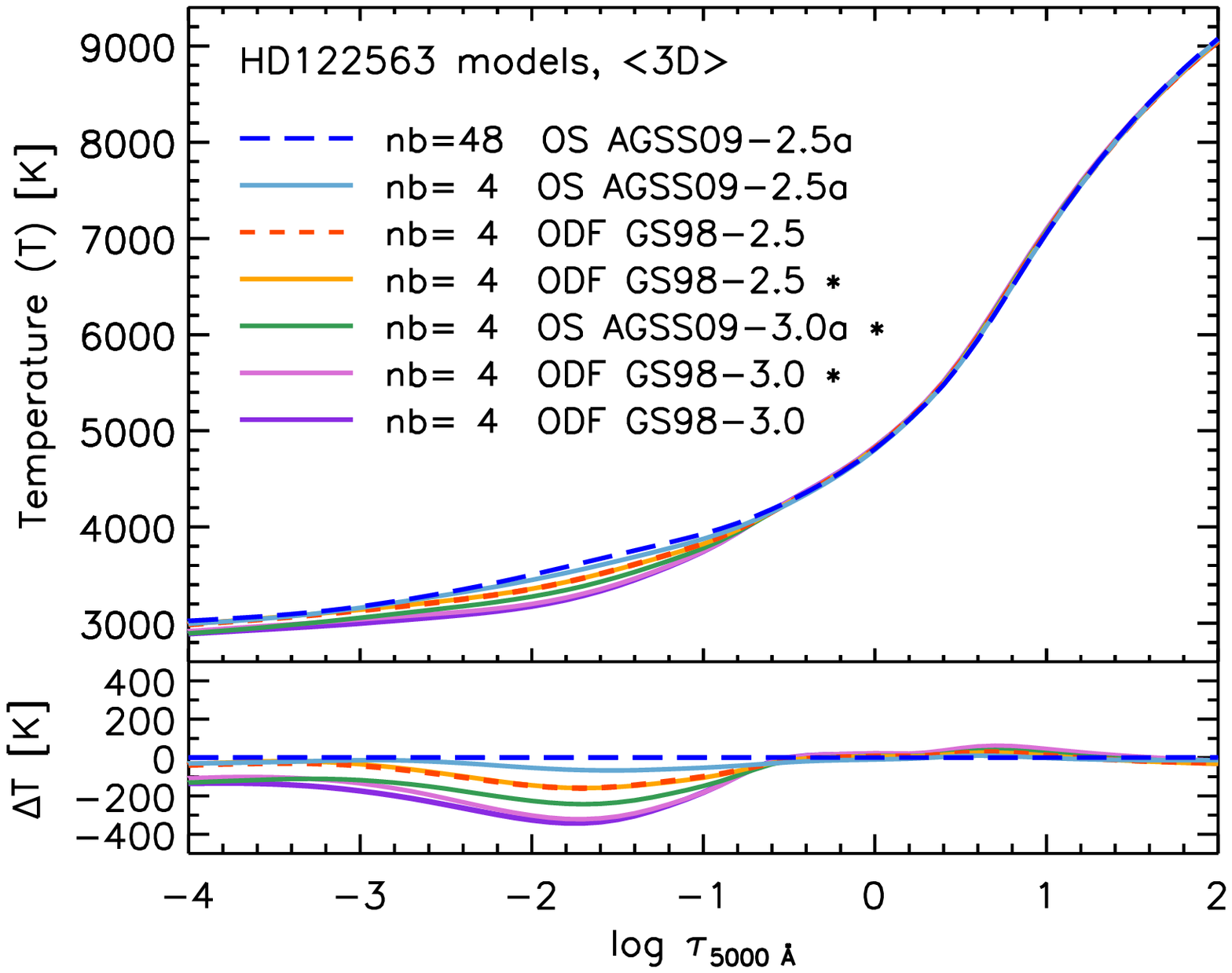} }
   \caption{Mean temperature stratification as a function of standard optical depth for various 3D surface convection simulations of {\hdstar} computed for different opacity binning realisations, chemical mixtures and line opacity data, and binning criteria for opacity strength. The mean stratification are constructed by averaging the gas temperature from the simulations on surfaces of constant optical depth $5000$\,{\AA}. \emph{Left}: comparison of the mean temperature stratifications from simulations generated for the scaled \citet{Asplund:2009} chemical composition with [Fe/H] $= -2.5$ and with [$\alpha$/Fe] $= +0.4$ but with different number of opacity bins. \emph{Dashed blue curve}: reference {\hdstar} simulation computed for 48 bins. \emph{Red line}: temperature profile from the  1D {\atmo} model atmosphere generated for the same stellar parameters.  \emph{Right}: comparison of mean temperature stratifications from simulations computed with four opacity bins but different chemical mixtures and input line opacity data (opacity sampling, OS, or opacity distribution functions, ODFs). The default measure of opacity strength in the opacity binning is the Rosseland optical depth of formation at a given wavelength. For the simulations marked with an \emph{asterisk}, the ratio of monochromatic to Rosseland opacity at formation height is adopted instead as binning criterion for opacity strength.}
   \label{fig:meanmod}
\end{figure*}

\subsection{3D simulations: temperature stratification}
\label{sect:temp3d}
Figure~\ref{fig:surfint} shows the spatially resolved emergent bolometric surface intensity from one snapshot of the high-resolution, 48-bin simulation of {\hdstar}, computed by integrating the contribution of all opacity bins to the vertical intensity. 
The granulation pattern with bright wide patches corresponding to warm upflows in the underlying convection zone and narrow dark lanes corresponding to cooler downflowing gas, typical of late-type stars, is evident.
While qualitatively the pattern resembles the one from observed and simulated solar-type granulation \citep[e.g.][]{Nordlund:2009}, there are some differences compared with the solar case.
The granules appear to possess a richer structure, particularly near the edges, which stems from the higher level of turbulence in the visible photospheric layers.
The granulation pattern predicted by the metal-poor red giant simulations generally shows a higher degree of inhomogeneity, with small cool patches forming more frequently within granules than in the solar case.

Figure~\ref{fig:lgtautt} (left panel) shows the temperature distribution as a function of standard optical depth at $5000$\,{\AA} in the atmospheric layers of the 48-bin $504^2{\times}252$ simulation of {\hdstar} computed for the scaled solar composition \citep{Asplund:2009} with [Fe/H] $= -2.5$ and [$\alpha$/Fe] $= +0.4$.
Compared with the stratification from the corresponding 1D {\atmo} model atmosphere computed for the same stellar parameters and with identical equation of state and opacity bins, the 3D simulation predicts on average cooler temperatures in the upper atmospheric layers.
This is a characteristic, well-established result from time-dependent 3D hydrodynamic surface convection simulations of late-type metal-poor stars \citep{Asplund:1999,Collet:2007,Collet:2011,Dobrovolskas:2013}.
In stationary, 1D, hydrostatic models, the assumptions of time independence and absence of velocity fields imply the temperature in the convectively stable upper layers of late-type stellar photospheres is solely controlled by radiative equilibrium.
The temperature in the upper atmospheric layers of 3D stellar convection simulations is instead determined primarily by the balance between radiative heating following absorption of radiation by spectral lines and adiabatic cooling associated with divergent gas flows above granules.
At solar metallicity, the line opacity in the upper photosphere is sufficiently high to provide enough radiative heating to keep the temperature stratification close to radiative-equilibrium levels as in 1D models.
The reduced line opacity in low-metallicity stellar atmospheres, on the other hand, implies the balance between radiative heating and mechanical cooling terms is reached at lower temperatures compared with the ones enforced by radiative equilibrium in 1D models.
Another characteristic difference is that the temperature in the layers of the 3D simulation immediately above the optical surface are slightly hotter on average than predicted by the corresponding 1D model atmosphere.
Compared with the mean 3D stratification, a slightly steeper temperature gradient is required in the 1D model near the sharp transition between the convective zone and the radiative-equilibrium dominated photosphere in order to carry the flux outwards.
Below the optical surface, the assumption of a 1D homogeneous stratification fails to capture the bimodal nature of the full 3D temperature-depth distribution, which is characterised by two main branches representing the stratification in warm upflows and cool downdrafts, respectively.

The fine details of the temperature stratification in the upper atmosphere are controlled by the assumed background chemical composition and by the choice of opacity package, opacity binning, and number of opacity bins.
The right panel of Fig.~\ref{fig:lgtautt} shows the temperature distribution resulting from another surface convection simulation of {\hdstar} that started from the same initial snapshot as the 48-bin simulation, but assumes instead a four-bin opacity binning realisation based on opacity distribution functions (ODFs) by \citet{Kurucz:1992,Kurucz:1993} that were computed for a scaled solar mixture by \citet{Grevesse:1998} with [Fe/H] $= -3.0$.
Also, the ratio of monochromatic to Rosseland opacity at the height where $\tau_\lambda = 1$ is used in this case as a measure of opacity strength to populate the bins.
As mentioned in Sect.~\ref{sect:binning}, this particular binning realisation is analogous to the one used for the surface convection simulation by \citet{Collet:2009}, allowing a direct comparison of the results from current and previous calculations.
We find that the older ODF data combined with the overall metal-poorer mixture effectively provide significantly lower line opacity than the present opacity sampling (OS) data.
This results in more pronounced cooling in the upper atmosphere of the 3D simulation and a flatter grey-like radiative-equilibrium stratification in the corresponding layers of the associated 1D model. 
The lower opacities imply the optical surface is also shifted slightly inwards, resulting in an increased effective temperature after relaxation compared to the 48-bin [Fe/H] $= -2.5$ simulation (${\Delta}{\teff}\approx 150$\,K).
The temperature stratification would therefore need to be scaled down at all depths in the four-bin [Fe/H] $= -3.0$ simulation in order to match {\hdstar}'s effective temperature, meaning that the upper atmospheric layers would end up being even cooler.

We have carried out a number of similar tests to study the response of the simulation's physical structure to different choices of number of opacity bins. 
Figure~\ref{fig:meanmod}, left panel, shows the temperature stratifications resulting from different opacity binning realisations based on the same opacity data as the 48-bin [Fe/H] $= -2.5$ simulation.
By carefully calibrating the binning, it is possible to achieve the same average stratification as the 48-bin simulation with only twelve bins.
However, we also show that a generic alternate binning realisation with twelve bins can lead to temperature differences of about $100$\,K in the uppermost atmospheric layers ($\log\tau_\text{5000\,\AA} \la -2.5$) with respect to our reference 48-bin  simulation.
The four-bin realisation provides a reasonably good agreement with both the 48-bin and twelve-bin cases, considering the inherently much more simplified opacity binning representation.
However, the four-bin simulation also results in cooler layers immediately above the optical surface ($-2.5 \la \log\tau_\text{5000\,\AA} \la 0.0$), hence a slightly steeper temperature gradient near continuum-forming regions, which, to first order, in LTE, would cause synthetic spectral lines to appear stronger.

Figure~\ref{fig:meanmod}, right panel, illustrates the results of a study of the response of the mean temperature stratification to changes in the assumed chemical mixture, line opacity data (OS or ODFs), and binning criterion for opacity strength.
in order to keep the analysis simple, we only consider opacity binning configurations with four bins in our tests, but nonetheless compare the results with the reference 48-bin [Fe/H] $= -2.5$ case.
All four-bin realisations lead to cooler stratifications in the upper photosphere and steeper temperature gradients around the continuum-forming region: it is clear that any four-bin scheme simply cannot capture in full the amount of radiative heating by spectral lines in the optically thin atmospheric layers.
As expected, temperature differences with respect to the 48-bin model are larger for the lower-metallicity mixtures, and are of the order of about $-200$\,K for a change $\Delta$[Fe/H] $= -0.5$ in metallicity. 
Also, as discussed earlier for the comparison with the \citet{Collet:2009} simulation, the four-bin realisations based on ODFs tend to underrepresent the level of line opacity in the upper atmosphere compared with the ones based on OS data and lead to cooler temperature stratifications by approximately $100$\,K for the same [Fe/H].\footnote{The ODF and OS data we use assume different scaled solar mixtures from \citet{Grevesse:1998} and \citet{Asplund:2009}, respectively. Nonetheless, the primary difference between the two data sets is not in the choice of reference solar abundances but in the effective level of line opacity coverage, which is intrinsically lower in the former set than in the latter.}
Finally, the adopted criterion for determining bin membership during binning --Rosseland optical depth of formation of a given wavelength versus ratio of monochromatic to Rosseland opacity at formation height-- can influence the temperature stratification, but the effect is generally small and, in practice, only appreciable in the lowest-metallicity cases (${\Delta}T \approx 50$\,K at most, in the optically thin layers).

\subsection{Fe lines}
\label{sect:fe_lines}

\begin{figure*}
  \centering
   \resizebox{\hsize}{!}{
   \includegraphics{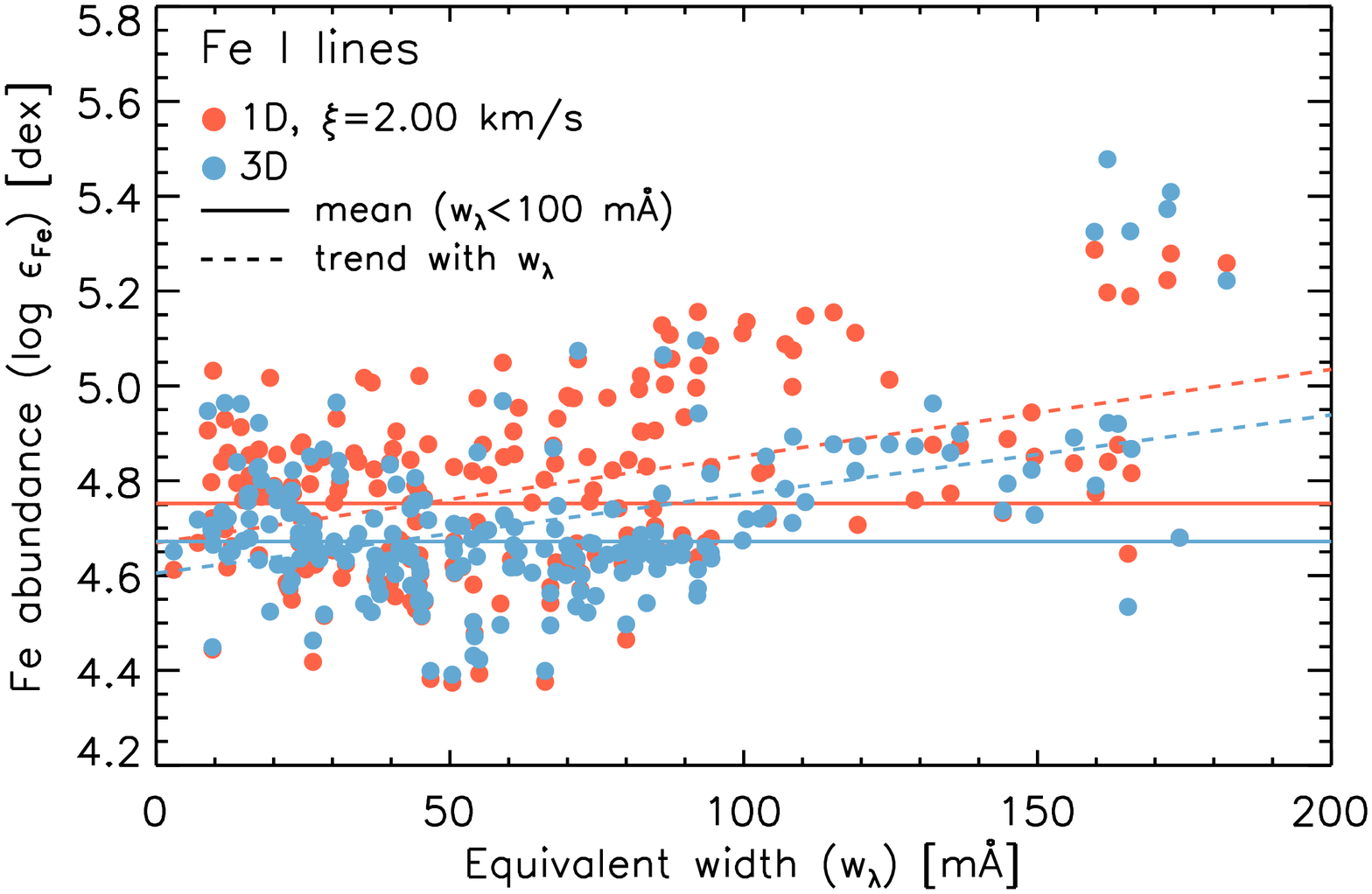}
   \includegraphics{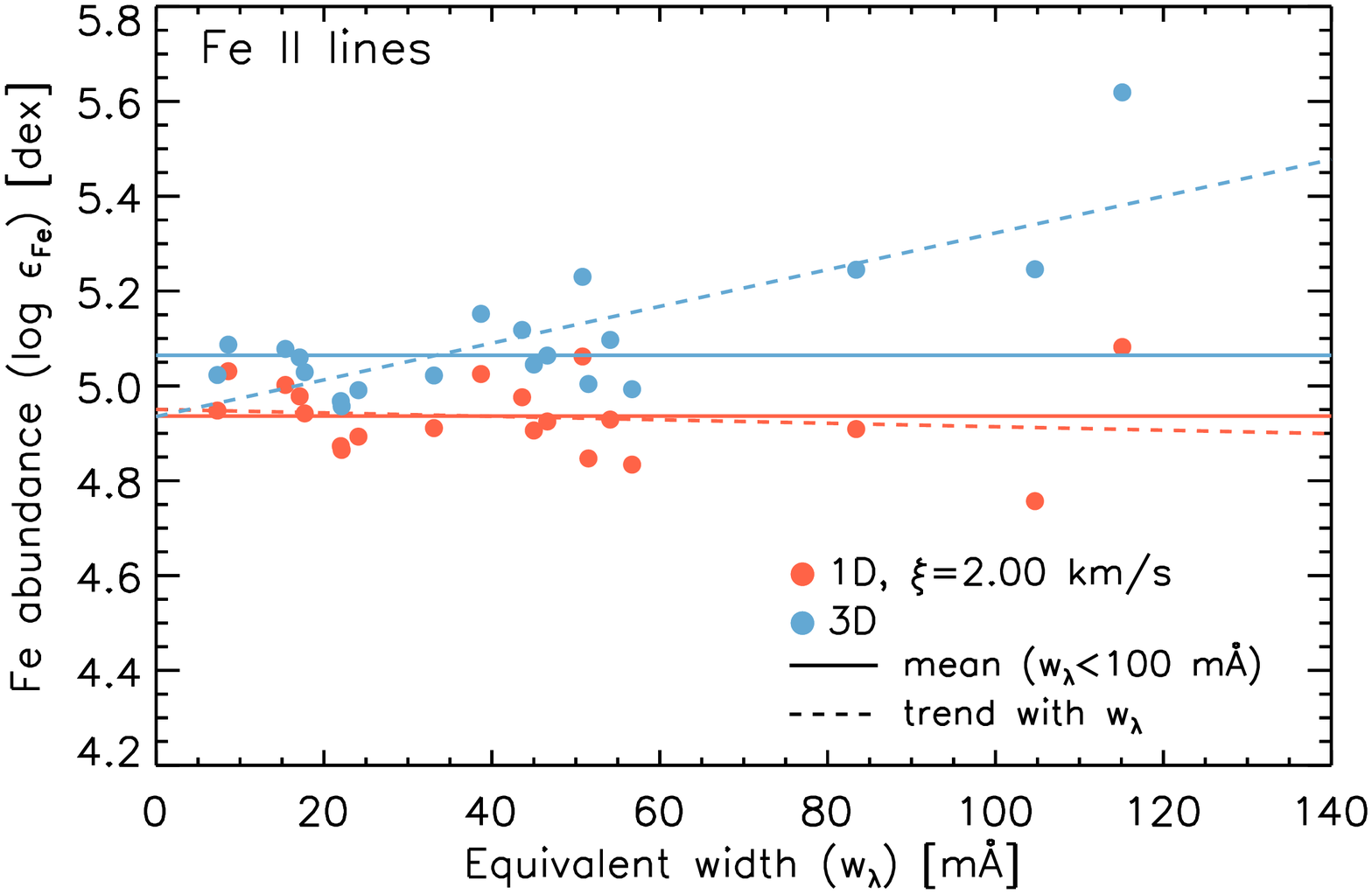} }
   \caption{Fe abundances as a function of equivalent width from \ion{Fe}{i} (\emph{left}) and \ion{Fe}{ii} lines (\emph{right}) as derived in LTE by the 3D (\emph{blue symbols}) and 1D (\emph{red symbols}) analyses. The \emph{continuous lines} indicate the average Fe abundance from lines weaker than $100$\,m{\AA}; the \emph{dashed lines} show the abundance trend with equivalent width. 
The cluster of points at high Fe abundances ($\abund{Fe,\,3D} > 5.2$\,dex) and large equivalent widths ($w_\lambda > 150$\,m{\AA}) on the left panel corresponds to strong low-excitation ($\chi_\text{l}$ < $1$\,eV) \ion{Fe}{i} lines  affected by blends. The strong \ion{Fe}{ii} line with $\abund{Fe}_\text{,\,3D} > 5.6$\,dex  on the right panel is also blended and therefore not a reliable for abundance analysis purposes.}
   \label{fig:fe_lines_eqw}
\end{figure*}

\begin{figure*}
  \centering
   \resizebox{\hsize}{!}{
   \includegraphics{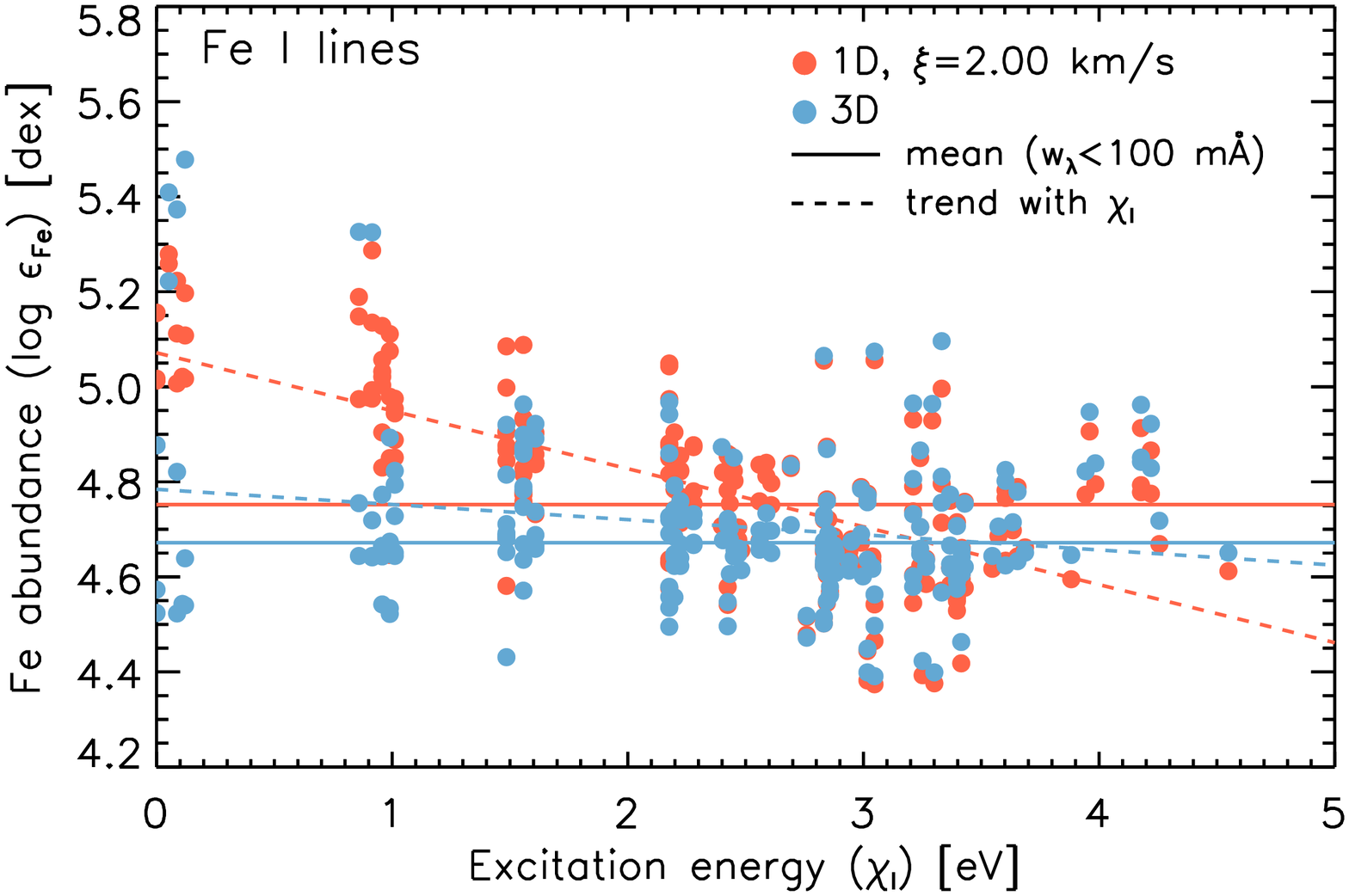}
   \includegraphics{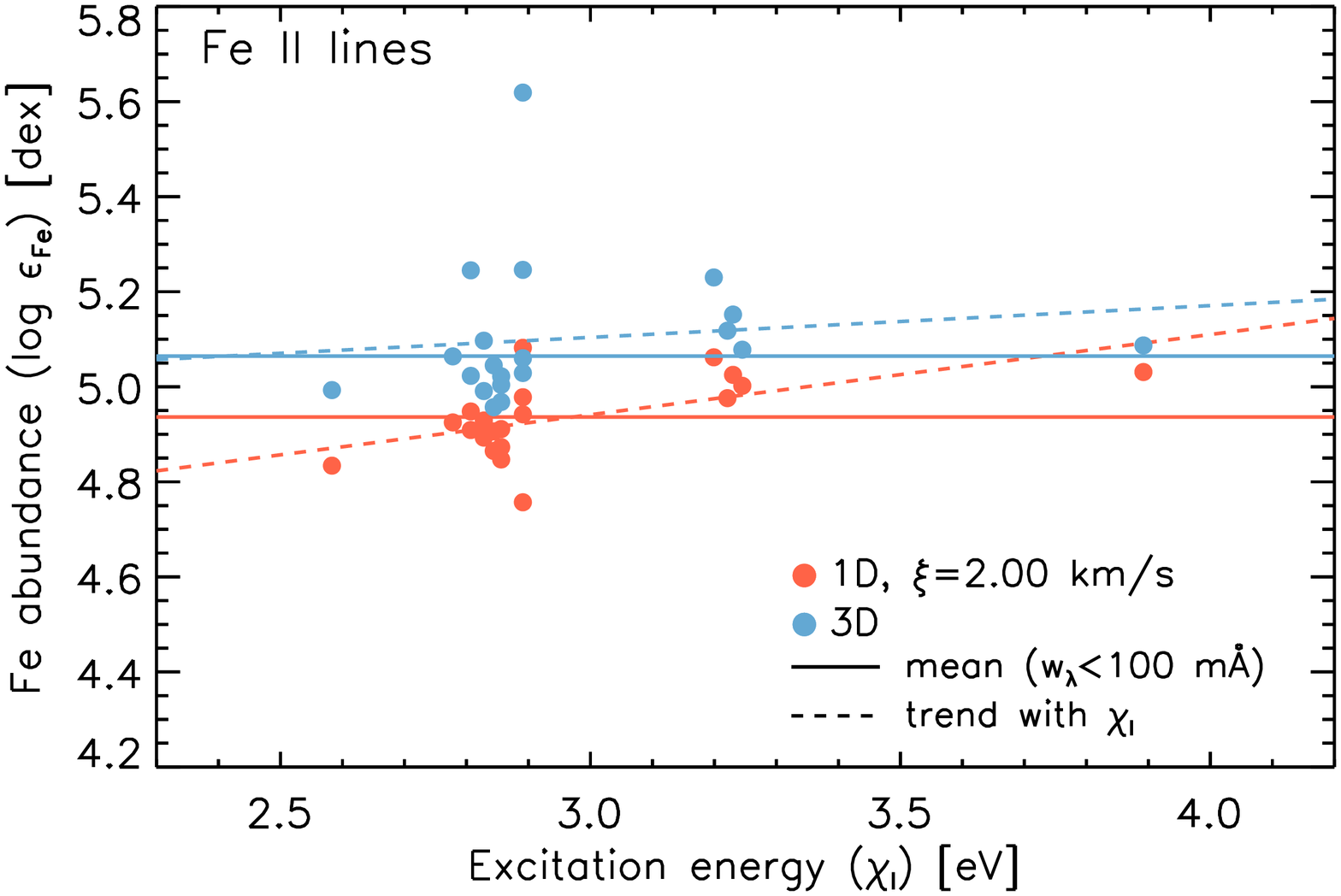} }
 \caption{Fe abundances as a function of lower-level excitation potential from \ion{Fe}{i} (\emph{left}) and \ion{Fe}{ii} lines (\emph{right}) as derived in LTE from the 3D (\emph{blue symbols}) and 1D (\emph{red symbols}) analyses. The \emph{continuous lines} indicate the average Fe abundance from lines with equivalent width smaller than $100$~m{\AA}; the \emph{dashed lines} show the abundance trend with excitation potential.}
   \label{fig:fe_lines_chi}
\end{figure*}

\begin{figure*}
  \centering
   \resizebox{\hsize}{!}{
   \includegraphics{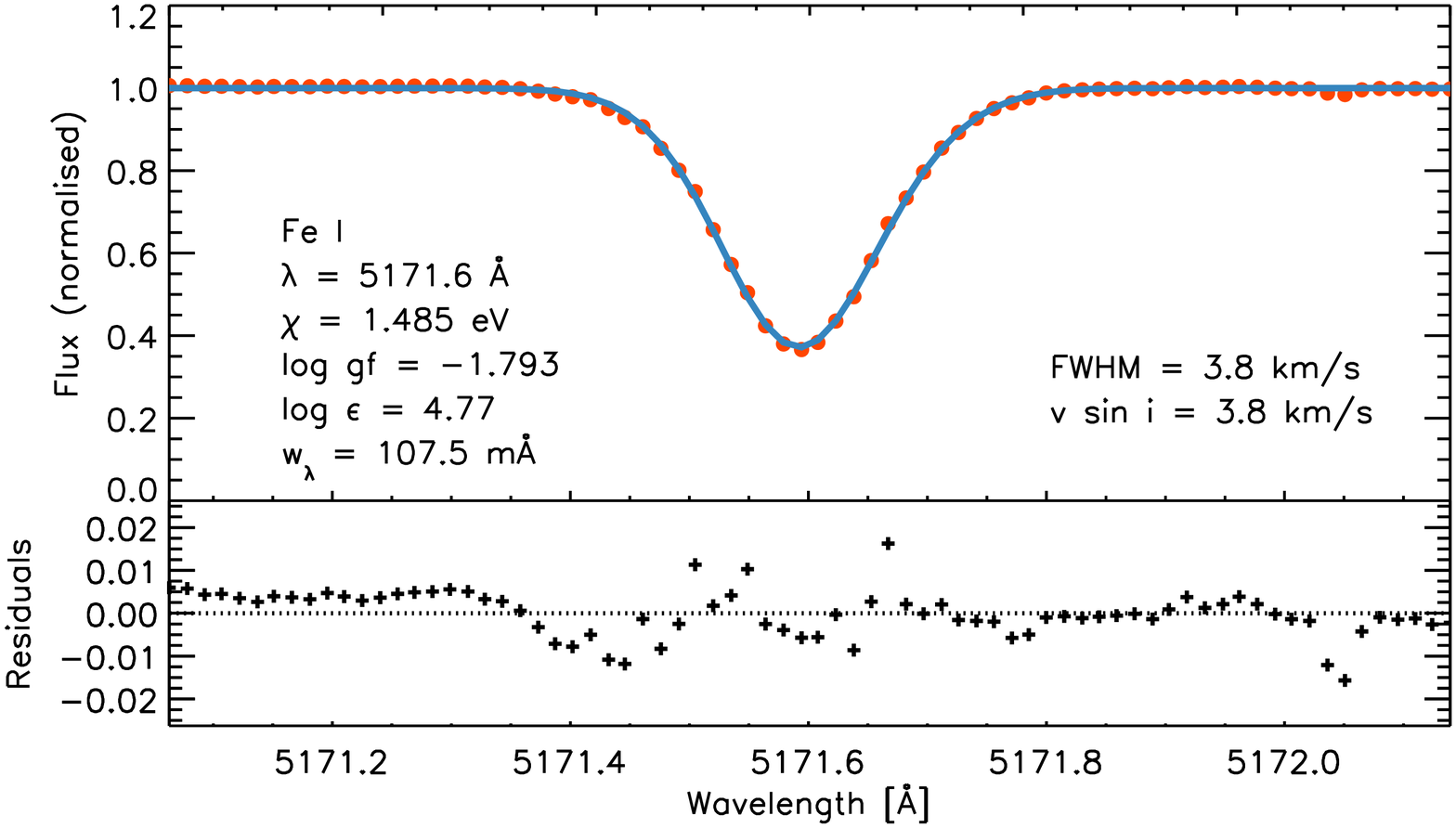}
   \includegraphics{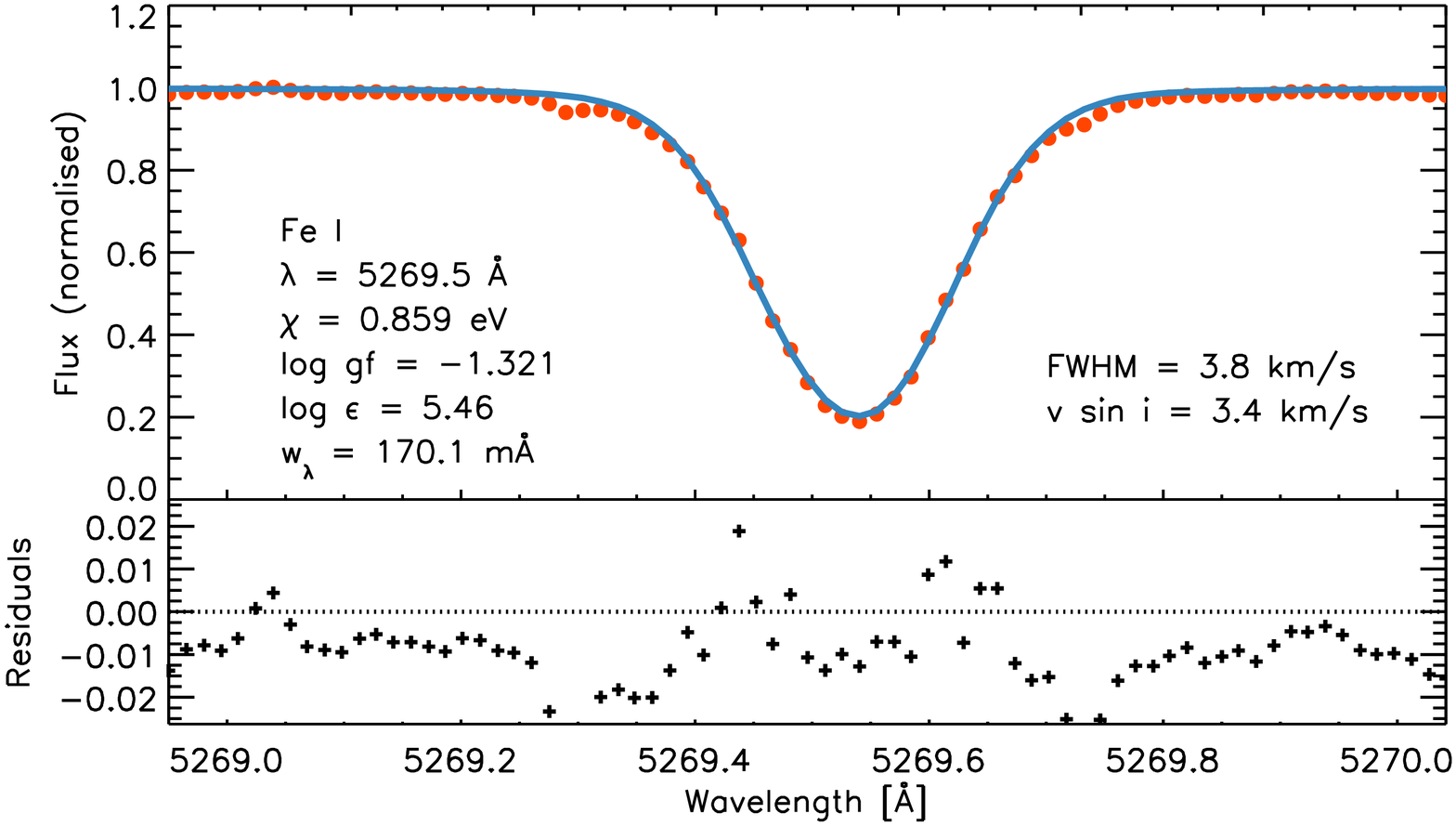} }
 \caption{Normalised flux profiles for two strong \ion{Fe}{i} lines. \emph{Red symbols}: observations (UVES POP spectrum); \emph{blue line}: best fitting line profiles; \emph{plus signs}: residuals (observed minus fit). For the fitting, we assumed a fixed Gaussian broadening of $3.8$~{\kms} and treated the equatorial rotational velocity projected along the line of sight, $v \sin i$ as a free broadening parameter. On the right panel, the relatively large deviations of the synthetic profile from the observed one are due to blends, which cause large uncertainties on the measured equivalent width of this very strong line. }
   \label{fig:fe_lines_prof}
\end{figure*}

Figure~\ref{fig:fe_lines_eqw} shows the iron abundances derived in LTE from \ion{Fe}{i} and \ion{Fe}{ii} lines as a function of equivalent width for the reference 48-bin 3D surface convection simulation of {\hdstar} and for the corresponding 1D {\atmo} model atmosphere computed for the same stellar parameters and with the same equation of state and opacity bins.

The Fe abundance values resulting from the 3D LTE analysis of \ion{Fe}{i} lines are generally lower than the ones determined in 1D LTE.
More specifically, from \ion{Fe}{i} lines weaker than $100$\,m{\AA}, we determine an iron abundance $\abund{Fe} \!=\! 4.67 \pm 0.12$ in 3D and $\abund{Fe} \!=\! 4.75 \pm 0.17$ in 1D.
Negative 3D$-$1D abundance corrections are expected for temperature-sensitive species such as \ion{Fe}{i} under the LTE approximation: the on-average cooler temperature stratification in the upper photospheric layers of the 3D model result in a larger fraction of neutral Fe compared with the 1D case.
The results of both 3D and 1D analyses of \ion{Fe}{i} lines, however, present a relatively large line-to-line scatter, which renders the derived average 3D$-$1D Fe abundance difference less significant.
Conversely, for the same reasons highlighted above, the analysis of \ion{Fe}{ii} lines returns moderately higher Fe abundance values in 3D than in 1D under the assumption of LTE. ($\abund{Fe} \!=\! 5.06 \pm 0.08$ and $\abund{Fe} \!=\! 4.94 \pm 0.07$ in 3D and 1D, respectively, for \ion{Fe}{ii} lines weaker than $100$\,m{\AA}.)
The presence of a difference between abundances derived from \ion{Fe}{i} and \ion{Fe}{ii} lines suggests possible residual uncertainties in the determination of the stellar parameters (surface gravity in particular) or significant departures of Fe ionisation equilibrium from LTE \citep{Amarsi:2016a}.

A slightly positive trend in the derived Fe abundance from \ion{Fe}{i} lines as a function of line strength is present in both the 3D and 1D analyses.
This points to some limitations in the analyses, as the derived Fe abundance should not depend on equivalent width.
The trend can be flattened in the 1D analysis by increasing the micro-turbulence parameter to $2.3$\,{\kms}; this solution is not satisfactory, however, as it has the effect to produce  a negative trend in the Fe abundance versus line strength derived from \ion{Fe}{ii} lines instead.
In 3D, we expressly avoid the use of any tuneable parameters like micro-turbulence to fit observed line strengths and artificially adjust the abundance trend as in the 1D analysis.

The presence of a residual positive trend of the Fe abundance with line strength in 3D could in principle be symptomatic of an insufficient numerical resolution of the 3D simulation which would result in an incomplete representation of the velocity field distribution particularly of the high-velocity tails, hence in too little turbulent broadening and smaller equivalent widths for the synthetic line profiles \citep[e.g.][]{Asplund:2000}.
On the other hand, with the reference $504^2{\times}252$ 3D simulation, we are able to reproduce the detailed shapes of flux profiles of strong Fe lines very well, with residuals of less than $1$\,\% near both line cores and wings (Fig.~\ref{fig:fe_lines_prof}). 
Also, line formation calculations carried out with the $1008^2{\times}504$-resolution simulation produce essentially the same outcome as with the $504^2{\times}252$-resolution simulation, indicating that for the purpose of spectral synthesis the velocity field in the reference 3D model atmosphere has effectively converged and suggesting that the reason for the positive Fe abundance trend with equivalent width is unlikely to be unresolved micro-turbulent-like flows in the simulations as in, e.g., \citet[][]{Allende-Prieto:2002a}.

A more likely possibility is that the trend may be due to inadequacies of the LTE approximation in reproducing the correct equivalent widths, particularly for the strong lines \citep{Amarsi:2016a}.
In general, very strong spectral lines are anyway less reliable than weak ones for the purpose of determining chemical compositions: first, their equivalent widths grow less than linearly with the number density of absorbers, making them only little sensitive to elemental abundances themselves; second, they are more likely to be affected by undetected blends that would result in too large measured equivalent widths hence erroneously high values of the derived elemental abundances.

Figure~\ref{fig:fe_lines_chi} shows line by line the derived Fe abundances as a function of the lower-level excitation potential $\chi_\text{l}$.
In 3D, the abundance trend with excitation potential from \ion{Fe}{i} lines is flat, albeit with a large scatter, particularly evident at $\chi_\text{l} \approx 3$\,eV. In the figure, the cluster of points with high Fe abundance ($\abund{Fe} > 5.2$\,dex) and low $\chi_\text{l}$ values corresponds to very strong \ion{Fe}{i} lines ($w_\lambda > 150$\,m{\AA}) that are affected by blends and for which the measured equivalent widths are therefore overestimated and unreliable. 
In 1D, on the contrary, the derived Fe LTE abundance from \ion{Fe}{i} lines shows a decreasing trend with lower-level excitation potential, even when only weak lines ($w_\lambda < 100$) are considered.
\ion{Fe}{ii} lines show a slightly increasing Fe abundances with excitation potential, with the trend being moderately steeper in 1D than in 3D and with an offset of about $0.1$~dex between 3D and 1D; however, as the selected \ion{Fe}{ii} lines span only a limited range of excitation potentials, the significance of these trend is relatively low. 
The outlying point with a derived Fe LTE abundance of $\abund{Fe} \approx 5.6$\,dex in Fig.\,\ref{fig:fe_lines_chi} corresponds again to a strong blended line (\ion{Fe}{ii} $5018.4$\,{\AA}). 
Our results are in excellent agreement with the Fe LTE abundance analysis by \citet{Amarsi:2016a}. The latter used our 1D and 3D reference model stellar atmospheres but adopted a different line list including only weak Fe lines with equivalent widths smaller than ${\sim}70$\,m{\AA}.
When restricting our analysis to only the Fe lines in common with \citet{Amarsi:2016a}, we can reproduce very closely their derived Fe LTE abundance trends with lower-level excitation potential and equivalent width as well as their low line-to-line scatter due to their careful selection of unblended features, with only negligible differences stemming from the independent choice of adopted line parameters (lower-level excitation potential $\chi_\text{l}$ and $\loggf$ values).
In particular, such differences are too small to explain the 3D and 1D Fe LTE abundance trends with excitation potential or the offset between Fe LTE abundances derived from \ion{Fe}{i} and \ion{Fe}{ii} lines.

The Fe LTE abundance offset between neutral and singly ionised Fe lines can be reduced by adopting different values for the star's $\teff$ and $\log g$. 
To examine the sensitivity of our results to the above parameters, we have scaled our reference 3D simulation and constructed two 3D models, one with a higher effective temperature, $\teff \approx 4\,750$\,K, and the other with a lower surface gravity, $\log g = 1.1$\,({\cmss}), and used them to re-derive Fe LTE abundances for a representative subset of weak \ion{Fe}{i} and \ion{Fe}{ii} lines.
We find that increasing the effective temperature of the 3D model by $\Delta\teff \approx 90$\,K causes the value of the Fe LTE abundance derived from \ion{Fe}{i} lines to increase by ${\sim}0.15$\,dex while leaving the one derived from \ion{Fe}{ii} lines virtually unaltered (with only a ${\sim}+0.02$\,dex change).
Decreasing the surface gravity from $\log g = 1.6$ to $1.1$\,({\cmss}) instead does not change in an appreciable way the Fe LTE abundance value derived from \ion{Fe}{i} lines but results in a ${\sim}0.20$\,dex-lower Fe LTE abundance as derived from \ion{Fe}{ii} lines.
In 1D, the response to changes in effective temperature and surface gravity is qualitatively similar, but weaker than in 3D.
Also, to first degree, the decreasing Fe LTE abundance trend with lower-level excitation potential derived for \ion{Fe}{i} lines in 1D is not rectified by changes in the surface gravity and is in fact exacerbated by the effective temperature increase.
A summary of the analysis of the response of the derived Fe LTE abundance to stellar parameter changes is given in Table\,\ref{tab:abund-params}.

In 3D, the combined effect of a higher effective temperature and lower surface gravity than the ones adopted for our reference model would allow to fulfil ionisation equilibrium between \ion{Fe}{i} and \ion{Fe}{ii} in terms of derived Fe LTE abundances. 
We caution however that from fundamental stellar parameter determinations there is presently no evidence in support of such a high value for {\hdstar}'s effective temperature.
Also, \citet{Amarsi:2016a} actually showed that the difference between the Fe abundances derived for the two ionisation stages can be significantly reduced by carrying out the analysis with full 3D non-local thermodynamic equilibrium (non-LTE) calculations.
Furthermore, they showed that in order to completely remove the above offset in the 3D non-LTE case one would also need to adopt a lower value of $\log g \approx 1.1$\,({\cmss}) for the {\hdstar}'s surface gravity, implying that the  value of the star's parallax derived from {\hipparcos} data may be significantly overestimated.  

\begin{table*}
\caption{Response of Fe, C, and O LTE abundances ($\log\epsilon$) derived from weak \ion{Fe}{i}, \ion{Fe}{ii}, CH, OH, and [\ion{O}{i}] lines to changes in effective temperature and surface gravity in the 3D and 1D reference models ($\teff\approx 4\,665$\,K, $\log g = 1.6$\,{\cmss}, [Fe/H]$=-2.5$). For the 1D line synthesis calculations, we adopt a micro-turbulence value of $\xi = 2.0$\,{\kms}.}
\begin{center}
\begin{tabular}{ccccccc}
\hline
 \multicolumn{2}{c}{3D model} & \multicolumn{5}{c}{$\Delta\log\epsilon$\,[dex]} \\
 $\Delta{\teff}$\,[K] & $\Delta{\log g}$\,[\cmss] & Fe (\ion{Fe}{i}) & Fe (\ion{Fe}{ii}) & C (CH) & O (OH) & O ([\ion{O}{i}]) \\
\hline
 $+90$ & $ 0  $ & $+0.15$ & $+0.02$ & $+0.21$ & $+0.26$ & $+0.08$ \\
 $  0$ & $-0.5$ & $+0.01$ & $-0.21$ & $+0.07$ & $+0.12$ & $-0.18$ \\
\hline
\multicolumn{2}{c}{1D model} & \multicolumn{5}{c}{$\Delta\log\epsilon$\,[dex]} \\
 $\Delta{\teff}$\,[K] & $\Delta{\log g}$\,[\cmss] & Fe (\ion{Fe}{i}) & Fe (\ion{Fe}{ii}) & C (CH) & O (OH) & O ([\ion{O}{i}]) \\
\hline
 $+90$ & $ 0  $ & $+0.11$ & $+0.00$ & $+0.18$ & $+0.20$ & $+0.06$ \\
 $  0$ & $-0.5$ & $-0.02$ & $-0.20$ & $+0.08$ & $+0.09$ & $-0.18$ \\
\hline
\end{tabular}
\end{center}
\label{tab:abund-params}
\end{table*}

\subsection{Atomic lines}
\label{sect:atom-res}

\begin{table*}
\caption{{\hdstar}'s elemental abundances for different neutral and singly ionised atomic species. Line list and equivalent widths are taken from \citet{Aoki:2007}, but with updated line parameters (lower-level excitation potential and $\loggf$ values). For some ions (e.g. \ion{Fe}{i}), very strong and obviously blended lines have been excluded from the analysis. Averages and standard deviations have been computed over all spectral lines selected for the analysis. Note that, in the case of Fe, restricting the analysis to Fe lines with equivalent widths $w_\lambda < 100$~\,m{\AA} would yield values of $\abund{Fe,\,3D}\!=\!4.67\!\pm\!0.12$\,dex and $5.06\!\pm\!0.08$\,dex ($\abund{Fe,\,1D}\!=\!4.75\!\pm\!0.17$\,dex and $4.94\!\pm\!0.07$\,dex, with $\xi=2.0$\,{\kms}) for the Fe abundance from neutral and singly ionised lines, respectively, as stated in the main text.}
\begin{center}
\begin{tabular}{lcrrrrrrr}
	\hline
	Element &
	Ion &
	$N_\text{lines}$ &
	\multicolumn{2}{c}{$\log\epsilon \pm \sigma$} &
	\multicolumn{2}{c}{$\log\epsilon \pm \sigma$} &
	\multicolumn{2}{c}{$\log\epsilon \pm \sigma$} \\
	{} &
	{} &
	{} &
	\multicolumn{2}{c}{3D} &
	\multicolumn{2}{c}{1D} &
	\multicolumn{2}{c}{1D} \\
	{} &
	{} &
	{} &
	\multicolumn{2}{c}{} &
	\multicolumn{2}{c}{$\xi = 2.0$\,{\kms}} &
	\multicolumn{2}{c}{$\xi = 2.3$\,{\kms}} \\
	\hline
    O  &   {\sc I}  &  $  2$ &   $ 6.77$  &  $ 0.03$  &  $ 6.80$  &  $ 0.03$  &  $ 6.80$  &  $ 0.03$  \\
   Na  &   {\sc I}  &  $  2$ &   $ 3.32$  &  $ 0.06$  &  $ 3.29$  &  $ 0.06$  &  $ 3.29$  &  $ 0.06$  \\
   Mg  &   {\sc I}  &  $  8$ &   $ 5.47$  &  $ 0.19$  &  $ 5.41$  &  $ 0.20$  &  $ 5.33$  &  $ 0.13$  \\
   Al  &   {\sc I}  &  $  2$ &   $ 3.33$  &  $ 0.56$  &  $ 3.57$  &  $ 0.45$  &  $ 3.36$  &  $ 0.47$  \\
   Si  &   {\sc I}  &  $  2$ &   $ 5.35$  &  $ 0.18$  &  $ 5.35$  &  $ 0.27$  &  $ 5.24$  &  $ 0.25$  \\
   Ca  &   {\sc I}  &  $ 20$ &   $ 3.84$  &  $ 0.12$  &  $ 3.84$  &  $ 0.12$  &  $ 3.81$  &  $ 0.12$  \\
   Sc  &  {\sc II}  &  $ 14$ &   $ 0.75$  &  $ 0.12$  &  $ 0.68$  &  $ 0.10$  &  $ 0.63$  &  $ 0.13$  \\
   Ti  &   {\sc I}  &  $ 27$ &   $ 2.03$  &  $ 0.12$  &  $ 2.36$  &  $ 0.08$  &  $ 2.33$  &  $ 0.09$  \\
   Ti  &  {\sc II}  &  $ 42$ &   $ 2.78$  &  $ 0.22$  &  $ 2.65$  &  $ 0.19$  &  $ 2.56$  &  $ 0.19$  \\
    V  &   {\sc I}  &  $  3$ &   $ 3.56$  &  $ 0.33$  &  $ 3.68$  &  $ 0.44$  &  $ 3.67$  &  $ 0.44$  \\
    V  &  {\sc II}  &  $  4$ &   $ 1.55$  &  $ 0.10$  &  $ 1.47$  &  $ 0.10$  &  $ 1.44$  &  $ 0.10$  \\
   Cr  &   {\sc I}  &  $ 13$ &   $ 2.36$  &  $ 0.16$  &  $ 2.58$  &  $ 0.12$  &  $ 2.52$  &  $ 0.19$  \\
   Mn  &   {\sc I}  &  $  7$ &   $ 2.26$  &  $ 0.20$  &  $ 2.34$  &  $ 0.23$  &  $ 2.23$  &  $ 0.13$  \\
   Fe  &   {\sc I}  &  $208$ &   $ 4.69$  &  $ 0.13$  &  $ 4.77$  &  $ 0.17$  &  $ 4.68$  &  $ 0.17$  \\
   Fe  &  {\sc II}  &  $ 20$ &   $ 5.10$  &  $ 0.15$  &  $ 4.93$  &  $ 0.08$  &  $ 4.88$  &  $ 0.11$  \\
   Co  &   {\sc I}  &  $  7$ &   $ 2.18$  &  $ 0.14$  &  $ 2.42$  &  $ 0.17$  &  $ 2.28$  &  $ 0.18$  \\
   Ni  &   {\sc I}  &  $ 15$ &   $ 3.57$  &  $ 0.12$  &  $ 3.58$  &  $ 0.12$  &  $ 3.53$  &  $ 0.18$  \\
   Zn  &   {\sc I}  &  $  2$ &   $ 2.11$  &  $ 0.01$  &  $ 2.00$  &  $ 0.00$  &  $ 1.99$  &  $ 0.00$  \\
   Sr  &  {\sc II}  &  $  2$ &   $ 0.25$  &  $ 0.51$  &  $ 0.08$  &  $ 0.39$  &  $-0.19$  &  $ 0.26$  \\
\hline
\end{tabular}
\end{center}
\label{tab:atom-res}
\end{table*}

Table~\ref{tab:atom-res} summarises the results of the 3D and 1D abundance analyses of atomic lines of various elements using the reference 48-bin 3D and 1D models and the assumption of local thermodynamic equilibrium.
For the derivation of the 1D abundances we consider two micro-turbulence values, $2.0$ and $2.3$\,{\kms}, as discussed in Sect.~\ref{sect:fe_lines}.
The quoted uncertainties for each tabulated elemental abundance are the line-to-line scatter.
The results of the 1D analysis are in overall very good agreement with the values reported by \citet{Aoki:2007}. 
Some differences exist between the results of the two 1D analyses, but are generally within the uncertainties and can otherwise be easily explained by the difference in adopted surface gravity (\citeauthor{Aoki:2007} use a value of $\log{g} = 1.1$), models, or atomic line parameters.
The derived elemental abundances for some ions such as \ion{Al}{i} and \ion{Sr}{i} present a large line-to-line scatter, which is owed to the fact that only few and very strong features subject to blends are available for the analysis of these species.
In the case of \ion{Fe}{i}, we have also excluded about ten of the strongest and most obviously blended lines from the analysis, thereby significantly reducing the line-to-line abundance scatter.

We would like to emphasise once again that both the 3D and 1D abundance analyses presented here have been carried out under the assumption of LTE.
We caution that departures from the latter may be important at this low metallicity, particularly but not exclusively for minority species. 
Owing to the steeper and cooler temperature stratification of the 3D model's upper photospheric layers, we expect that non-LTE effects on derived elemental abundances may be more significant in 3D than 1D \citep{Asplund:2005a}, possibly also contributing to explain the large line-to-line abundance scatter for some of the ions relative to the one obtained through the 1D analysis with micro-turbulence $\xi=2.3$\,{\kms}.
We defer the analysis of non-LTE effects for such ions to a future work.

\begin{figure*}
  \centering
   \resizebox{\hsize}{!}{
	   \includegraphics{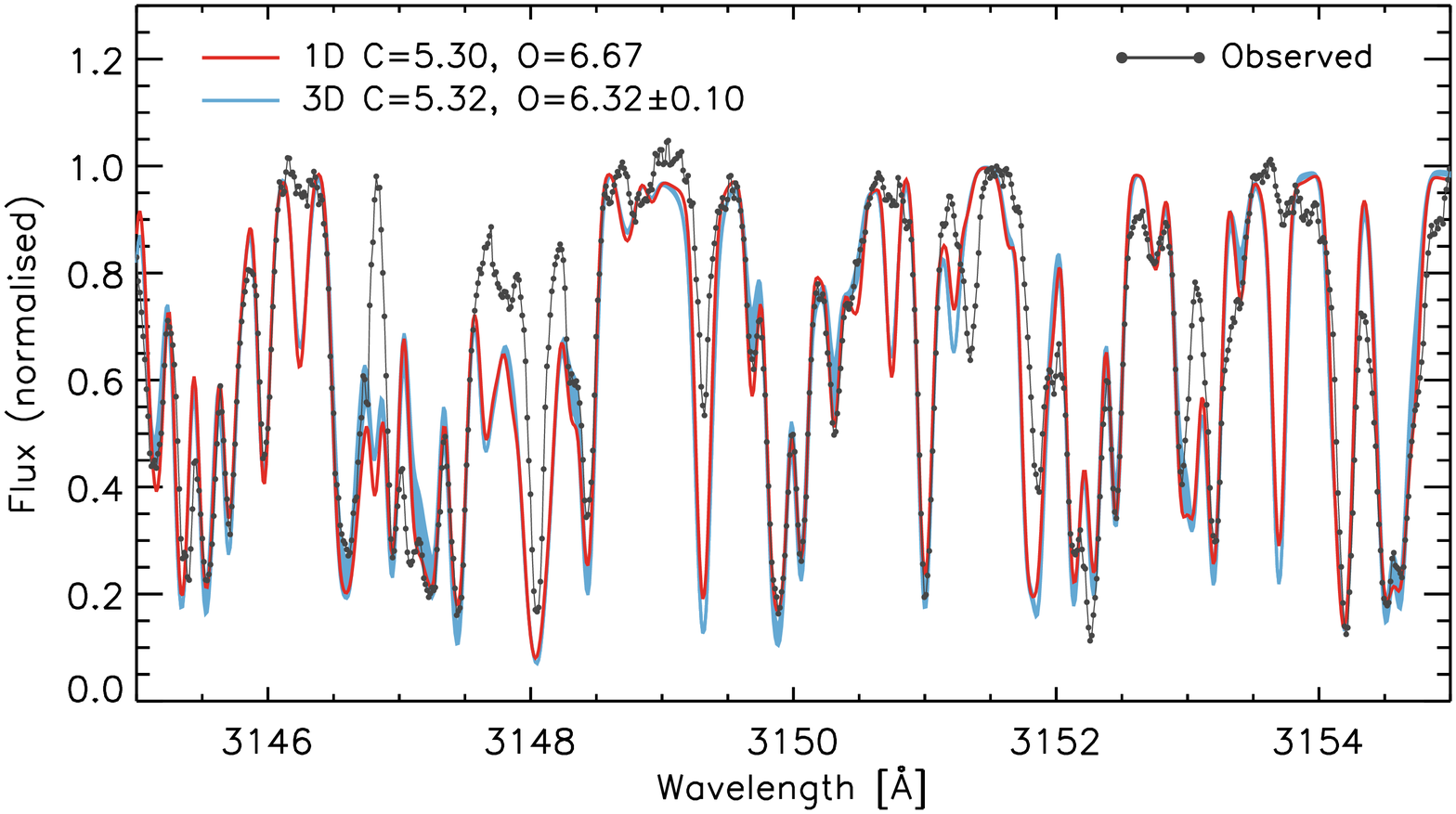}
	   \includegraphics{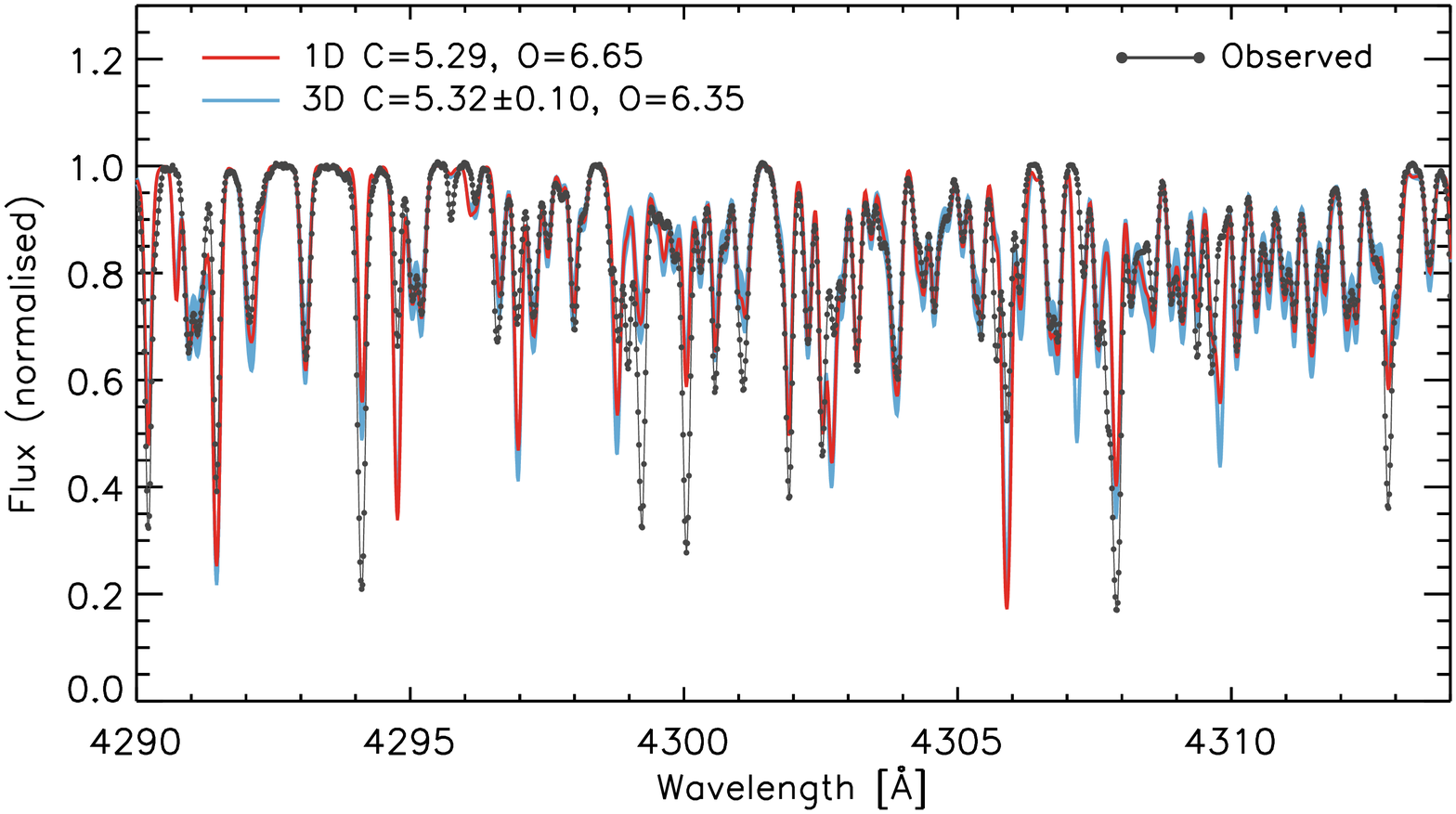} 
	}
   \resizebox{\hsize}{!}{
	   \includegraphics{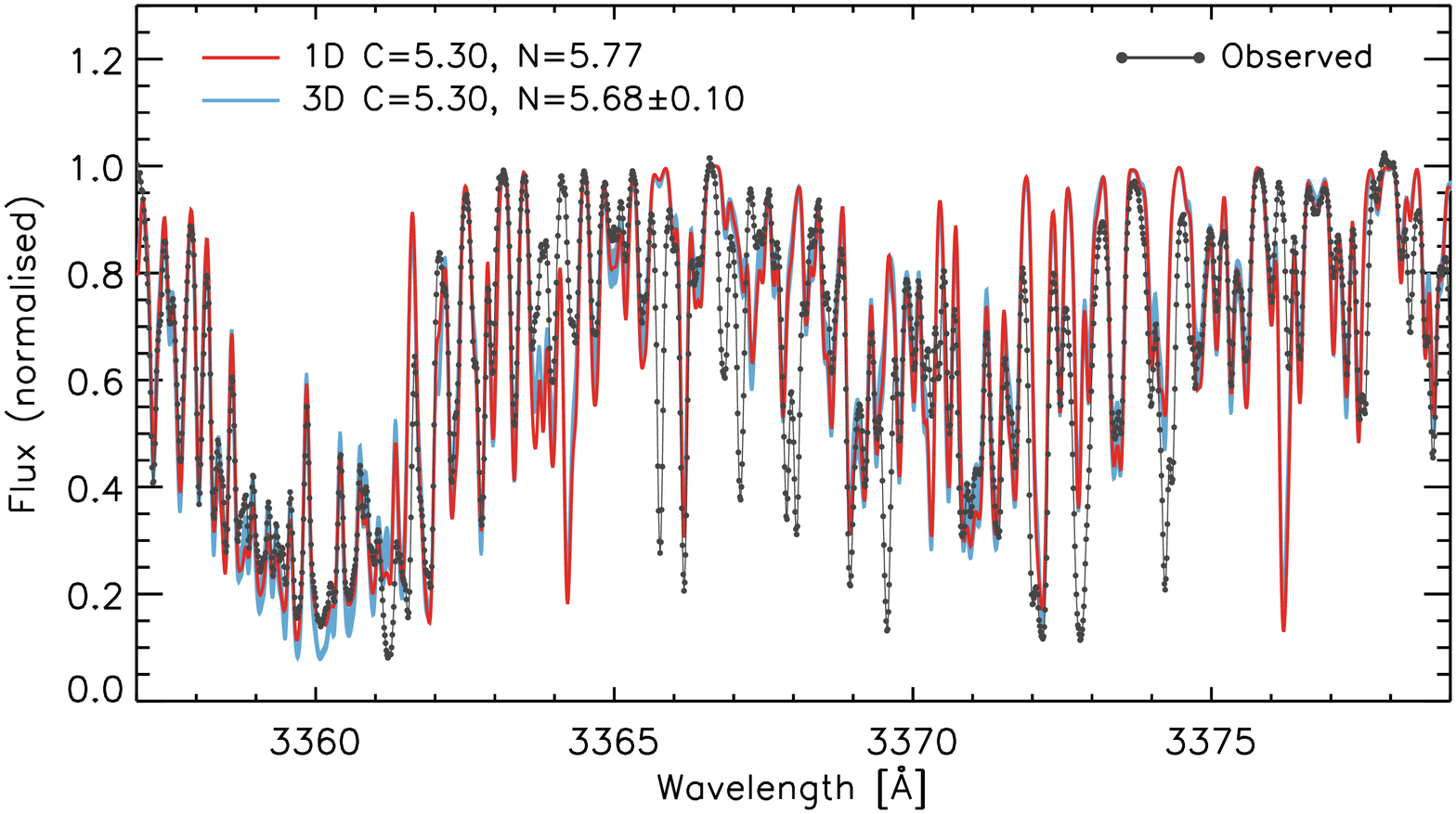}
	   \includegraphics{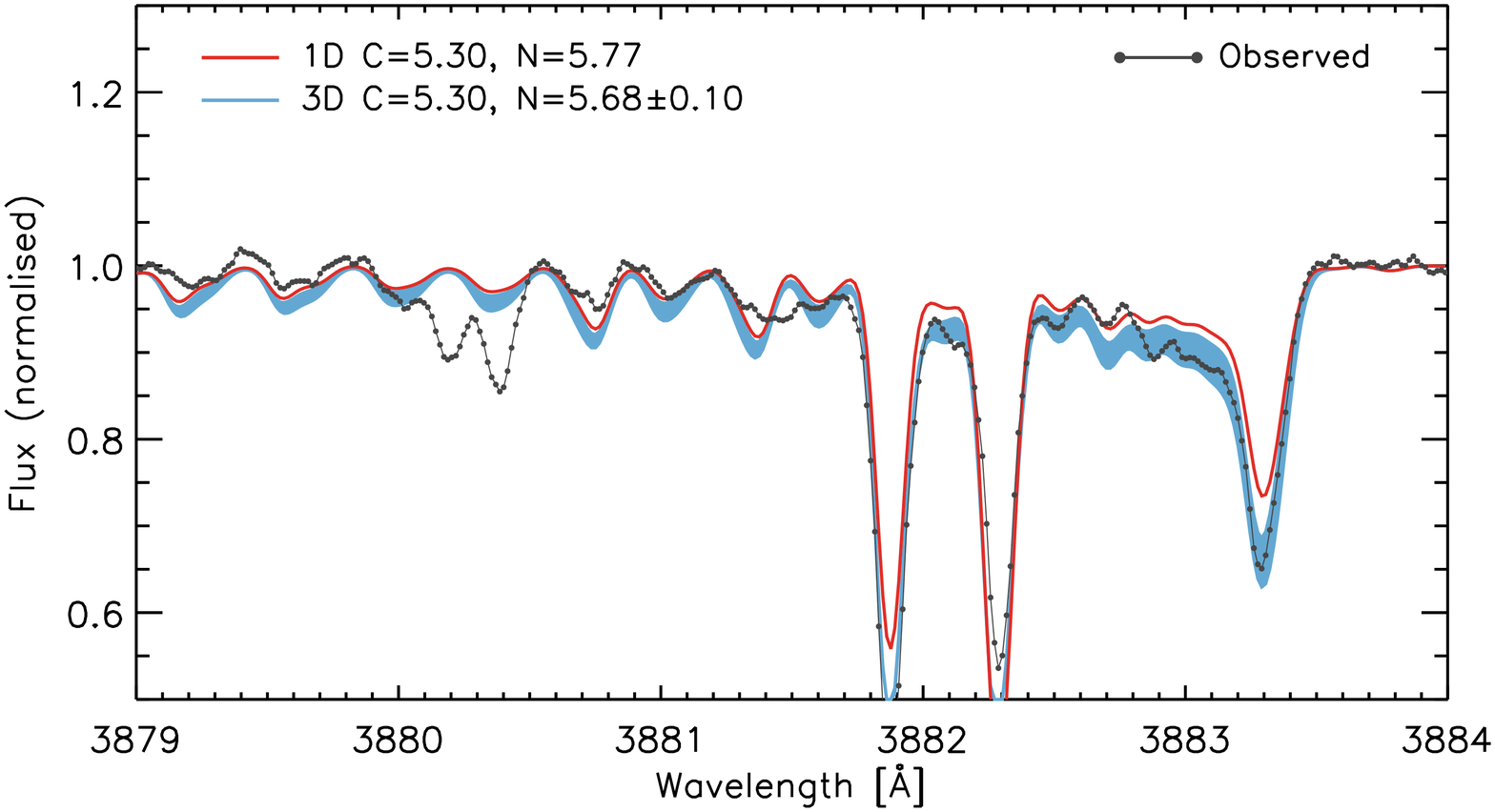}
	}
   \caption{Portions of {\hdstar}'s spectral regions including lines from the OH UV band (\emph{upper left}), CH lines in the G band (\emph{upper right}), NH UV lines (\emph{lower left}), and CN lines from the violet system (\emph{lower right}). \emph{Filled circles (connected)}: observed spectrum; \emph{red lines}: best fitting 1D synthetic spectra; \emph{blue lines}: best fitting 3D synthetic spectra;}
   \label{fig:mol_bands}
\end{figure*}

\subsection{Carbon and oxygen abundances from molecular lines}
\label{sect:carboxyg}

\begin{figure*}
  \centering
   \resizebox{\hsize}{!}{
   \includegraphics{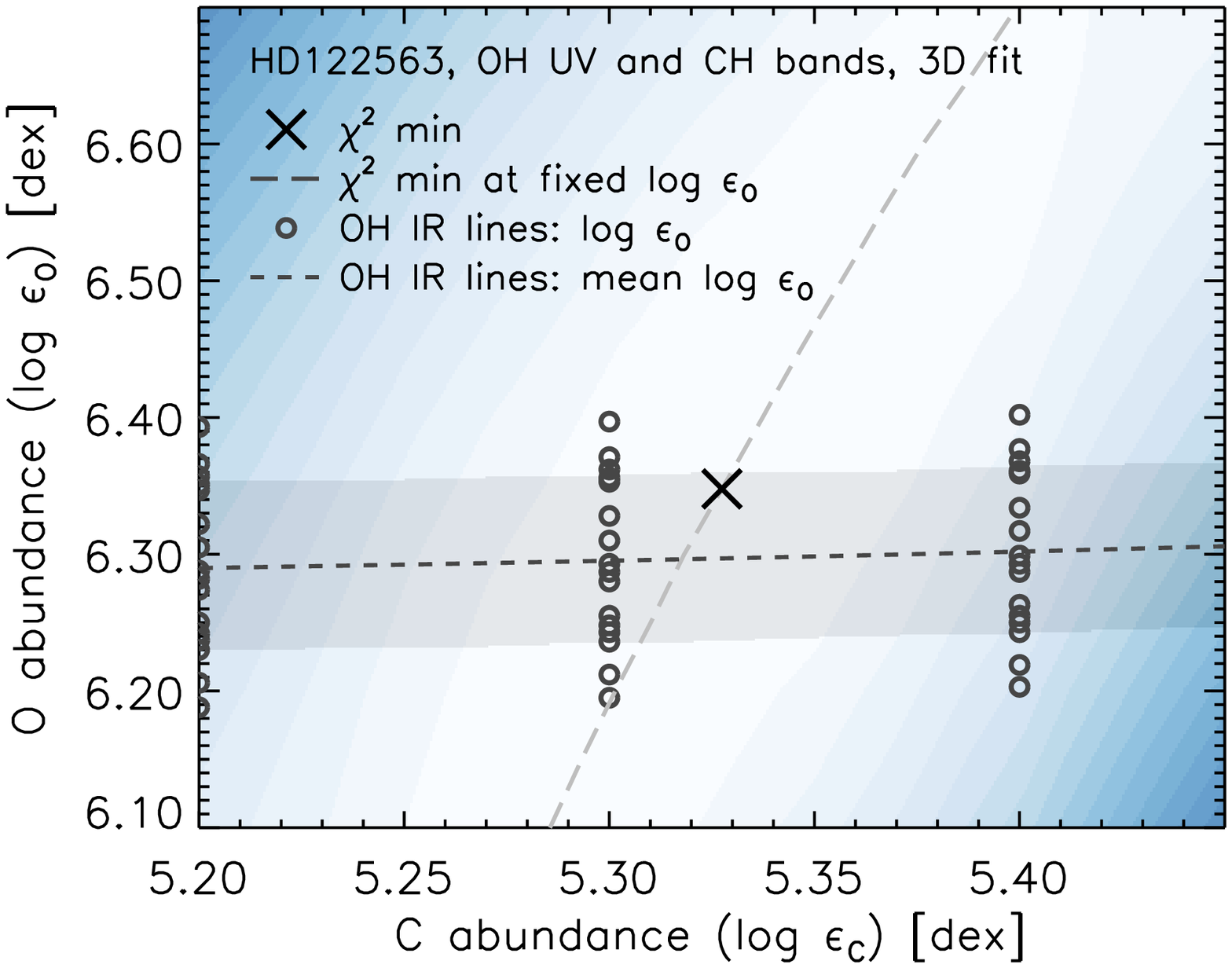}
   \includegraphics{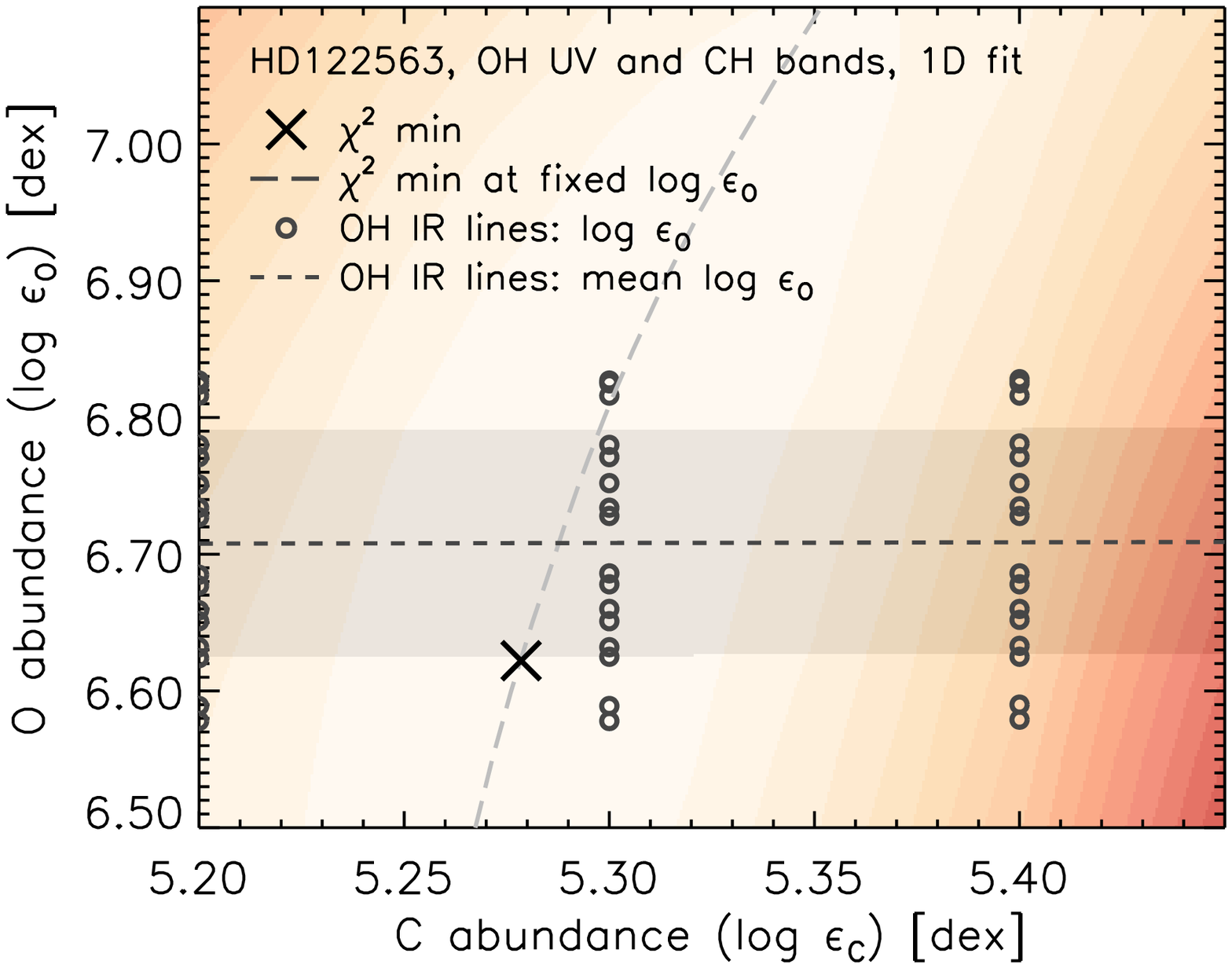} }
   \caption{Contours of constant $\chi^2$ from the fitting of OH and CH bands in the UV and optical. \emph{Blue} contours refer to the 3D analysis (\emph{left panel}), \emph{red} ones to the 1D analysis (\emph{right panel}). \emph{Dotted lines} show the location of the $\chi^2$ minimum in the C-O abundance plane. \emph{Long-dashed lines} trace the loci of the $\chi^2$ minima at constant oxygen abundance. \emph{Filled circles} represent the O abundance values derived from individual OH IR lines at constant C abundance. The \emph{short-dashed lines} indicate the average O abundances derived from OH IR lines as a function of the C abundance. The \emph{grey bands} overlain on the contours represent the 1-$\sigma$ dispersion ranges around the average values.}
   \label{fig:ch_oh_fit}
\end{figure*}

\begin{figure}
  \centering
   \includegraphics[width=\columnwidth]{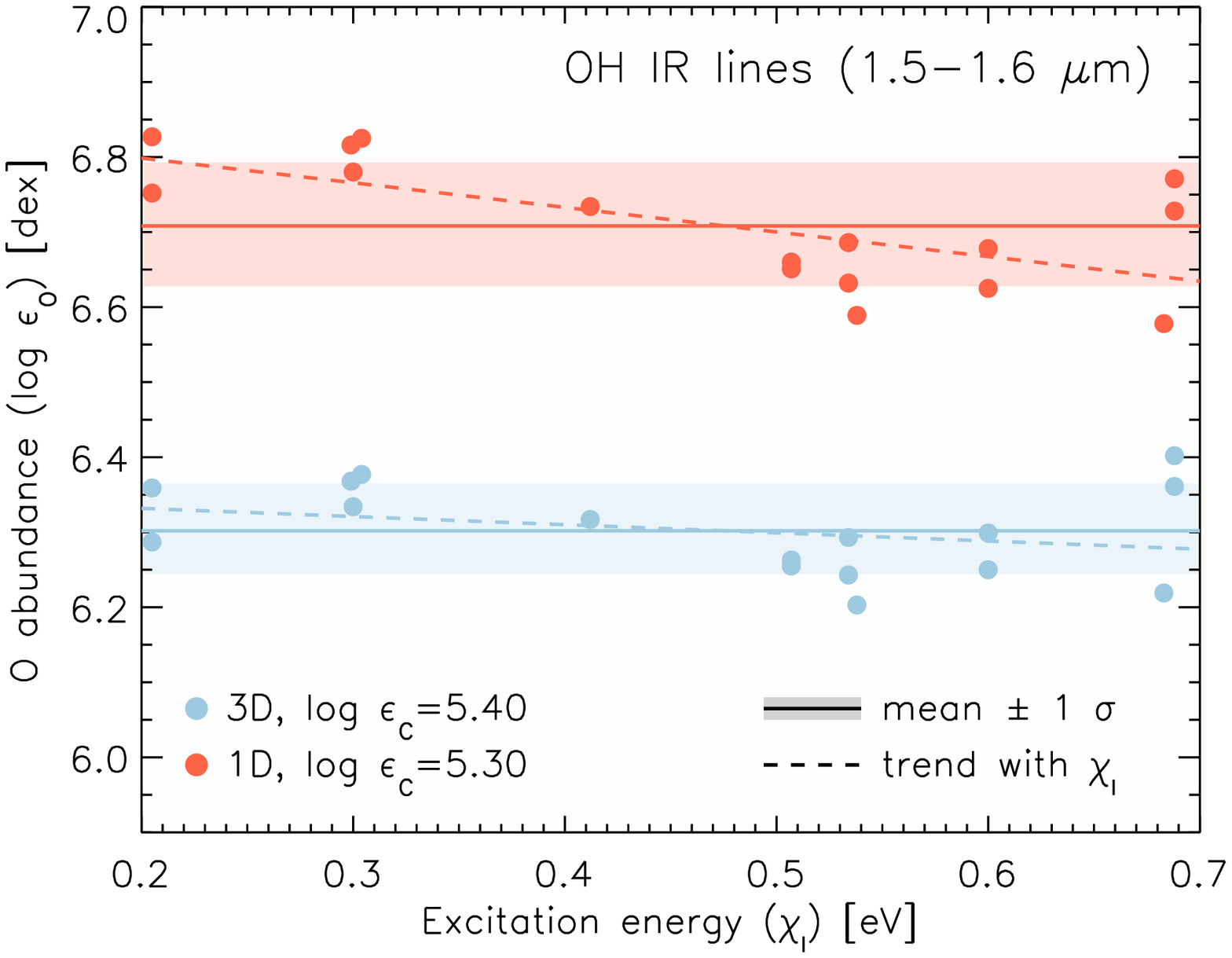}
   \caption{\emph{Filled circles}: oxygen abundances derived from individual OH IR lines as a function of lower-level excitation potential. \emph{Blue symbols}: results from the 3D analysis; \emph{red symbols}: results from the 1D analysis. \emph{Continuous lines}: average oxygen abundances. \emph{Dashed lines}: abundance trends with excitation potential. \emph{Coloured bands}: 1-$\sigma$ dispersion around the averages.}
   \label{fig:oh_ir_lines}
\end{figure}

We first determine the C and O abundances by minimising the $\chi^2$ from the simultaneous fit of the OH and CH bands in the ultraviolet (UV) and visible.
Figure~\ref{fig:ch_oh_fit} shows the $\chi^2$ contours as a function of C and O abundances for both the 3D and 1D analyses.
The $\chi^2$ minimisation yields values of $\abund{C} = 5.33 \pm 0.01$ and $\abund{O} = 6.35 \pm 0.03$ for the best fitting C and O abundances in the 3D case, and $\abund{C} = 5.28 \pm 0.01$ and $\abund{O} = 6.62 \pm 0.05$ in the 1D case.
From the $\chi^2$ contour plot, it is apparent that the derived C and O abundances are strongly correlated: low $\chi^2$ values are located along and near a diagonal line in the C-O abundances plane, in both the 3D and 1D cases, and a concurrent increase (or decrease) of the C and O abundances therefore produces a better fit than a random perturbation of similar magnitude of the two abundances.
Such correlation is due to the coupling between the chemical equilibrium number densities of atomic and molecular carbon and oxygen via CO formation.
With the C/O ratio being less than one in the atmosphere of the {\hdstar}, an increase in C abundance would to first order cause the additional carbon to be locked primarily in CO molecules and only to a lesser degree in CH molecules; the overall strength of CH lines would therefore increase, but only slightly.
At the same time, the formation of CO would also reduce the amount of oxygen available for forming other molecules. 
As a result, OH lines would become weaker, making it necessary to increase the O abundance in the spectral synthesis calculations in order to match their observed strength. 
The increase in O abundance would in turn induce further CO formation, subtracting carbon from the reservoir of particles available for CH formation, therefore weakening again the CH lines.
Quantitatively, a good fit to the OH and CH can be maintained with a relative O/C abundance variation of $\Delta \abund{O} /  \Delta \abund{C} \approx 5$ in 3D and $\Delta \abund{O} /  \Delta \abund{C} \approx 11$ in 1D.
Hence, in practice, the derived C abundance is significantly less sensitive to coupling via CO, and its determination more robust than the O abundance, both in 3D and 1D.

CO coupling and non-linearities in molecule formation also explain the apparently counter-intuitive result of a slightly positive 3D$-$1D carbon abundance correction determined from the fitting of the CH G band.
In the upper layers of a highly oxygen-rich atmosphere such as {\hdstar}'s (C/O $\la 0.1 \ll 1$), most of the carbon is locked in CO molecules and only a small fraction of it is atomic form or in CH molecules: this is due to CO having the highest dissociation energy among all molecules and radicals ($D_\text{CO}=11.108$\,eV compared with $D_\text{CH}=3.465$\,eV).
Because of such disparity between CO and CH dissociation energies, at the temperatures encountered in the upper layers of the 3D model, which are lower on average than the ones in the corresponding 1D model, molecular equilibrium favours the formation of considerably more CO at the expense of atomic carbon and CH molecules, weakening the absorption lines from the latter; as a consequence, in 3D, a higher C abundance than in the 1D case is necessary to fit the observed CH features.

In order to provide an additional constraint on the C and O abundances, we also fit the observed equivalent widths of individual OH lines in the infrared (IR).
In Fig.~\ref{fig:ch_oh_fit}, we have plotted the O abundances derived from such lines for different fixed values of the C abundance over the $\chi^2$ iso-contours from the fitting of the UV and visible bands.
The O abundances derived from individual OH IR features exhibit a line-to-line scatter of about $0.1$\,dex in 1D, significantly larger than the corresponding spread of $0.05$\,dex in 3D.
Figure~\ref{fig:oh_ir_lines} also indicates that a clear decreasing trend with lower-level excitation potential is present in the O abundances derived in 1D, while no significant trend is found in the 3D analysis.

Unlike OH features in the UV, OH IR lines appear to be virtually independent of the adopted value of the C abundance, yielding a value of $\abund{O} = 6.30 \pm 0.06$ in 3D and $\abund{O} = 6.71 \pm 0.09$ in 1D.
The lower sensitivity on C abundance compared with the OH UV band can be ascribed to the fact that OH IR lines form deeper into the stellar photosphere \citep[see also][]{Dobrovolskas:2015,Prakapavicius:2017}, that is over a different range of depths characterised by a lower degree of CO coupling.
We average the results of the analyses of the UV and visible bands and of the IR lines to find an O abundance of $\abund{O} = 6.33 \pm 0.07$ in 3D and $\abund{O} = 6.67 \pm 0.10$ in 1D. 
With the above constraints on the O abundance, we can then minimise the $\chi^2$ at fixed $\abund{O}$ to retrieve a value for the C abundance of $\abund{C} = 5.32$ in 3D and $\abund{C} = 5.29$ in 1D.
Our determination of the 3D oxygen abundance from molecular features is compatible with the results by \citet{Dobrovolskas:2015} for OH IR lines ($\abund{O,\,3D}=6.39 \pm 0.11$) and by \citet{Prakapavicius:2017} for OH UV lines ($\abund{O,\,3D}=6.23 \pm 0.13$).
For illustration purposes, in Fig.~\ref{fig:mol_bands}, upper panels, we show portions of the observed OH UV and CH G bands and of the 1D and 3D synthetic spectra computed for the best fitting C and O abundances.
Deviations of the synthetic spectra from observations are primarily due to occasional shortcomings with the modelling of atomic lines; the wavelength windows around problematic lines are however masked prior to the fitting in order to minimise systematic effects on the C and O abundance determinations.

Besides deriving C and O abundances from molecular lines, we have also determined the oxygen abundance from the analysis of the [\ion{O}{i}] line at $6300.3$\,{\AA}.
Using the 3D model, we infer a value of $\abund{O} = 6.79$, while with the corresponding 1D model we derive a value of $\abund{O} = 6.82$. 
While the 1D oxygen abundance derived from the [\ion{O}{i}] $6300.3$\,{\AA} differs by about $0.1$\,dex (approximately 1 $\sigma$) from that inferred from molecular lines, in 3D the deviation is more significant and non-negligible. 
A similar discrepancy between molecular and atomic oxygen abundance indicators in {\hdstar} has also been reported by \citet{Prakapavicius:2017}.
We discuss in more detail the possible causes for the difference between the oxygen abundace values returned by the various spectroscopic indicators in Sect.~\ref{sect:discussion}.

\subsection{Nitrogen abundance}
\label{sect:nitr}

\begin{figure*}
  \centering
   \resizebox{\hsize}{!}{
   \includegraphics{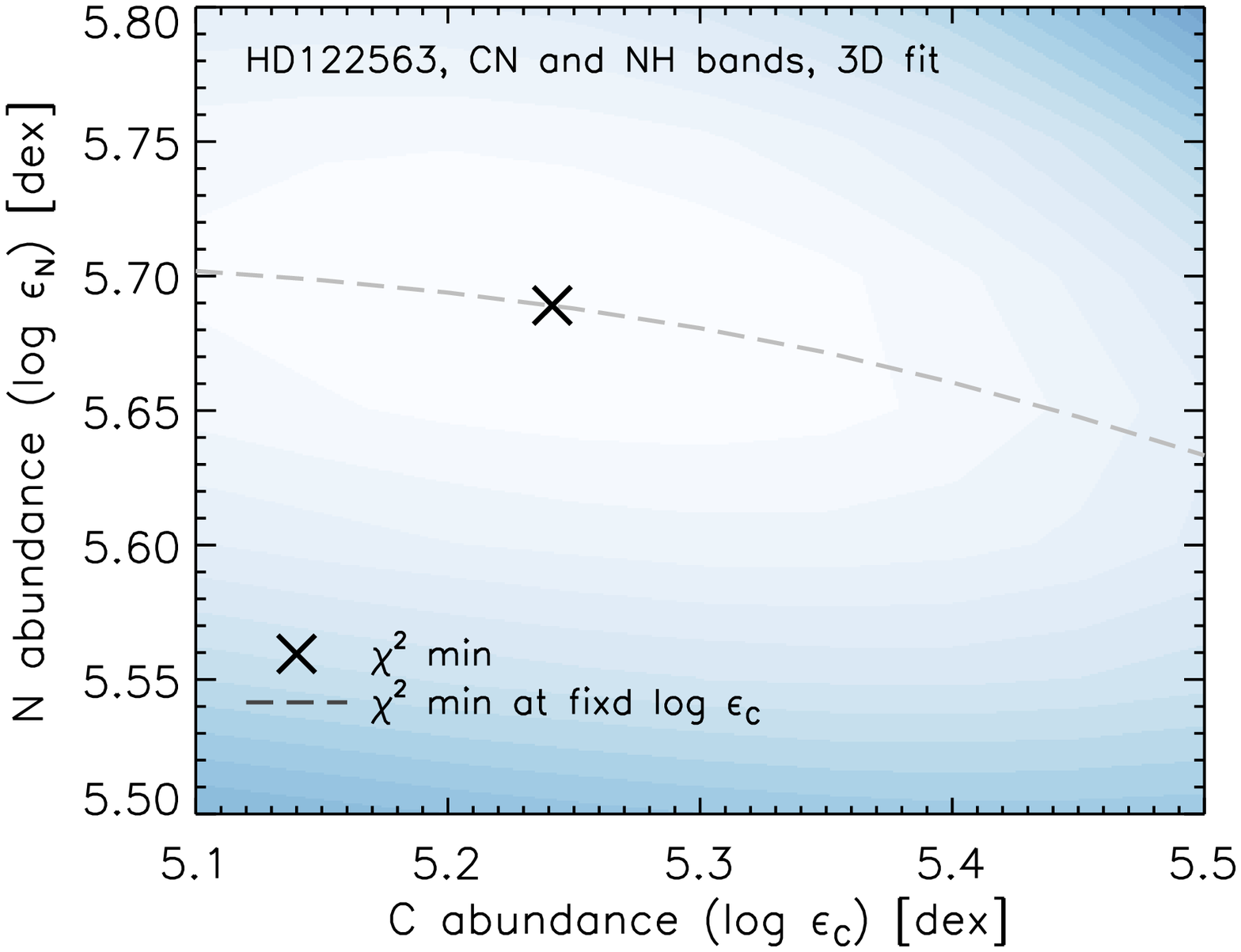}
   \includegraphics{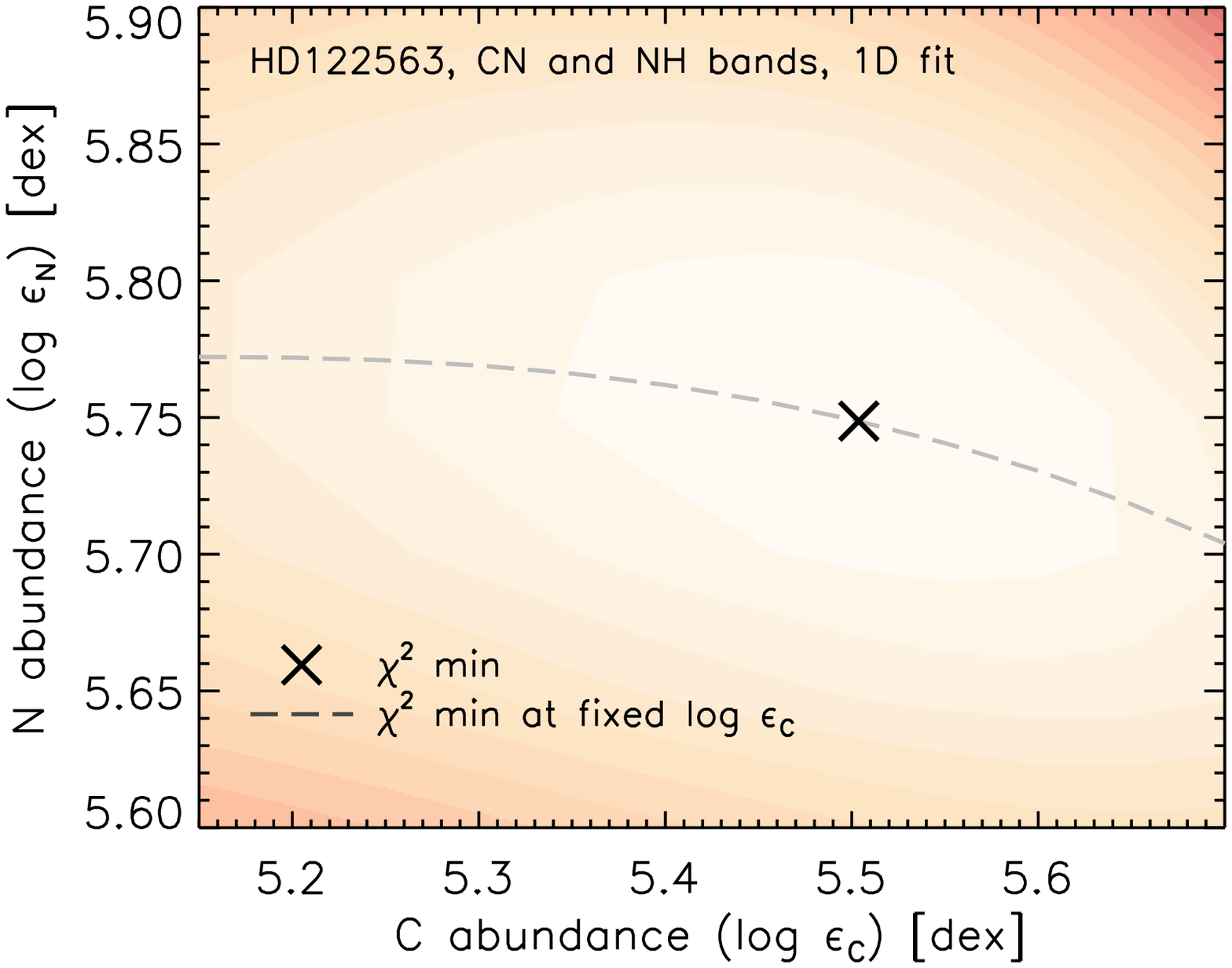} }
   \caption{Contours of constant $\chi^2$ from the simultaneous fitting of NH and CN bands in the UV and violet. \emph{Blue} contours refer to the 3D analysis (\emph{left panel}), \emph{red} ones to the 1D analysis (\emph{right panel}). \emph{Dotted lines} show the location of the $\chi^2$ minimum in the C-N abundance plane. \emph{Long-dashed lines} trace the loci of the $\chi^2$ minima at constant carbon abundance.}
   \label{fig:nh_cn_fit}
\end{figure*}

We determine the N abundance by minimising the $\chi^2$ from the simultaneous fit of the NH and CN bands in the UV and violet.
Figure~\ref{fig:nh_cn_fit} shows that the $\chi^2$-minimisation procedure returns well-defined values of the best-fitting N and C abundances: $\abund{N} = 5.69 \pm 0.01$ and $\abund{C} = 5.24 \pm 0.05$ for the 3D case and $\abund{N} = 5.75 \pm 0.02$ and $\abund{C} = 5.50 \pm 0.10$ for the 1D one.
The above values of the C abundance differ from the ones inferred from the analysis of the OH and CH bands by approximately $-0.1$\,dex and $+0.2$\,dex in 3D and 1D, respectively.
The CN band, however, is not as robust an indicator of C abundance as the CH lines in the UV and visible.
Looking at the contours of constant $\chi^2$ it is apparent that the goodness of the fit is primarily sensitive to the N abundance and only moderately dependent on the C abundance.
We have therefore chosen to fix the value of the latter using the results from Sect.~\ref{sect:carboxyg}; with this constraint, new $\chi^2$ minima are found for a N abundance of $\abund{N} = 5.68$ and $5.75$ for the 3D and 1D cases, respectively.

\section{Discussion}
\label{sect:discussion}
Three-dimensional hydrodynamic simulations of convection at the surface of a metal-poor red giant result in a cooler upper-photosphere temperature stratification compared with the one predicted by stationary 1D hydrostatic model atmospheres.
Under the assumption of local thermodynamic equilibrium (LTE), this implies higher atmospheric number densities of neutral atomic species and molecules in 3D models of {\hdstar} compared with their 1D counterparts and, consequently, lower values of the elemental abundances derived from those indicators.

Our reference 3D {\hdstar} surface convection simulation with 48 opacity bins is based on a considerably finer representation of atmospheric opacities than most 3D models used for spectral line formation and abundance analysis purposes until now.
In Sect.~\ref{sect:temp3d}, we have highlighted the main differences among simulations using different line opacity packages, variants of the opacity binning method, opacity binning representations, and numbers of opacity bins. 
We have shown that there is a systematic tendency for models using a small number of opacity bins (four) and at the same time no binning in wavelength to produce cooler upper photospheric layers and steeper temperature gradients near and immediately above the continuum-forming region.
In addition, significantly lowering the number of opacity bins results in a higher emergent radiative flux, meaning that that the temperature stratification of the model needs to be scaled down further in order to match the effective temperature of the reference 48-bin model. 
Consequently, 3D$-$1D corrections to elemental abundances derived from neutral atomic and molecular lines with the current reference 48-bin model are overall significantly smaller than previously determined with models adopting  very few bins grouping opacities based only on opacity strength \citep[e.g.][]{Collet:2007}.
Our calculations show on the other hand that it is possible to reproduce the main statistical properties of the 48-bin simulation sequence (e.g. mean stratification and standard deviation of physical variables such as temperature, density, and pressure as a function of optical depth) using only twelve opacity bins.
However, the ability to closely reproduce the results from the reference simulation depends on how the twelve opacity bins are selected. 
Our tests indicate that, with a limited number of bins, binning opacities primarily based on wavelength leads to a poorer agreement with the reference simulation compared with sorting opacities in terms of both strength and wavelength.
The resulting differences between the stratifications of the two 12-bin simulations should serve as an indication of the uncertainties to be expected when using the opacity binning approach with a limited number of bins.
Obviously, without carrying out more extensive tests, it is not possible to generalise this conclusion to other simulations at this stage, as the effective distribution of opacity strengths as a function of wavelength varies significantly depending on stellar parameters and metallicity.

\subsection{Oxygen abundance}
\label{sect:oxyg_abund}

\begin{figure}
  \centering
   \resizebox{\hsize}{!}{\includegraphics{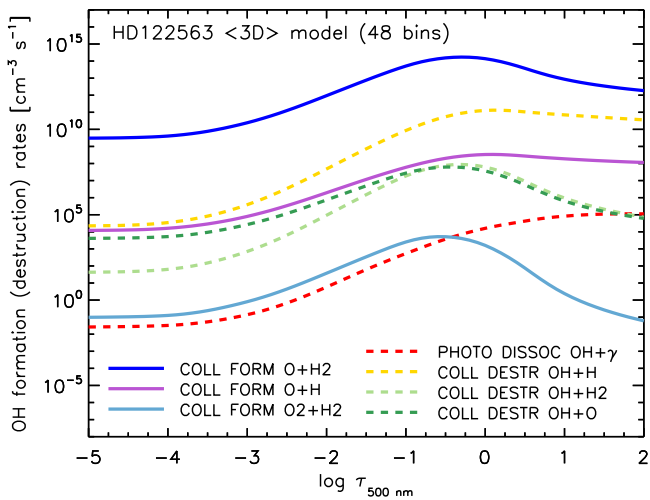}}
   \caption{OH formation and destruction rates for the main collisional and photodissociation processes as a function of optical depth in {\hdstar}'s mean 3D model atmosphere stratification.}
   \label{fig:photo_dissoc}
\end{figure}

\begin{table*}
\centering
\caption{Collisional OH formation and destruction processes considered for the present analysis: numerical parameters for the Arrhenius form of the rate coefficients and temperature range of validity. Collisional formation processes are listed in the upper part of the table, destruction processes in the lower. The data are taken from the UMIST Database for Astrochemistry (UDfA).}
\label{tab:coll-proc}
\begin{tabular}{lcccc}
	\hline
	Process  & $\alpha$ & $\beta$ & $\gamma$ & T range [K] \\
	\hline
	H$_2$ + O $\rightarrow$ OH + H \hspace{1cm}     & $3.14 \cdot 10^{-13}$ &  $2.7$  & $3\,150$ & $297$--$3\,532$ \\
	H + O     $\rightarrow$ OH + photon &  $9.9 \cdot 10^{-19}$ & $-0.38$ & $0$    & $0$--$300$ \\
	H + O$_2$ $\rightarrow$ OH + O      & $2.61 \cdot 10^{-10}$ &  $0.0$  & $8\,156$ & $0$--$4\,000$ \\
	H$_2$ + O$_2$ $\rightarrow$ 2 OH    & $3.16 \cdot 10^{-10}$ &  $0.0$  & $21\,890$ & $300$--$2\,500$ \\
	\hline
	OH + H     $\rightarrow$ O + 2 H       & $6.0 \cdot 10^{-9}$  & $0.0$  & $50\,900$ & $1\,696$--$41\,000$ \\
	OH + H$_2$ $\rightarrow$ O + H$_2$ + H & $6.0 \cdot 10^{-9}$  & $0.0$  & $50\,900$ & $1\,696$--$41\,000$ \\
	OH + O     $\rightarrow$ O$_2$ + H     & $3.69 \cdot 10^{-11}$ & $-0.27$ & $12.9$ & $10$--$500$ \\
	\hline
\end{tabular}
\end{table*}

Our results show that we can achieve a very satisfactory agreement between synthetic and observed spectra and a consistent determination of the C, N, and O abundances through the simultaneous fit of OH, CH, NH and CN bands, especially with 3D modelling of spectral lines.
With regards to oxygen, while we have found excellent agreement between the O abundances inferred from OH UV and OH IR lines, the analysis of the [\ion{O}{i}]~$6\,300.3$~{\AA} line yields a significantly higher value, particularly in 3D.
The discrepancy between the O abundances derived from OH features and the [\ion{O}{i}]~$6\,300.3$~{\AA} line cannot simply be explained by the well-known blend with a \ion{Ni}{i} line \citep[e.g.][]{Allende-Prieto:2001}, which we have included in the analysis and which we find to be effectively unimportant in the case of {\hdstar}.
One possible explanation is that the adopted value of the surface gravity of {\hdstar} may be overestimated because of systematic errors on the stellar parallax or other fundamental stellar parameters. 

To test this hypothesis, we have recomputed molecular lines from the OH UV and CH G bands as well as the [\ion{O}{i}] line using our two scaled 3D models with lower surface gravity ($\log g = 1.1$\,{\cmss}) and with higher effective temperature ($\teff \approx 4\,665$\,K), respectively.
We find that in order to fit the observed line strengths with the scaled 3D model with $\log g = 1.1$\,({\cmss}), we need to increase the C and O abundance by about $0.07$\,dex and $0.12$\,dex, respectively, for molecular lines and decrease the O abundance by about $0.18$\,dex for the [\ion{O}{i}] line.
This translates into a reduction of the 3D O abundance difference between the atomic and molecular indicators  to about $0.16$\,dex.
Increasing instead the effective temperature of the 3D reference model by ${\sim}90$\,K results in higher C and O abundances from molecular indicators by about $0.21$\,dex and $0.26$\,dex, respectively, as well as in a $0.06$\,dex higher O abundance from the [\ion{O}{i}] line. This corresponds to a reduced 3D O abundance difference between molecular and atomic indicators of about $0.26$\,dex, which is still significant.
We caution, however, that these results are based on non-fully relaxed 3D model atmospheres and that a more accurate  calculation of the response of C and O abundances to changes in stellar parameters would require a more in-depth analysis.
The results of the study of the C and O abundance sensitivity to changes in the 3D model's effective temperature and surface gravity are summarised in Table\,\ref{tab:abund-params}.
For reference, we have also carried out an analysis of the dependence of the derived C and O abundances on the above parameters using 1D model stellar atmospheres. The results are qualitatively similar to the ones for the 3D case, but the response of the abundances to changes in both ${\teff}$ and $\log g$ is generally weaker.
However, the O abundance difference between [\ion{O}{i}] and OH lines inferred with the 1D reference model is only $0.15$\,dex and lowering the surface gravity by ${\sim}0.3$\,dex would be sufficient to bring the abundances determined from the two indicators into agreement in 1D.

Another possible explanation for the discrepancy between molecular and atomic O abundance indicators may be related to the neglect of molecular photodissociation processes in the spectral synthesis calculations. 
Efficient photodissociation of OH molecules would imply intrinsically weaker OH lines and therefore result in a higher value of the O abundance inferred from these indicator.
In order to test this hypothesis, we have compared the rates of OH photodissociation and of OH formation and destruction by various important collisional processes, which, for simplicity, we estimated for the mean stratification of {\hdstar}'s 3D model atmosphere.
With regards to the OH photodissociation rates, we considered only the contribution of the UV transition at $\lambda \!=\! 1\,570$~{\AA} from the ground state $X^2\,\Pi$ to the repulsive state $1^2\,\Sigma^-$ of the molecule \citep{van-Dishoeck:1983};
other photodissociation channels are also possible, but, according to our tests, this radiative transition is the dominant one at the physical conditions in {\hdstar}'s atmosphere.
The photodissociation rates per OH particle in the ground state for the above channel were computed by evaluating the integral
\begin{equation}
R_\text{pd,OH} = 4 \pi \, 
\int \frac{\sigma^\text{pd}_\nu}{h\nu} \, J_\nu \, \diff \nu
\end{equation}
where $\sigma^\text{pd}_\nu$ is the photodissociation cross-section for the transition as a function of photon frequency $\nu$ and $J_\nu$ is the mean radiation field.
We approximated $\sigma^\text{pd}_\nu$ with a Gaussian profile centred at $h\nu \!=\! 7.9$~eV ($\lambda \!=\! 1\,570$~{\AA}), with a FWHM of $1.0$~eV and an amplitude at maximum of $3.3 \, \cdot 10^{-18}$~cm$^2$ 
\citep{van-Dishoeck:1983}.
For the collisional OH formation and destruction processes, we assumed two-body reactions of the kind $A + B \rightarrow C$ where the reactants can be atoms or molecules.
We estimated the rates of change of number density of the reaction products C in the atmospheric layers of {\hdstar}'s 3D model's mean stratification using the expression
\begin{equation}
\frac{\diff n_\text{C}}{\diff t} = k \, n_\text{A} \, n_\text{B}.
\label{eq:rate}
\end{equation}
The rate coefficients $k$ are parametrised and calculated using the so-called \emph{Arrhenius} form
\begin{equation}
k \, [\text{cm}^3\,\text{s}^{-1}] = \alpha \left( \frac{T\,\text{[K]} }{300} \right)^\beta \, \exp (-\gamma \, / \, T\,\text{[K]})
\label{eq:arrhenius}
\end{equation}
with coefficients $\alpha$, $\beta$, and $\gamma$ taken from the UMIST Database for Astrochemistry \citep[UDfA,\footnote{\url{http://udfa.net}}][]{McElroy:2013}.
The list of collisional formation and destruction processes considered for the present analysis and the corresponding $\alpha$, $\beta$, and $\gamma$ coefficients is given in Table~\ref{tab:coll-proc}.
The table also provides the temperature ranges recommended by UDfA over which the parametrisation is valid. 
In several cases, the nominal range of validity is limited to rather low temperatures;
nevertheless, for lack of better alternatives, we elected to apply the parametrisation of the rates of all collisional processes to the entire mean atmospheric stratification.
In any case, the range of validity for the dominant collisional formation and destruction processes extends over the whole upper atmosphere, i.e. over the layers that are most relevant for OH molecule formation.

Figure~\ref{fig:photo_dissoc} shows the comparison of the rates of OH photodissociation and collisional formation and destruction for the processes that we have included in our analysis.
It is apparent that collisional formation and destruction rates dominate over photodissociations at all heights in the atmosphere.
In particular, the rates of collisional formation through the H$_2$ + O $\rightarrow$ OH + H channel are several orders of magnitude higher than the main OH photodissociation channel.
This result suggests that photodissociation processes have a negligible impact on OH molecular number densities and may be safely disregarded in the present analysis.
However, we caution that this conclusion is based on a limited and by no means complete analysis, assuming a 1D-like stellar atmosphere stratification:
a more comprehensive study of the role of photodissociations and collisional formation and destruction processes with the a full 3D model would therefore be necessary to verify our finding.

As a final possibility, the discrepancy between molecular and atomic oxygen abundance indicators may be ascribed to departures of OH line formation from LTE.
\citet{Asplund:2001} carried out calculations to estimate non-LTE effects on the formation of OH UV lines with 1D model atmospheres of dwarfs and turn-off stars using a two-level approximation with complete redistribution for the source function as proposed by \citet{Hinkle:1975}.
They found indication of possible radiatively driven departures from LTE at low metallicity that would lead to decreased line strengths of the OH UV lines, hence larger derived oxygen abundances.
However, as the effective temperature and surface gravity of {\hdstar} are outside the stellar parameter range considered in their work, the magnitude of non-LTE effects on OH UV lines and the derived oxygen abundance in the metal-poor giant may be significantly different, so their results are not directly applicable here.
Also, we note that, in order to resolve the above discrepancy, such non-LTE effects would need to
be of the same order of magnitude for both OH UV and OH IR bands, i.e. for different systems of energy levels of the OH molecule.
A detailed study of non-LTE effects on molecular line formation goes beyond the scope of the present paper and we defer it to a future work.

\section{Conclusions}
\label{sect:conclusions}
We have presented an extensive abundance analysis of atomic lines and molecular features in the spectrum of the low-metallicity red giant star {\hdstar} based on high-resolution state-of-the-art time-dependent 3D hydrodynamic model stellar atmospheres including non-grey radiative transfer through opacity binning.
We have explored the effects of different opacity binning configurations with varying number of opacity bins on the resulting atmospheric temperature stratification of the 3D simulations.
In line with previous surface convection simulations of metal-poor late-type stars, our 3D model atmospheres of {\hdstar} exhibit cooler mean stratifications than corresponding stationary 1D hydrostatic models constructed for the same stellar parameters.
This is due to the different heating and cooling mechanisms effectively at play in the outer layers of the two kinds of models: cooling following adiabatic expansion of gas above granules in the 3D simulations lowers the equilibrium upper photospheric temperature compared with classical stationary 1D hydrostatic models where this mechanism is completely absent.
In particular, our reference 3D simulation with 48 opacity is on average $\sim 300 - 500$\,K cooler in the upper photospheric layers ($-5 \la \log\tau_\text{Ross} \la 0$) than its 1D counterpart.
This temperature difference is significant yet lower by a few hundred K than determined with previous simulations \citep[e.g.][]{Collet:2009}.
Also, the temperature gradient with optical depth near the optical surface is slightly shallower in the current  48-bin simulation than in previous ones.
Using four-bin simulations, we are able to fully explain such differences in the temperature stratification through a combination of changes in opacity binning method, line opacity source data (opacity sampling or opacity distribution functions), and adopted chemical mixture.

The cooler temperature stratification of the reference 3D model atmospheres of {\hdstar} compared with its 1D counterpart imply that a 3D-based local thermodynamic equilibrium (LTE) analysis of spectral lines from neutral and singly ionised atoms generally yields lower and higher abundances, respectively, than the corresponding 1D analysis.
From the LTE analysis of \ion{Fe}{i} lines, in particular, we determine an iron abundance of $\abund{Fe} = 4.67 \pm 0.12$ in 3D and $4.75 \pm 0.17$ in 1D. 
From \ion{Fe}{ii} lines, we determine instead $\abund{Fe} = 5.06 \pm 0.08$ with the 3D analysis and $4.94 \pm 0.07$ with the 1D one.
The 3D LTE analysis results in a generally lower line-to-line scatter and a significantly shallower gradient for the distribution of derived line-by-line Fe abundance with respect to lower excitation potential compared with 1D. 
Yet, the relatively large difference (${\sim}0.4$\,dex) between the Fe abundance determined from \ion{Fe}{i} and \ion{Fe}{ii} lines suggests that departures from LTE may be important, as also pointed out by \citet{Amarsi:2016a}.
Alternatively, the difference between Fe abundance determinations from \ion{Fe}{i} and \ion{Fe}{ii} lines may also be ascribed to some extent to uncertainties in the fundamental stellar parameters of {\hdstar}, especially surface gravity, whose estimated value may suffer from systematic errors on the stellar parallax.

We have also explored the impact of our reference 3D model atmosphere of {\hdstar} on the formation of molecular bands and lines. 
In particular, we have determined the abundances of carbon, nitrogen, and oxygen by simultaneously fitting OH, CH, NH, and CN molecular bands and lines synthesised with the 3D and 1D models to observed features in the ultraviolet (UV), visible, and infrared (IR) portions of {\hdstar}'s spectrum.
In both the 3D and 1D case, we are able to get a consistent solution for the best fitting C, N, and O abundances across all features.
The 3D solution is more precise in this respect, even if only marginally so, with a better agreement between the O abundance determination from OH features in the UV and IR, a more consistent C abundance determination between CH and CN features, and no trend of line-by-line O abundance with excitation potential in the case of individual OH IR lines.
Similarly to the case of neutral atomic lines, the cooler temperature stratification of the 3D model atmosphere favours molecule formation, generally resulting in the prediction of intrinsically stronger molecular features than with the 1D model at a given composition, hence implying lower values of the abundances of trace elements determined from such features.
Indeed, from the analysis of molecular bands and lines, we have determined negative 3D$-$1D abundance corrections for nitrogen ($\abund{N,\,3D} = 5.68$ and $\abund{N,\,1D} = 5.75$) and oxygen ($\abund{O,\,3D} = 6.33$ and $\abund{O,\,1D} = 6.67$). 
However, we find the 3D$-$1D abundance correction for carbon to be positive although only slightly so ($\abund{C,\,3D} = 5.32$ and $\abund{C,\,1D} = 5.29$).
We explain this result through non-linear effects in molecule formation and coupling of the equilibrium number densities of atomic and molecular carbon and oxygen through CO formation in an oxygen-rich atmosphere (C/O $\la 0.1 \ll 1$).

Finally, we have also determined the oxygen abundance from the analysis of the [\ion{O}{i}] $6\,300.3$\,{\AA} line and found it to be significantly higher than the one determined from the analysis of molecular features, namely $\abund{O,\,3D} = 6.79$ and $\abund{O,\,1D} = 6.82$ for the atomic line. 
We have considered various possible causes of the discrepancy, which has also been reported by \citet{Prakapavicius:2017}, including systematic uncertainties on {\hdstar}'s surface gravity and effects of molecular photodissociation processes on OH formation and on the predicted strength of OH lines.
While we are unable to provide a definitive explanation for the discrepancy at the present stage, we rule out the importance of OH photodissociation for the O abundance determination from molecular features, and note that, according to preliminary test calculations, lowering the surface gravity from to $\log g = 1.6$ to $1.1$ ({\cmss}) would reduce the discrepancy between the oxygen abundance values determined from molecular and atomic indicators to about $0.16$~dex.

\section*{Acknowledgments}
RC acknowledges partial support from the Australian Research Council (ARC) through a Discovery Early Career Researcher Award (DECRA) grant (project DE120102940).
Funding for the Stellar Astrophysics Centre is provided by The Danish National Research Foundation (Grant DNRF106).
MA gratefully acknowledges funding through ARC Laureate Fellowship FL110100012.
RT acknowledges funding from NASA grant NNX15AB24G.
This research was undertaken with the assistance of resources provided at the NCI National Facility systems at the Australian National University through the National Computational Merit Allocation Scheme supported by the Australian Government.
Finally, the authors would like to thank the anonymous referee for the constructive feedback, which helped improve the manuscript.



\bibliographystyle{mnras}
\input{hd122563.bbl}





\bsp	
\label{lastpage}
\end{document}